\documentclass[longauth,useAMS,usenatbib]{aa}  

\usepackage{graphicx}
\usepackage{txfonts}
\usepackage{hyperref}
\usepackage{longtable}
\usepackage{appendix}
\usepackage{xcolor}

\begin{document}

\title{Diving into the planetary system of Proxima with NIRPS\thanks{Based (in part) on guaranteed time observations collected at the European Southern Observatory under ESO programme 1102.C-0744 by the NIRPS Consortium.} \thanks{The data used in this paper is available in electronic form at the CDS via anonymous ftp to cdsarc.u-strasbg.fr (130.79.128.5) or via http://cdsweb.u-strasbg.fr/cgi-bin/qcat?J/A+A/}}

\subtitle{Breaking the meter per second barrier in the infrared}

\author{
Alejandro Su\'arez Mascare\~no\inst{1,2,*},
\'Etienne Artigau\inst{3,4},
Lucile Mignon\inst{5,6},
Xavier Delfosse\inst{6},
Neil J. Cook\inst{3},
Fran\c{c}ois Bouchy\inst{5},
Ren\'e Doyon\inst{3,4},
Jonay I. Gonz\'alez Hern\'andez\inst{1,2},
Thomas Vandal\inst{3},
Izan de Castro Le\~ao\inst{7}
% ...
}

\authorrunning{A.~Su{\'a}rez Mascare{\~n}o et al.}
    
\institute{
\inst{1}Instituto de Astrof\'isica de Canarias (IAC), Calle V\'ia L\'actea s/n, 38205 La Laguna, Tenerife, Spain\\
\inst{2}Departamento de Astrof\'isica, Universidad de La Laguna (ULL), 38206 La Laguna, Tenerife, Spain\\
\inst{3}Institut Trottier de recherche sur les exoplan\`etes, D\'epartement de Physique, Universit\'e de Montr\'eal, Montr\'eal, Qu\'ebec, Canada\\
\inst{4}Observatoire du Mont-M\'egantic, Qu\'ebec, Canada\\
\inst{5}Observatoire de Gen\`eve, D\'epartement d’Astronomie, Universit\'e de Gen\`eve, Chemin Pegasi 51, 1290 Versoix, Switzerland\\
\inst{6}Univ. Grenoble Alpes, CNRS, IPAG, F-38000 Grenoble, France\\
\inst{7}Departamento de F\'isica Te\'orica e Experimental, Universidade Federal do Rio Grande do Norte, Campus Universit\'ario, Natal, RN, 59072-970, Brazil\\
\inst{8}Instituto de Astrof\'isica e Ci\^encias do Espa\c{c}o, Universidade do Porto, CAUP, Rua das Estrelas, 4150-762 Porto, Portugal\\
\inst{9}Departamento de F\'isica e Astronomia, Faculdade de Ci\^encias, Universidade do Porto, Rua do Campo Alegre, 4169-007 Porto, Portugal\\
\inst{10}Department of Physics, University of Toronto, Toronto, ON M5S 3H4, Canada\\
\inst{11}Department of Physics \& Astronomy, McMaster University, 1280 Main St W, Hamilton, ON, L8S 4L8, Canada\\
\inst{12}Department of Physics, McGill University, 3600 rue University, Montr\'eal, QC, H3A 2T8, Canada\\
\inst{13}Department of Earth \& Planetary Sciences, McGill University, 3450 rue University, Montr\'eal, QC, H3A 0E8, Canada\\
\inst{14}Departamento de F\'isica, Universidade Federal do Cear\'a, Caixa Postal 6030, Campus do Pici, Fortaleza, Brazil\\
\inst{15}Centro de Astrobiolog\'ia (CAB), CSIC-INTA, ESAC campus, Camino Bajo del Castillo s/n, 28692, Villanueva de la Ca\~nada (Madrid), Spain\\
\inst{16}Centre Vie dans l’Univers, Facult\'e des sciences de l’Universit\'e de Gen\`eve, Quai Ernest-Ansermet 30, 1205 Geneva, Switzerland\\
\inst{17}European Southern Observatory (ESO), Karl-Schwarzschild-Str. 2, 85748 Garching bei M\"unchen, Germany\\
\inst{18}Space Research and Planetary Sciences, Physics Institute, University of Bern, Gesellschaftsstrasse 6, 3012 Bern, Switzerland\\
\inst{19}Consejo Superior de Investigaciones Cient\'ificas (CSIC), E-28006 Madrid, Spain\\
\inst{20}Bishop's Univeristy, Dept of Physics and Astronomy, Johnson-104E, 2600 College Street, Sherbrooke, QC, Canada, J1M 1Z7\\
\inst{21}Department of Physics and Space Science, Royal Military College of Canada, PO Box 17000, Station Forces, Kingston, ON, Canada\\
\inst{22}Instituto de Astrof\'isica e Ci\^encias do Espa\c{c}o, Faculdade de Ci\^encias da Universidade de Lisboa, Campo Grande, 1749-016 Lisboa, Portugal\\
\inst{23}Departamento de F\'isica da Faculdade de Ci\^encias da Universidade de Lisboa, Edif\'icio C8, 1749-016 Lisboa, Portugal\\
\inst{24}Centre of Optics, Photonics and Lasers, Universit\'e Laval, Qu\'ebec, Canada\\
\inst{25}Herzberg Astronomy and Astrophysics Research Centre, National Research Council of Canada\\
\inst{26}Aix Marseille Univ, CNRS, CNES, LAM, Marseille, France\\
\inst{27}Center for Space and Habitability, University of Bern, Gesellschaftsstrasse 6, 3012 Bern, Switzerland\\
\inst{28}European Southern Observatory (ESO), Av. Alonso de Cordova 3107,  Casilla 19001, Santiago de Chile, Chile\\
\inst{29}Plan\'etarium de Montr\'eal, Espace pour la Vie, 4801 av. Pierre-de Coubertin, Montr\'eal, Qu\'ebec, Canada\\
\inst{30}Lund Observatory, Division of Astrophysics, Department of Physics, Lund University, Box 118, 221 00 Lund, Sweden\\
\inst{31}York University, 4700 Keele St, North York, ON M3J 1P3\\
\inst{32}University of British Columbia, 2329 West Mall, Vancouver, BC, Canada, V6T 1Z4\\
\inst{33}Western University, Department of Physics \& Astronomy and Institute for Earth and Space Exploration, 1151 Richmond Street, London, ON N6A 3K7, Canada\\
\inst{34}Light Bridges S.L., Observatorio del Teide, Carretera del Observatorio, s/n Guimar, 38500, Tenerife, Canarias, Spain\\
\inst{35}University Observatory, Faculty of Physics, Ludwig-Maximilians-Universit\"at M\"unchen, Scheinerstr. 1, 81679 Munich, Germany\\
\inst{36}Hamburger Sternwarte, Gojenbergsweg 112, D-21029 Hamburg, Germany\\
\inst{37}Subaru Telescope, National Astronomical Observatory of Japan (NAOJ), 650 N Aohoku Place, Hilo, HI 96720, USA\\
\inst{38}Department of Astronomy \& Astrophysics, University of Chicago, 5640 South Ellis Avenue, Chicago, IL 60637, USA\\
\inst{39}Laboratoire Lagrange, Observatoire de la C\^ote d’Azur, CNRS, Universit\'e C\^ote d’Azur, Nice, France\\
\inst{*}\email{asm@iac.es}
}

\date{Written August 2024 - January 2025}
 
\abstract
   {We obtained 420 high-resolution spectra of Proxima, over 159 nights, using the Near Infra Red Planet Searcher (NIRPS). We derived 149 nightly-binned radial velocity measurements with a standard deviation of 1.69 m$\cdot$s$^{-1}$ and a median uncertainty of 55 cm$\cdot$s$^{-1}$, and performed a joint analysis combining radial velocities, spectroscopic activity indicators, and ground-based photometry, to model the planetary and stellar signals present in the data, applying multidimensional Gaussian process regression to model the activity signals. 
   
   We detect the radial velocity signal of Proxima b in the NIRPS data. All planetary characteristics are consistent with those previously derived using visible light spectrographs. In addition, we find evidence of the presence of the sub-Earth Proxima d in the NIRPS data. When combining the data with the HARPS observations taken simultaneous to NIRPS, we obtain a tentative detection of Proxima d and parameters consistent with those measured  with ESPRESSO. By combining the NIRPS data with simultaneously obtained HARPS observations and archival data, we confirm the existence of Proxima d, and demonstrate that its parameters are stable over time and against change of instrument. We refine the planetary parameters of Proxima b and d, and find inconclusive evidence of the signal attributed to Proxima c (P = 1900 d) being present in the data. We measure Proxima b and d to have minimum masses of 1.055 $\pm$ 0.055 M$_{\oplus}$, and 0.260 $\pm$ 0.038 M$_{\oplus}$, respectively. 

   Our results show that, in the case of Proxima, NIRPS provides more precise radial velocity data than HARPS, and a more significant detection of the planetary signals. The standard deviation of the residuals of NIRPS after the fit is $\sim$ 80 cm$\cdot$s$^{-1}$, showcasing the potential of NIRPS to measure precise radial velocities in the near-infrared.}


   \maketitle

\section{Introduction}

The discovery and characterization of exoplanets orbiting nearby stars has revolutionised our understanding of planetary systems and their potential habitability. Proxima, an M5.5 dwarf star located at a distance of 1.3 parsec \citep{Innes1915}, is the closest stellar neighbor to the Sun and has become a prime target for exoplanet studies. The system hosts a confirmed terrestrial-mass planet within its habitable zone, Proxima b \citep{AngladaEscude2016}, along with two additional planet candidates, Proxima c and Proxima d \citep{Damasso2020,Masca2020,Faria2022}. Thanks to its proximity, the Proxima system offers a unique opportunity to characterise the atmosphere of its habitable zone planet using reflected light, or thermal emission, with future facilities such as RISTRETTO \mbox{\citep{Lovis2017,Blind2022,Blind2024}}, the ArmazoNes high Dispersion Echelle Spectrograph (ANDES, \citealt{Marconi2022}), for the Extremely Large Telescope (ELT), the Large Interferometer for Exoplanets (LIFE, \citealt{Quanz2022}), or the Planetary Camera and Spectrograph (PCS, \citealt{Kasper2021}), for the ELT, and to study planetary formation and dynamics in the context of low-mass stars.

While the radial velocity (RV) method has proven successful in detecting exoplanets, its application to M dwarfs poses significant challenges. Most high-precision radial velocity spectrographs operate in visible light (VIS). However, M dwarfs emit most of their energy in the near-infrared (NIR). This limits the signal-to-noise ratio (S/N), and therefore the RV precision, achievable by medium-sized telescopes (up to 4m), which restricts the observations to the closest and brightest stars. While it is possible to obtain RV time series with 8-meter class telescopes (e.g., ESPRESSO, \citealt{Pepe2021}), the pressure on them makes it challenging to obtain the appropriate cadence needed to characterise the planetary and stellar signals in many stars. Even with large telescopes, it is still very difficult to obtain accurate measurements for lower-mass M dwarfs, and Ultra Cool Dwarfs (UCDs) remain out of reach because their VIS flux is too low. In addition, the stellar activity of M dwarfs typically introduces signals in the RV data that can mimic planetary signatures, sometimes coinciding with the periods of putative planets orbiting within the habitable zone of the star (e.g., \citealt{Queloz2001, Huelamo2008,Figueira2010, Robertson2014,SuarezMascareno2017}). 

Observing M dwarfs in the NIR to obtain RV data has significant advantages. Their increased brightness allows the use of smaller telescopes without sacrificing RV precision, and lets us reach spectral types near the hydrogen burning limit. In addition, the amplitudes of flux-dominated activity-induced signals and noise from magnetic phenomena, such as starspots and flares, are reduced relative to optical wavelengths \citep{Reiners2018, Carmona2023}. However, it might be that some stars show larger jitter in the NIR, due to Zeeman splitting \citep{Reiners2018}. By leveraging high-resolution NIR spectroscopy, we can improve the precision of RV measurements of M dwarfs and thus enhance our sensitivity to planetary signals.

In this work, we analyse RV data obtained from the Near Infrared Planet Searcher (NIRPS) spectrograph, a state-of-the-art $YJH$-band instrument designed for detecting exoplanets orbiting M dwarfs (\citealt{Wildi2022,Bouchy2025}). NIRPS, located at the 3.6-m ESO telescope at the La Silla observatory, has been built to be operated simultaneously with the High Accuracy Radial Velocity Planet Searcher (HARPS,  \citealt{Mayor2003}). The primary goals of this study are: (i) to demonstrate the potential of NIRPS to obtain high precision RV measurements in the infrared ($YJH$ bands); (ii) to highlight the advantages of using NIR data in the search for exoplanets around M dwarfs; (iii) to refine the orbital parameters of Proxima b; (iv) to assess the planetary nature of Proxima c and Proxima d. For the last point, we combine the NIRPS dataset with the complete historical archival dataset. 

This paper is organised as follows. Section ~\ref{obs_data} describes the observations, data reduction, and RV extraction. In Section ~\ref{prop_proxima}, we present the current knowledge about the Proxima system. Section ~\ref{sec_analysis} details our RV analysis methods and the techniques used to model stellar activity. Section ~\ref{sec_results} presents our findings on the orbital characteristics of Proxima's planets, and Section ~\ref{sec_disc} discusses the implications of these results.

\section{Observations and data} \label{obs_data}

\subsection{NIRPS observations}

\begin{figure*}[!ht]
    \centering
	\includegraphics[width=16cm]{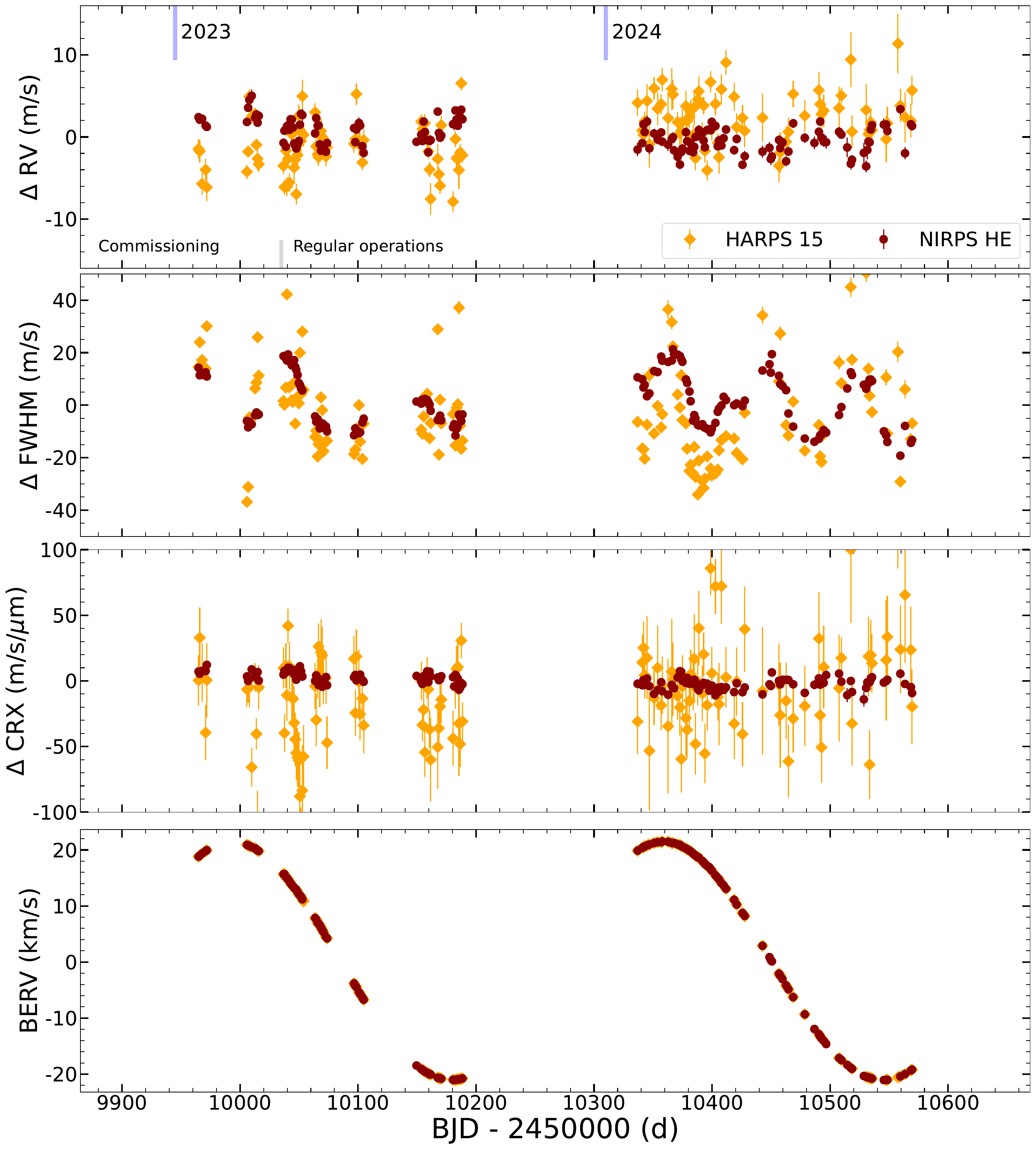}
	\caption{\textbf{NIRPS GTO data.} The upper panel shows the nightly-binned RV data of NIRPS and HARPS, obtained by the NIRPS GTO. The upper-middle panel shows the time series of FWHM of the same spectra. The lower-middle panel shows the CRX data. The lower panel shows the BERV at which the measurements were taken. All data has its median value subtracted.}
	\label{data_nirps}
\end{figure*}

NIRPS (\citealt{Wildi2022, Artigau2024, Bouchy2025}) is a pressure- and temperature-stabilised fibre-fed high resolution echelle spectrograph installed at the 3.6 m ESO telescope in La Silla Observatory, Chile. It uses a VIS-NIR dichroic splitter to split the telescope beam between NIRPS and HARPS. The visible light goes straight into HARPS, enabling simultaneous observations. It has two modes of observation, with different resolutions and efficiencies. The High Accuracy (HA) mode has a resolving power of $R\sim 90\,000$, using a 0.4$\arcsec$ octagonal fibre. The High Efficiency (HE) mode has a resolving power of $R\sim 75\,000$, using a 0.9$\arcsec$ octagonal fibre, which is sliced in two halves at the pupil and fed into a rectangular fibre. The spectrum is projected to a \hbox{4096 $\times$ 4096} pixels Hawaii-4RG detector.

We obtained 420 NIRPS spectra, taken over 159 nights, as part of the NIRPS commissioning program (ID 60.A-9109), and the Guaranteed Time Observations program (IDs 111.251P.001, 112.25NZ.001/002/003; PI: F. Bouchy). We used the HE mode, which provides the most precise RV measurements due to lower modal noise, and a typical observing strategy of 2-6 exposures of 200 seconds each (depending on operational constraints) per visit. The data were reduced using two data reduction softwares (DRS). The NIRPS-DRS, adapted from the publicly available ESPRESSO DRS \citep{Pepe2021}, and the APERO pipeline \citep{Cook2022}, originally developed for the NIR spectograph SPIRou \citep{Donati2020}. In this work, we used the radial velocities obtained using spectra reduced with the APERO pipeline, as it provided data with better precision.

\subsection{HARPS observations}

Simultaneous to the NIRPS observations, we obtained 292 HARPS spectra. HARPS is a fibre-fed high resolution echelle spectrograph with a resolving power of $R\sim 115\,000$ over a spectral range from $\sim$380 to $\sim$690 nm and has been designed to attain very high long-term RV precision. It is contained in temperature- and pressure-controlled vacuum vessels to avoid spectral drifts due to temperature and air pressure variations, thus  ensuring its stability. HARPS is equipped with its own pipeline providing extracted and wavelength-calibrated spectra, as well as RV measurements and other data products such as cross-correlation functions and their bisector profiles. Observations were taken with a typical exposure time of 400 s, and 1--3 exposures per visit, matching the NIRPS total exposure time per visit. It is important to mention, that the observing strategy and exposure times where adjusted for NIRPS observations, which resulted in lower signal-to-noise ratio HARPS data, and therefore worse RV precision, compared to previous campaigns. We used the HARPS DRS version 3.5 to process all spectra.

\subsection{Archival data}

\begin{figure*}[!ht]
    \centering
	\includegraphics[width=16cm]{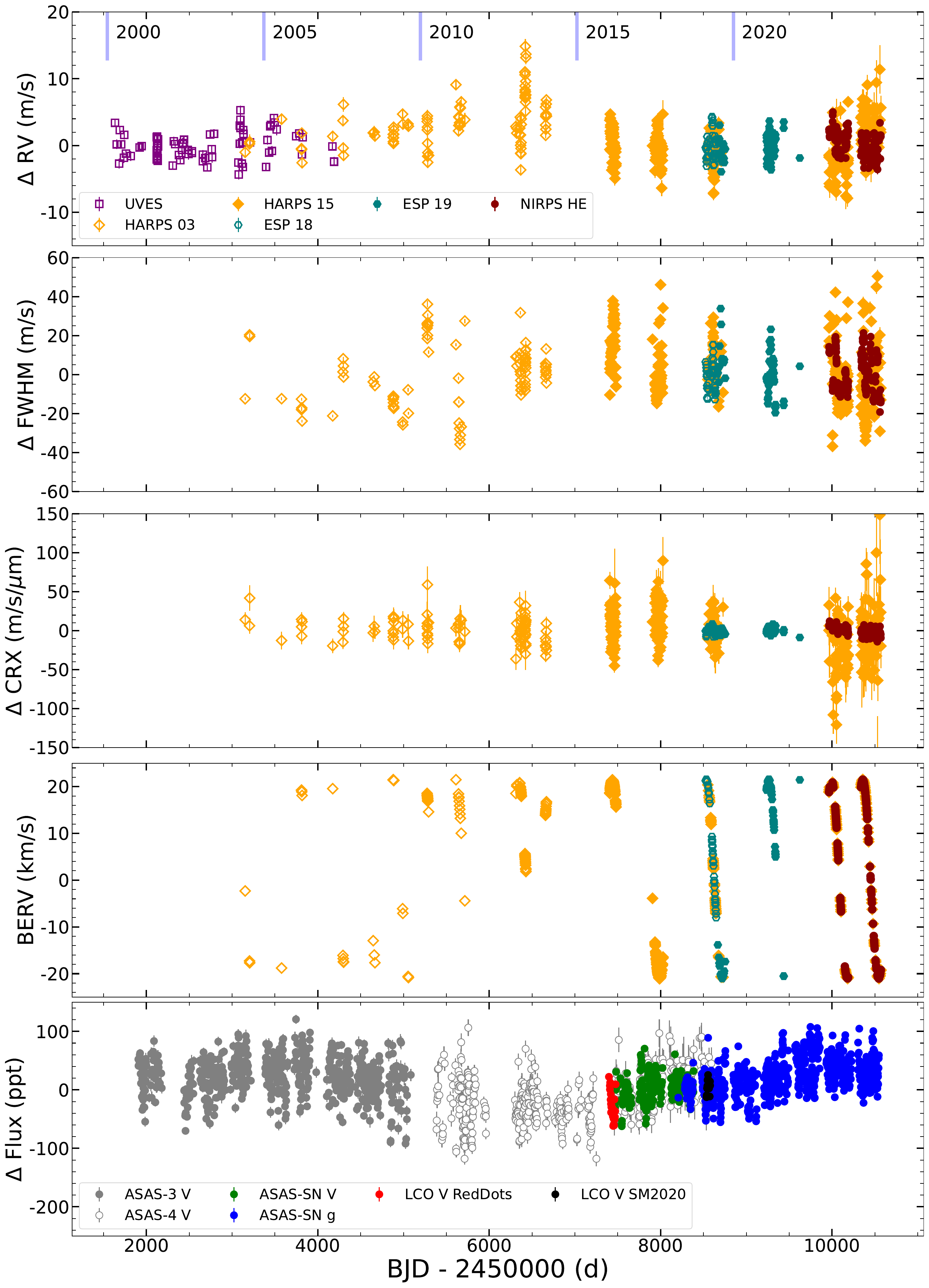}
	\caption{\textbf{All data.} The upper panel shows the combined RV time series of Proxima. The upper-middle panel shows the time series of FWHM of the same spectra. The middle panel shows the CRX data. The lower-middle panel shows the BERV at which the measurements were taken. The lower panel shows the photometric data used in this work.}
	\label{data_all}
\end{figure*}

In combination with the newly acquired NIRPS and HARPS data, we included all the publicly available spectroscopic data of Proxima, as well as long-term photometric observations. Proxima has been extensively observed over the years, going as far back as the year 2000. 

\subsubsection{ESPRESSO}
We use 116 observations obtained within the GTO of the {\'E}chelle SPectrograph for Rocky Exoplanets and Stable Spectroscopic Observations (ESPRESSO), as part of programme IDs 1102.C-0744, 1102.C-0958, 106.21M2, 1104.C0350, 108.2254 (PI:~F. Pepe). These spectra were previously analysed in \citet{Masca2020} and \citet{Faria2022}. In June 2019, ESPRESSO underwent an intervention to update its fibre link \citep{Pepe2021}, which introduced an RV offset. We consider separate datasets before and after, denoted ESPRESSO-18 (E18) and ESPRESSO-19 (E19). We reduced the data with the ESPRESSO DRS, version 3.3.0. 

\subsubsection{HARPS archival data}
We used 442 HARPS spectra obtained between 2003 and 2022 as part of the Geneva/Grenoble survey \citep{Bonfils2013}, and the RedDots project \citep{AngladaEscude2016}, under programmes 072.C-0488, 082.C-0718, 183.C-0437, 191.C-0505, 096.C-0082, 099.C-0205, 099.C-0880, and 1102.C-0339. Part of these spectra were used in the original detection of Proxima b \citep{AngladaEscude2016}, to propose the candidate planet Proxima c \citep{Damasso2020}, and in the confirmation of Proxima b \citep{Masca2020}. The HARPS set of fibres was upgraded in May 2015, introducing an offset and a slightly different behaviour before and after \citep{LoCurto2015}. We split the HARPS data in two datasets, denoted HARPS-03 (H03) for the original HARPS data, and HARPS-15 (H15) for the HARPS post-upgrade data. All data were reduced with DRS version 3.5. 

\subsubsection{UVES}

We use 77 observations from the Ultraviolet and Visual Echelle Spectrograph (UVES, \citealt{Dekker2000}), obtained between 2000 and 2007. The UVES data were obtained in one of the early RV surveys for planets around M dwarfs under ESO programme IDs: 65.L-0428, 66.C-0446, 267.C-5700, 68.C-0415, 69.C-0722, 70.C-0044, 71.C-0498, 072.C-0495, 173.C-0606, and 078.C-0829 (PI: M. Kürster). These data were previously analysed in \citet{Damasso2020} and \citet{Masca2020}.

\subsection{Telemetry data}

Modern RV spectrographs are designed to minimise instrumental effects caused by the changes in their environments. However, small effects either linked to the stability of both instruments or to the extraction of the velocities can be present in the data. We used the temperatures of the optical elements of HARPS, ESPRESSO, and NIRPS, as a proxy of the instruments' behaviour and used them to evaluate the quality of the data.

\subsection{Radial velocities and derived products}

We extracted all NIRPS, HARPS, and ESPRESSO, RVs using the Line-By-Line (\texttt{LBL}\footnote{version 0.65.001, lbl.exoplanets.ca}) code developed by \citet{Artigau2022}, and based on \citet{Dumusque2018}. The \texttt{LBL} algorithm performs an outlier-resistant template matching to each individual line in the spectra. For non-telluric corrected spectra, it produces its own telluric correction. In addition to the velocity, other quantities are also derived, the most important being a differential line-width (dLW, similar to \citealt{Zechmeister2018}), which is obtained from the second derivative of the template and can be understood (for a Gaussian line) as a change in the line full width at half maximum (FWHM).  We will use these measured changes in the FWHM as our main activity indicator. Using the dLW variations, and an estimation of the average FWHM of the lines, the \texttt{LBL} algorithm estimates a variation of the FWHM of the lines. In addition, the \texttt{LBL} algorithm derives a chromatic RV slope over the full spectral range (chromatic index, CRX,  similar to \citealt{Zechmeister2018}). In addition, we will use the Barycentric Earth Radial Velocity (BERV) to track potential leftover effects from the telluric correction. This algorithm, similar to other template matching codes (e.g. \texttt{SERVAL}, \citealt{Zechmeister2018}), performs significantly better than the widely used cross correlation method, when extracting RVs of M dwarfs.

The individual raw NIRPS RV measurements show a root mean square (RMS) of 1.88 m$\cdot$s$^{-1}$, and a typical uncertainty of 75 cm$\cdot$s$^{-1}$. After rejecting the data showing anomalous telemetry measurements (e.g. large deviations in temperature) and nightly binning, we obtain 149 RV measurements with an RMS of 1.69 m$\cdot$s$^{-1}$, and a typical uncertainty of 0.55 m$\cdot$s$^{-1}$. The HARPS RV measurements show an RMS of 4.2 m$\cdot$s$^{-1}$, and a typical uncertainty of 1.3 m$\cdot$s$^{-1}$. After a similar cleaning, and binning process, we obtain 393 RV measurements with an RMS of 3.20 m$\cdot$s$^{-1}$, and a typical uncertainty of 84 cm$\cdot$s$^{-1}$. The main reason behind the reduction of the RMS being the rejection of a small number of large outliers (deviations > 10 m$\cdot$s$^{-1}$). For the ESPRESSO we measured an RV RMS of 1.98 m$\cdot$s$^{-1}$, and RV internal uncertainties of 15 cm$\cdot$s$^{-1}$. 

The UVES RV data were obtained directly from \citet{Damasso2020}. The data reduction and RV measurement is described in \citet{Butler2019}. They consist in 77 nightly-binned measurements, with an RMS of 2.00 m$\cdot$s$^{-1}$, and a typical uncertainty of 0.64 m$\cdot$s$^{-1}$. 

Figure~\ref{data_nirps} shows the data obtained during the NIRPS GTO. Figure~\ref{data_all} shows the complete set of dataset used in this study, including the archival data. Table~\ref{tab:spec_data} summarises the distribution of data per instrument, reduction software (DRS), RV extraction method, dispersion, and typical uncertainty. Interestingly, before performing any correction on stellar activity or other time-dependent effects, the NIRPS RV data shows the lowest RMS of all the individual time series.

\begin{table*}
    \begin{center}
        \caption{RV data used in this work \label{tab:spec_data}}
        \begin{tabular}[center]{l l l l l l l l}
            \hline
            Instrument & DRS & RV extraction & N. data & Baseline [d] & RMS RV [m$\cdot$s$^{-1}$] & $\sigma$RV [m$\cdot$s$^{-1}$] & Ref. \\
            \hline 
            NIRPS & APERO & LBL & 149 & 604 & 1.69 & 0.55 & 0\\
            ESPRESSO & ESPRESSO DRS 3.2.5 & LBL & 103 & 1099 & 1.98 & 0.15 &  0\\
            HARPS    & HARPS DRS 3.5 & LBL & 393 & 7417 & 3.20 & 0.84 &0\\
            UVES & & Iodine cell & 77 & 2175 & 2.00 & 0.64 & 1\\

            \hline

            \hline
        \end{tabular}
    \end{center}
    \textbf{References:} 0 - This work, 1 -  \citet{Damasso2020}
\end{table*}

\subsection{Photometric data}

The flux variations of Proxima have been intensively monitored over the years by various photometric surveys. We complemented the spectroscopic data with photometric time series obtained by the All Sky Automated Survey (ASAS; \citealt{Pojmanski1997}), the All Sky Automated Survey for SuperNovae (ASAS-SN; \citealt{Kochanek2017}) and with data obtained using the Las Cumbres Observatory Global network (LCOGT, \citealt{Brown2013}). The full dataset consists of several thousands of epochs. To make it more manageable, we binned all photometric data every 2 days, resulting in 2212 binned measurements. 

The ASAS data is composed of the observations taken by the ASAS-3 and ASAS-4 surveys. Both are $V$-band data taken over long baselines. We obtained the ASAS-3 data from the ASAS website~\footnote{https://www.astrouw.edu.pl/asas/}. The light curves include the photometric measurements obtained using four different apertures. We averaged over the four apertures to obtain the final photometric series, reducing the short-term scatter of the data. The ASAS-4 data was obtained from \citet{Damasso2020}. The ASAS-4 data has error bars much larger than its short-term scatter. After an initial model, we concluded they are significantly over-estimated. For our analysis, we opted to scale them down to quarter. Even then, the excess white noise we measure remains comparable with the other photometric sources. The ASAS-3 data spans 8.7 years, between the years 2000 and 2009. It has an RMS of 33 ppt, and a median uncertainty of 13 ppt. The ASAS-4 data spans 8.8 years, between the years 2010 and 2019, with an RMS of 39 ppt, and a median uncertainty of 14 ppt.

The ASAS-SN data includes $V$-band and $g$-band data, taken between 2000--2007, and 2010--2024, respectively. We measured comparable RMS between the two. A preliminary photometric analysis showed indistinguishable behaviour between the two, which led us to combine the g-band data with the rest of the V-band data, to simplify the analysis. The $V$-band photometry has an RMS of 21.5 ppt, and a median uncertainty of 2.9 ppt. The g-band data has an RMS of 29.6 ppt, and a median uncertainty of 2.9 ppt. 

There are two campaigns of V-band observations taken with the LCOGT. One obtained as support observations of the RedDots project, which were published alongside the original discovery of Proxima b \citep{AngladaEscude2016}. The other obtained as support observations of the ESPRESSO GTO, used in the confirmation of Proxima b with ESPRESSO \citep{Masca2020}. These campaigns produced time series with RMS of 23.0 ppt, and 10.59 ppt, respectively, and median uncertainties of 1.6 ppt, and 0.15 ppt, respectively. 

The full photometric dataset can be found in Figure~\ref{data_all}.

\section{Properties of Proxima} \label{prop_proxima}

Proxima, the closest star to the Sun, is one of the most well known stars. It is an M5.5 star, part of the triple system Alpha Centauri~\citep{Kervella2017}. It has a mass of 0.12 M$_{\odot}$~\citep{Mann2015}, a radius of 0.14 R$_{\odot}$~\citep{Boyajian2012}, an effective temperature of 2900 K~\citep{Pavlenko2017}, and a luminosity of 0.00151 $\pm$ 0.00008 L$_{\odot}$~\citep{Ribas2017}. Using these values, we estimate the limits of the habitable zone (HZ) of the star to be 0.03731 $\pm$ 0.0075 au to 0.088 $\pm$ 0.017 au, following \citet{Kopparapu2014, Kopparapu2017}, for the runaway greenhouse to early-Mars limits. These limits correspond to orbital periods of 7.5 $\pm$ 2.2 days to 27.3 $\pm$ 7.8 days. From polarimetric measuremets, Proxima's rotation axis has been estimated to be tilted 47 $\pm$ 7$^{\circ}$ with respect to the line of sight  \citep{Klein2021}. This value is reinforced by the report of a dust belt in the system, with a tilt of $\sim$ 45$^{\circ}$ \citep{Anglada2017}. Table ~\ref{tab:parameters} shows the full list of parameters used in this study. 

\begin{table}
    \begin{center}
        \caption{Stellar properties of Proxima. \label{tab:parameters}}
        \begin{tabular}[center]{l l l}
            \hline
            Parameter & Value & Ref. \\ \hline
            RA [J2000] & 14:29:42.9461 & 1 \\
            DEC [J2000] & --62:40:46.1647 & 1\\
            $\mu \alpha \cdot \cos\delta$ [mas yr$^{-1}$]& --3781.741 & 1 \\
            $\mu \delta$ [mas yr$^{-1}$]& --20.578 & 1 \\
            Parallax [mas] &  	768.067 $\pm$ 0.050 & 1\\
            Distance [pc] & 1.3012 $\pm$ 0.0003 & 1\\
            $m_{B}$	 [mag] & 12.95 $\pm$ 0.01  & 2 \\
            $m_{V}$	 [mag] & 11.13 $\pm$ 0.01 & 2 \\
            Spectral Type  & M5.5V & 3\\
            L$_{*}$ [L$_{\odot}$] & 0.00151 $\pm$ 0.00008 & 4 \\
            T$_{eff}$ [K] & 2900 $\pm$ 100 & 5 \\    
            M$_{*}$ [M$_{\odot}$] & 0.1221  $\pm$ 0.0022 & 6 \\
            R$_{*}$ [R$_{\odot}$] & 0.141  $\pm$ 0.021 & 4 \\   
            i [$^{\circ}$]  & 47 $\pm$ 7 & 7 \\
            P$_{\rm rot ~GP}$ [days] & 83.2 $\pm$ 1.6 & 0 \\
            P$_{\rm cycle}$ [days] & 6560 $\pm$ 85 & 0 \\
            HZ inner edge [au] & 0.03731 $\pm$ 0.0075 & 8\\
            HZ outer edge [au] & 0.088 $\pm$ 0.017 & 8\\
            HZ inner period [days] & 7.5 $\pm$ 2.2& 8 \\
            HZ outer period [days] & 27.3 $\pm$ 7.8 & 8 \\
            \hline
        \end{tabular}
    \end{center}
    \textbf{References:} 0 - This work, 1 -  \citet{GaiaEDR3}, 2 - \citet{Joao2014}, 3 - \citet{Bessell1991}, 4 - \citet{Ribas2017}, 5 - \citet{Pavlenko2017}, 6 - \citet{Mann2015}, 7 - \citet{Klein2021}, 8 - Estimated following \citet{Kopparapu2014, Kopparapu2017}
\end{table}
    
\subsection{Magnetic cycle}

Proxima is among the select group of fully-convective M dwarfs with a known magnetic cycle. The cycle has been detected in photometry and X-ray \citep{Masca2016, Wargelin2017}, with a reported period be between 6.8 and 7.8 years, respectively. It has been reported to have an amplitude of 15--20 mmag in photometry, and its effects in RV and spectroscopic indicators have not been studied in-depth. More recently,~\citet{Wargelin2024} re-analysed the photometric data, using a dataset very similar as the one used in this work, reporting a period of 7.99 $\pm$ 0.17 years.

\subsection{Rotation}

Proxima rotates every 83--86 days, with variations easily detectable in photometry, spectroscopic indicators, and radial velocity. \citet{Wargelin2017} measured variations in these range of periods likely related to differential rotation. These rotational variations have been characterised many times over the years. The shape of the rotation curve is quite sinusoidal in photometry, with asymmetries that create some power at half the rotation period. \citet{Masca2020} showed that the relationship between the photometric and RV variations followed the concepts stated in \citet{Aigrain2012}, with the activity-induced RV variations showing a good correlation with the gradient of the flux. The RV variations would often have the dominant signal at half the rotation period. 

\subsection{Planetary system}

The planetary system of Proxima consists of one confirmed planet (Proxima b), and two candidates (Proxima c and d), that we know of. 

Proxima b is an Earth-mass planet orbiting the habitable zone of the star. It has an orbital period of 11.1868 $\pm$ 0.0031 days, and a minimum mass of 1.07 $\pm$ 0.06 M$_{\oplus}$ \citep{Faria2022}. t was announced by \citet{AngladaEscude2016}, using HARPS and UVES data, with the result being validated by \citet{Damasso2017}. The planet was later independently confirmed by \citet{Masca2020}, using ESPRESSO data. As it is one of the first terrestrial planets for which we expect to be able to survey the atmosphere, its potential habitability has been extensively discussed, with arguments in favour and against it \citep{Ribas2016,Turbet2016, Dong2017, Howard2018, Meadows2018, Turbet2023}.

Proxima c is a candidate long-period super-Earth. It was proposed by \citet{Damasso2020}, after a re-analysis of the HARPS and UVES datasets. It orbits the star with a period of \hbox{1900 $\pm$ 96 days}, and has a minimum mass of 5.8 $\pm$ 1.9 M$_{\oplus}$. A subsequent analysis of the HARPS data, using a different RV extraction method, could not find evidence of the presence of the planet \citep{Artigau2022}. 

Proxima d is a candidate short-period sub-Earth. The presence of the signal was originally proposed by \citet{Masca2020}, using ESPRESSO data, and solidified as a candidate by \citet{Faria2022}, after an additional observation campaign. It orbits with a period of 5.122 $\pm$ 0.036 days and has a minimum mass of 0.26 $\pm$ 0.05 M$_{\oplus}$~\citep{Faria2022}.  

\section{Analysis} \label{sec_analysis}

We performed a global analysis of the data, including always the full set of photometric data, along with spectroscopic activity proxies (FWHM and CRX), and the radial velocities. Both the spectroscopic proxy and the RV were separated between NIR and VIS data. Every time series includes a zero point per instrument, a model for the cycle, and a model for the stellar rotation. The cycle model is defined as a set of sinusoidal signals, with a common period and phase for all time series, and independent amplitudes. The stellar rotation is modelled as a multi-dimensional Gaussian Process. The RV data, in addition, includes polynomial terms against the CRX and BERV, and a circular/Keplerian model for each planet in the model.

We performed models that restricted the spectroscopic data to NIRPS, then models using the NIRPS and HARPS data from the NIRPS GTO, and models using the full spectroscopic dataset. 

The full model is defined as:   

\begin{equation} \label{eq_full_model}
    \begin{split}
    \Delta ~Flux = & ~V0 + Cycle + Rot ~,\\
    \Delta ~FWHM_{NIR} = & ~V0 + Cycle + Rot~, \\
    \Delta ~FWHM_{VIS} = & ~V0 + Cycle + Rot~, \\
    \Delta ~RV_{NIR} = & ~V0 + f_{CRX} + f_{BERV} + Cycle + Rot + Planets~,\\
    \Delta ~RV_{VIS} = & ~V0 + f_{CRX} + f_{BERV} + Cycle + Rot + Planets\\
\end{split}
\end{equation}

\noindent where V0 represents the zero-point of each time series, with priors $\mathcal{N}[0,100]$ ppt for the photometric data, $\mathcal{N}[0,40]$ m$\cdot$s$^{-1}$ for the FWHM, and $\mathcal{N}[0,10]$ m$\cdot$s$^{-1}$ for the RVs. The rest of the components are described in the following subsections.

\subsection{Cycle model}

Based on the results of a photometric-only analysis (see Appendix~\ref{append_photonly}), and the test of the presence of the cycle in spectroscopic data (see Appendix~\ref{append_cyclespec}), the cycle component is defined as: 

\begin{equation} \label{eq_full_model}
    \begin{split}
    \Delta ~y = &~ -A_{1} \cdot sin(2\pi (t - t_{1})/P_{\rm cyc})\\
    &~ -A_{2} \cdot sin(4\pi (t - t_{2})/P_{\rm cyc})  \\
    &~ -A_{3} \cdot sin(6\pi (t - t_{3})/P_{\rm cyc})  \\
    &~ -A_{4} \cdot sin(8\pi (t - t_{4})/P_{\rm cyc})  
\end{split}
\end{equation}

\noindent with:

\begin{equation} \label{eq_full_model_par}
    \begin{split}
    &t_{1} = t_{0} + P_{\rm cyc} \cdot (\phi_{1} - 1) \\
    &t_{2} = t_{0} + P_{\rm cyc} \cdot (\phi_{2} - 1) / 2\\
    &t_{3} = t_{0} + P_{\rm cyc} \cdot (\phi_{3} - 1) / 3\\
    &t_{4} = t_{0} + P_{\rm cyc} \cdot (\phi_{4} - 1) / 4
\end{split}
\end{equation}

\noindent with $t_{0} =  10560$ (BJD -- 2\,450\,000) being the integer date of the last NIRPS observation. The cycle has independent amplitudes for photometry, FWHM, and RV. The photometry-only analysis revealed the length of the cycle to be much longer than previously thought (see Appendix~\ref{append_photonly}). Based on these results, the cycle period has a prior $\mathcal{N} [6400 , 300]$ days. The phases have priors $\mathcal{U} [-0.5 , 0.5]$. The amplitudes in photometry are restricted to be positive ($\mathcal{U} [0 , 50]$ ppt), while the amplitudes in FWHM and RV can be either positive or negative, to account for opposition of phase ($\mathcal{N} [0 , 10]$ m$\cdot$s$^{-1}$ for the FWHM, and $\mathcal{N} [0 , 5]$ m$\cdot$s$^{-1}$ for the RV).

This function will be applied to the photometric data, and to the visible-light FWHM and RV data, when using the complete dataset. In the case of the NIRPS data, the baseline of observations is too short to contribute to the determination of the cycle parameters ($\sim$ 600 days). The same is true for the HARPS data from the NIRPS GTO, when analysed alone. For the NIRPS data, and the HARPS data when restricted to the GTO observations, we simply account for a linear trend (eq.~\ref{eq_gto_model}). 

\begin{equation} \label{eq_gto_model}
    \begin{split}
    \Delta ~FWHM = &~ a \cdot (t - t_{mid})\\
    \Delta ~RV = &~ a \cdot (t - t_{mid})\\
\end{split}
\end{equation}

\noindent with $t_{mid} =  10264$ (BJD -- 2450000) being the mid-point of the NIRPS observing baseline. We set a prior $\mathcal{N} [0 , 1]$ m$\cdot$s$^{-1}$d$^{-1}$ for both the FWHM and RV slopes. 

When analysing the NIRPS data together with the complete visible-light dataset, we will use a scaling factor of the complete visible-light model. \textbf{We follow the formula}:

\begin{equation} \label{eq_nirps_all}
    \begin{split}
    \Delta ~FWHM_{NIR} = &~ a \cdot \Delta ~FWHM_{VIS}\\
    \Delta ~RV_{NIR} = &~ a \cdot \Delta ~RV_{VIS}\\
\end{split}
\end{equation}

\noindent where the scaling factor has a prior ln~$a~\mathcal{U} [-5,3]$. We assume the sign of the variations in NIR data to be the same as in the visible-light data.

Given the length of the cycle, and its shape, all models always include the full photometric dataset. 
 
\subsection{Stellar rotation model}

To model the rotation we opted to work within the multidimensional Gaussian Processes (GP) framework \citep{Rajpaul2015}. The GP framework has become the cornerstone method in the analysis of stellar activity in RV time series (e.g. \citealt{Haywood2014}). The stellar noise is described by a covariance with a prescribed functional form and the parameters attempt to describe the physical phenomena to be modelled. The GP framework can be used to characterise the activity signal without requiring a detailed knowledge of physical parameters of the underlying source of the variability. One of the biggest advantage of GPs is that they are flexible enough to effortlessly model quasi-periodic signals, and account for changes in the amplitude, phase, or even small changes in the period of the signal. This flexibility is also one of their biggest drawbacks, as they can easily overfit the data, suppressing potential planetary signals. 

The multidimensional GP differs from the traditional 1D GP analysis by including the assumption that there exists an underlying function governing the behaviour of the stellar activity in all time series, which we denote $G(t)$. $G(t)$ manifests in each time series (Photometry, RV, etc.) as a linear combination of itself and its gradient, $G'(t)$, with a set of amplitudes for each time series, following the idea of the $FF'$ formalism \citep{Aigrain2012}, as described in equation~\ref{eq_gp_grad}.

\begin{equation} \label{eq_gp_grad}
\begin{split}
&\Delta ~TS_{1} = A_{1} \cdot G(\tau) + B_{1} \cdot G'(\tau)~, \\
&\Delta ~TS_{2} = A_{2} \cdot G(\tau) + B_{2} \cdot G'(\tau)~, \\
&...
\end{split}
\end{equation}

\noindent with $\Delta~TS_{i}$ representing the variations of each time series, and $A_{i}$ and $B_{i}$ the scaling coefficients of the underlying function $G(t)$ and its gradient, $G'(t)$, respectively. 

The main advantage of the multidimensional GP over the traditional 1D GP, even if the parameters are trained in activity proxies, is the inclusion of all the data in the same covariance matrix. This imposes the behaviour of all time series to be correlated, rather than simply following qualitatively similar relationships. 

We used the \texttt{S+LEAF} code \citep{Delisle2022} \footnote{\url{https://gitlab.unige.ch/delisle/spleaf}}, which extends the formalism of semi-separable matrices introduced with \texttt{celerite} \citep{Foreman-Mackey2017} to allow for fast evaluation of GP models even in the case of large datasets. This computational efficiency is key when evaluating datasets that include hundreds, or thousands, of measurements. The \texttt{S+LEAF} code supports a wide variety of GP kernels with fairly different properties. Based on the tests performed on the photometry-only model, we opted for a combination of two stochastic harmonic oscillators (SHO), one with a period equal to the rotation period, and the second at half the rotation period. This model resulted in the lowest amount of overfitting of long-period signals (see Appendix~\ref{append_photonly}). The kernel is defined as: 

\begin{equation} \label{act_model}
    \begin{split}
    \fontsize{8}{11}\selectfont
     k(\tau, P_{\rm rot}, L) = k_{SHO ~\rm 1}(\tau, \alpha_{1}, P_{1}, Q_{1}) + k_{SHO ~\rm 2}(\tau, \alpha_{2}, P_{2}, Q_{2}) \\
     + (\sigma^2 (t) + \sigma^2_{j}) \cdot \delta_{\tau}
     \end{split}
\end{equation}

\noindent where $k$ denotes the GP term, $\alpha_{i}$ is the standard deviation of the process,  $P_{i}$ is the period of the oscillator, related to the rotation period, and $Q_{i}$ is the quality factor of the oscillator, related to the timescale of evolution $L$. These parameters are defined as: 

\begin{equation} \label{eq_params}
\begin{split}
&\alpha_{1} = 1 ~; P_{1} = P_{\rm rot} ~; Q_{1} = {\pi {L}\over{P_{\rm rot}}} \\
&\alpha_{2} = 1 ~; P_{2} = 0.5 \cdot P_{\rm rot} ~; Q_{2} = {2\pi {L}\over{P_{\rm rot}}} \\
\end{split}
\end{equation}

\noindent Equation~\ref{act_model} also includes a term of uncorrelated noise ($\sigma$), independent for every instrument, added quadratically to the diagonal of the covariance matrix to account for all unmodelled sources of variation, such as uncorrected activity, instrumental instabilities, or additional planets. $\delta_{\tau}$ is the Kronecker delta function, and $\tau$ represents a time interval between two measurements, $t-t'$. The white noise model uses log-normal priors, centered around the dispersion of the data and with a sigma similar to the dispersion of the data. This parametrization makes it difficult for the model to converge to very low white noise values, while not forbidding them, which helps preventing overfitting. For the photometric data we used a prior ln~$\sigma~\mathcal{N}[3.5,3.5]$ (centred at \hbox{$\sim$ 33 ppt}), for the FWHM ln~$\sigma~\mathcal{N}[2.3,2.3]$ ($\sim$ 10 m$\cdot$s$^{-1}$), and for the RVs ln~$\sigma~\mathcal{N}[0.5,0.5]$ ($\sim$ 1.65 m$\cdot$s$^{-1}$). For the case of the RV data, this parametrization makes it that deviating 3$\sigma$ from the median value of the prior is equivalent to 37 cm$\cdot$s$^{-1}$, and deviating 5$\sigma$ corresponds to 14 cm$\cdot$s$^{-1}$.

The amplitudes $\alpha_{i}$ are related with the amplitude of the underlying function, not to any of the specific time series. These amplitudes are degenerate with the amplitudes of the component at each time series. We fixed them to 1, with the amplitudes of every component governed by the parameters A$_{i}$ and B$_{i}$ shown in equation~\ref{eq_gp_grad}. 

We used a prior $\mathcal{N}[85,5]$ days for the rotation period, and a prior ln~$L$ $\mathcal{U}[3,10]$ days for the timescale of evolution. 

We used the photometric data as main dataset guiding the GP, which meant not including a gradient amplitude for it. The GP model, over all time series, is described as: 

\begin{equation} \label{eq_full_gp}
    \begin{split}
    \Delta ~Flux = &~A_{11}~k_{1} + A_{12}~k_{2} ~, \\
    \Delta ~FWHM_{NIR} = &~A_{21}~k_{1} + B_{21}~k'_{1} + A_{22}~k_{2} + B_{22}~k'_{2} ~, \\
    \Delta ~FWHM_{VIS} = &~A_{31}~k_{1} + B_{31}~k'_{1} + A_{32}~k_{2} + B_{32}~k'_{2} ~, \\
    \Delta ~RV_{NIR}   = &~A_{41}~k_{1} + B_{41}~k'_{1} + A_{42}~k_{2} + B_{42}~k'_{2} ~, \\
    \Delta ~RV_{VIS}   = &~A_{51}~k_{1} + B_{51}~k'_{1} + A_{52}~k_{2} + B_{52}~k'_{2} ~, \\
    \end{split}
    \end{equation}

\noindent The scaling factor of the photometric component $A_{1i}$ have a prior $\mathcal{U}[0,50]$. The scaling factors of the FWHM and RV linear components $A_{ii}$ have priors $\mathcal{U}[-50,50]$, while the scaling factors of the gradient components $B_{ii}$ have priors $\mathcal{U}[-100,100]$.

\subsection{Correlation with the chromatic index}

The RV data of some of the instruments under study showed a correlation with the CRX (see Appendix~\ref{append_crx}). Our initial tests found that this correlation survived the GP model and remained in the residuals. To account for this effect, we included a slope term between the RVs and the CRX, with different slopes for different instruments. Due to the lack of available activity indicators, the UVES data was not detrended. 

The correlation takes the form of: 
\begin{equation} \label{rv_crx_correl}
    \Delta RV = a \cdot CRX
\end{equation}

\noindent with $a$ being the slope of the relationship, with a prior $\mathcal{N}[0,0.3]$. 

\subsection{Correlation with the BERV}

There are several effects potentially affecting the spectra, such as telluric contamination, ghosts, or CCD stitching, anchored at the detector reference frame \citep{Dumusque2015, Coffinet2019, Cretignier2021}. Combined with the movement of the Earth around the Sun, this results in periodic shifts with 1-year periodicity. All our spectra have been corrected from telluric contamination. However, there could be leftover effects which could still cause small RV shifts. To account for potential contamination, we included a polynomial term against the BERV (see Fig. 3 of~\citealt{Dumusque2015}, and Appendix~\ref{append_berv}). This term is defined as shown in eq.~\ref{rv_berv_correl}, with different parameters for each instrument. Due to the lack of available data, the UVES data was not detrended. 

The correlation takes the form of: 
\begin{equation} \label{rv_berv_correl}
    \Delta RV = a \cdot BERV + b \cdot BERV^2
\end{equation}

\noindent with $a$ and $b$ being the parameters of the polynomial, with priors $\mathcal{N}[0,0.3]$. 

\subsection{Planetary model}

All previously stated components of the model define our null-hypothesis model, i.e. a model aimed to account for stellar activity and instrumental effect. On top of them, we include a planetary model where the number of planets will vary. The planetary signals are defined as circular for the detection tests, and later as full Keplerians to characterise their eccentricities. This choice avoids the potential pitfall of having a poorly sampled eccentric signal mimicking two circular signals and significantly cuts down on computing time. 

RV variations due to circular planetary orbits are defined as:
\begin{equation} \label{eq_circular}
    y(t)=-K \cdot \sin(2 \pi \cdot (t - t_{0})/P_{pl})
\end{equation} 

\noindent where $t_{0} = 10560 + P_{pl} \cdot (\phi_{pl} - 1)$. This parametrization ensures that our $t_{0}$ coincides to the inferior conjunction of the planets, and is within the baseline of observations. 

When conducting a blind search for planets, the orbital period is parametrised as angular frequency $\omega = 2\pi/P_{pl}$. We use a prior $\mathcal{U}[2\pi/200$, $2\pi/2$] when analysing the NIRPS GTO data, as 200 days would be the largest period in which we could sample three full orbits. For the complete dataset, we use a prior $\mathcal{U}[2\pi/3000$, $2\pi/2$]. When performing a guided search using the published solutions, we directly sample the periods and use $\mathcal{N}$ priors centred around the published solution. The phases $\phi_{pl}$ are parametrised as $\mathcal{U}[-0.25, 0.75]$. 

We parameterise the planetary amplitude $K$ as ln $K$ with a prior $\mathcal{U}[-5,2]$ m$\cdot$s$^{-1}$ (which keeps 50\% of the prior space below 22 cm$\cdot$s$^{-1}$). This parametrization expands the parameter space around amplitudes consistent with zero, reducing potential biases towards large posteriors in noise dominated data, as described in \citet{Rajpaul2024}. 

RV variations due to planetary elliptical orbits are defined as: 
\begin{equation} \label{eq_kepler}
    y(t)=K \left(\cos(\eta+\omega) + e \ \cos\omega\right)
\end{equation} 
  
\noindent where the true anomaly $\eta$ is related to the solution of the Kepler equation, which depends on the orbital period of the planet $P_{\rm orb}$ and the orbital phase $\phi$. This phase corresponds to the periastron time, which depends on the mid-point transit time $T_{0}$, the argument of periastron $\omega$, and the eccentricity of the orbit $e$.

We parametrise $e = (\sqrt{e} ~cos(\omega))^{2} + (\sqrt{e} ~sin(\omega))^{2}$ and $\omega = \arctan^{2}(\sqrt{e} ~sin(\omega),\sqrt{e} ~cos(\omega))$. We then sample $\sqrt{e} ~cos(\omega)$ and $\sqrt{e} ~sin(\omega)$ with priors $\mathcal{N}$[0, 0.3]. This parametrization favours low eccentricities, which are expected for short-period signals.

The parameters of the planets are the same for the VIS and NIR time series.

\subsection{Inference}

We optimised the parameters of the models using Bayesian inference through the nested sampling \citep{Skilling2004, Skilling2006} code \texttt{Dynesty}~\citep{Speagle2020, kosopov2023}.We sampled the parameter space using random slice sampling, which is well suited for the high-dimensional spaces \citep{Handley2015a,Handley2015b} resulting from modelling several time series at once. We used a number of live points of $20 \times N_{free}$ in models with narrow priors in period/frequency, and $100 \times N_{free}$ in models with wide priors, to ensure the discoverability of the narrow frequency posteriors. 

\section{Results} \label{sec_results}

\subsection{NIRPS-only RVs} \label{nirps_only}

The NIRPS GTO programme accumulated 149 nightly binned RV spectroscopic observations. The data spans a baseline of 604 days, with a median cadence of one visit per 1.1 nights (average one visit every 4.1 nights), corresponding with nightly observations with large gaps in between campaigns. The root mean square (RMS) of the RV data is of 1.69 m$\cdot$s$^{-1}$, with a median uncertainty of 55 cm$\cdot$s$^{-1}$. The RMS of the FWHM data is of 10.21  m$\cdot$s$^{-1}$, with a median uncertainty of 64 cm$\cdot$s$^{-1}$. We analysed these data together with the complete photometric dataset, to guide the GP model.

\begin{figure}[!ht]
    \centering
	\includegraphics[width=9cm]{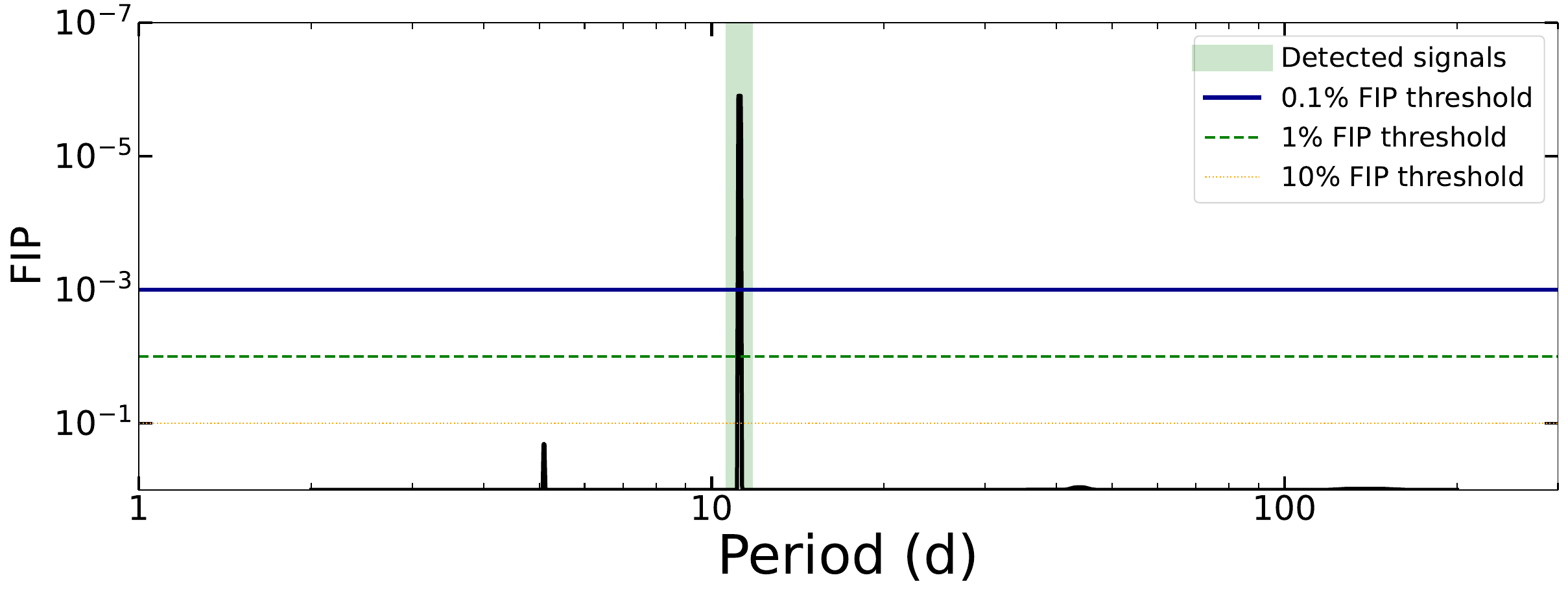}
	\caption{\textbf{FIP periodogram of the NIRPS-only model.} The highlighted peak corresponds to the period of 11.19 days of Proxima b.}
	\label{fip_periodogram}
\end{figure}

We first performed a blind search of planetary signals. We used the model, as described in the previous section, with the spectroscopic data restricted to the NIRPS dataset. We include two planetary components in the model. Given the baseline of the observations, we do not attempt to survey the presence of Proxima c in the NIRPS data alone. We assessed the significance of the detection of the signals using the False inclusion probability (FIP) framework \mbox{\citep{hara2022a}}. This framework uses the posterior distribution of the nested sampling run, and compute the probability of having a planet within a certain orbital frequency interval based on the relative weight of the samples within each frequency interval. The FIP framework has been demonstrated to minimise both false detections and missed detections, compared to other detection criteria such as the False Alarm Probability, or the Bayesian evidence. Based on an extensive set of simulations, \citet{hara2022a} suggested an optimistic FIP threshold of 50\%, that maximises detections, and a conservative threshold of 1\%, that minimises false positives. Figure~\ref{fip_periodogram} shows the FIP periodogram of the data after analysing the posterior distribution of the 2-planet model. The FIP periodogram shows the significant detection of a signal with a period of 11.19 days, corresponding to Proxima b, with a FIP < 0.001\%, far below the conservative 1\%. This strong detection supports the presence of Proxima b. There is an additional peak at 5.12 days, that passes  the optimistic threshold but falls significantly short of the conservative threshold. This can be considered a tentative detection of the signal of Proxima d, or a hint of its presence in the data.

\begin{figure*}[!ht]
    \centering
	\includegraphics[width=17cm]{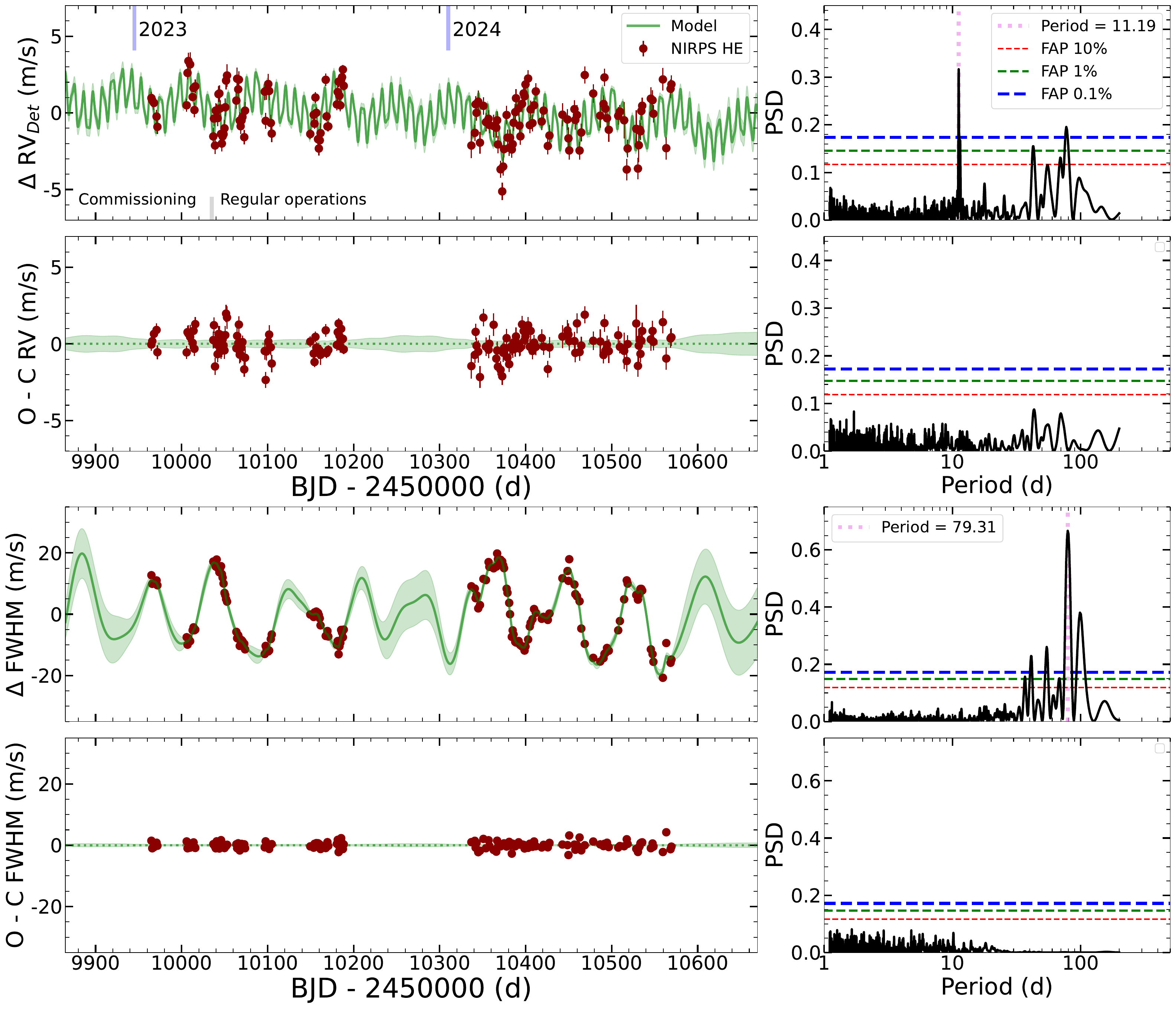}
	\caption{\textbf{Data and model of the NIRPS RV and DLW time series.} The two top panels show the NIRPS RV data (detrended from CRX and BERV), with the best model fit (top; GP + 2 planets), and the residuals after the fit (bottom), along with the periodograms of both (right). The two bottom panels show the same for the FWHM. The shaded region shows the standard deviation of the GP model.}
	\label{model_nirps}
\end{figure*}

\begin{figure}[!ht]
    \centering
	\includegraphics[width=9cm]{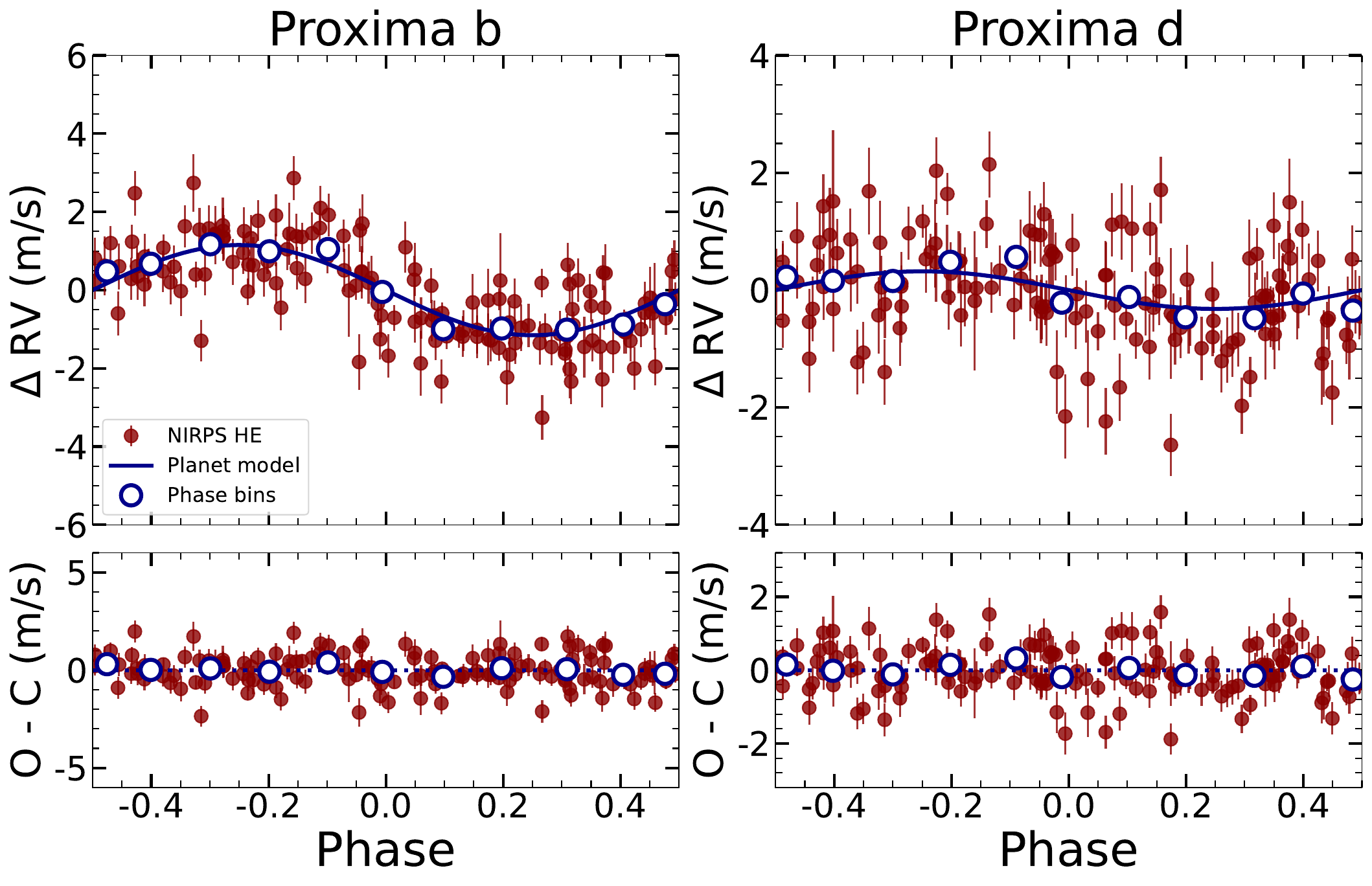}
	\caption{\textbf{NIRPS phase-folded plots of the planetary-induced RV signals.} Left panels show the phase-folded RV variations induced by Proxima b (top), and the residuals after the fit (bottom). Right panels show the same for Proxima d. The blue hollow points represent the data binned by phase.}
	\label{model_nirps_phase}
\end{figure}

\begin{figure}[!ht]
    \centering
	\includegraphics[width=9cm]{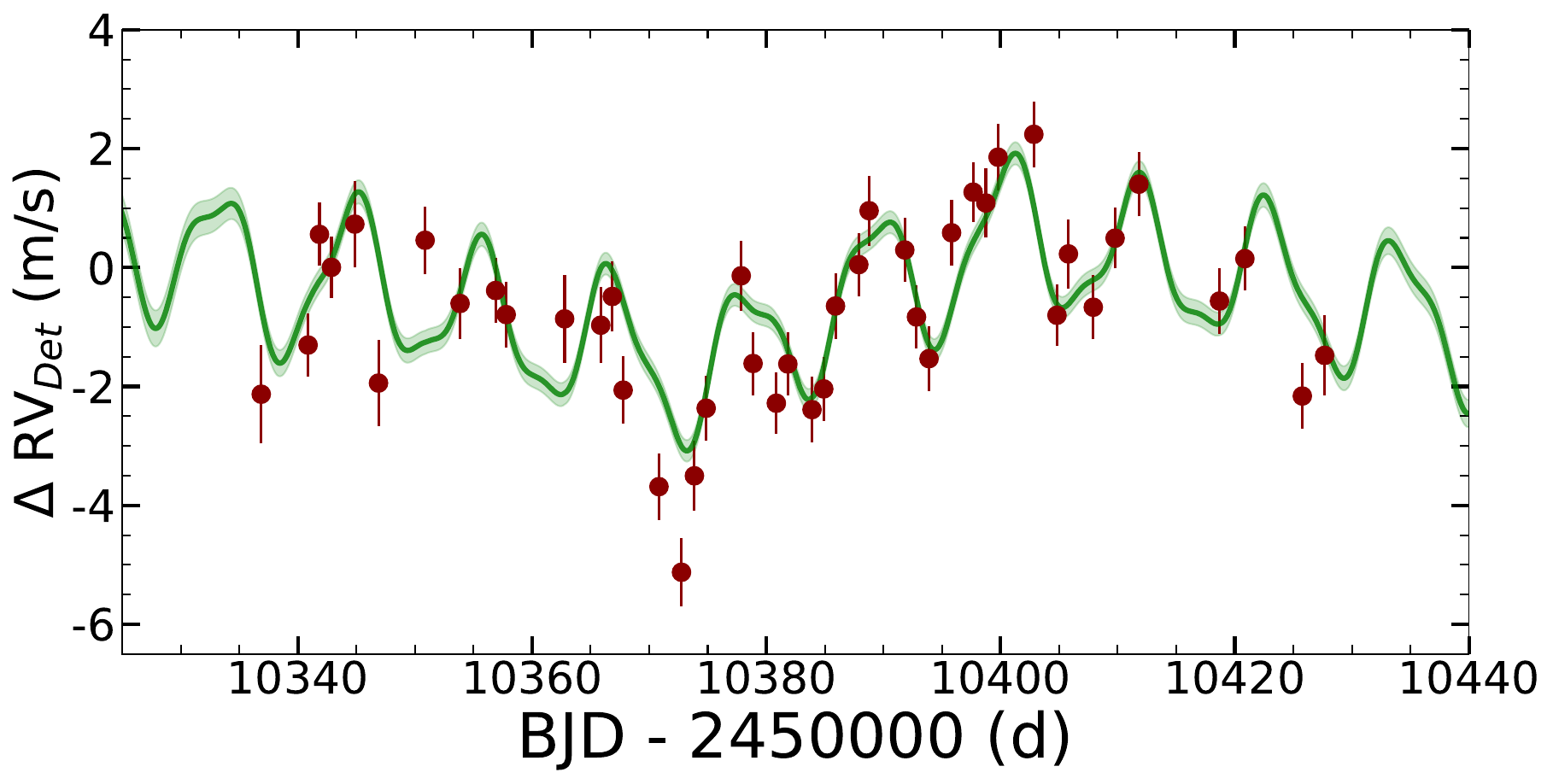}
	\caption{\textbf{Zoom to a high cadence RV campaign.} NIRPS RV data of Proxima, detrended from CRX and BERV, along with the best model fit. The shaded region shows the standard deviation of the GP model. }
	\label{model_nirps_zoom}
\end{figure}

After the blind search, we perform a guided search, using the results obtained by \citet{Faria2022} as priors for the periods of the planetary signals. We use normal priors around the period of the solution of \citet{Faria2022}, with a standard deviation of \hbox{$\sim$ 1\%} of the period. This corresponds to a prior $\mathcal{N}[11.1868,0.1]$ for Proxima b, and $\mathcal{N}[5.122,0.05]$ for Proxima d. The priors for the RV amplitude and phase remain the same. We measure RV amplitudes of 1.15 $\pm$ 0.11 m$\cdot$s$^{-1}$ for Proxima b, and 32 $\pm$ 14 cm$\cdot$s$^{-1}$ for Proxima d, in complete agreement with the ESPRESSO results. The ephemerides of both are consistent with all previous publications. While not a formal detection, this is evidence of a signal consistent with the properties of Proxima d being present in the data. Figure~\ref{model_nirps} shows the model of the NIRPS RV, detrended from CRX, and of the FWHM data, along with their respective residuals, and Generalised Lomb-Scargle (GLS) periodograms \citep{Zechmeister2009}. The combination of the GP, the detrending, and planetary models accurately describes the variations in the data, leaving residuals with an RMS of 81 cm$\cdot$s$^{-1}$, a 52\% reduction with respect to the original data. In addition, the model accurately describes the variations in FWHM, which clearly show an $\sim$ 80-day periodicity, linked to stellar rotation. The shape of the variations hints at the presence of several active regions, rotating at different rates. Figure~\ref{model_nirps_phase} shows the phase-folded plots of the RV component caused by both planets.

Figure~\ref{model_nirps_zoom} shows a zoom to a section with high cadence observations taken at the beginning of the 2024 campaign. The model and the data show very good agreement. The visible variations are caused by the signals of Proxima b and d with the GP accounting only for low-frequency variations (timescales $\sim$ > 40 days) linked to stellar rotation. 

Regarding the stellar activity model, we measure the length of the stellar cycle to be 6425 $\pm$ 71 days, based entirely on the photometric data, with amplitudes of 34.0 $\pm$ 2.1 ppt, 5.0 $\pm$ 1.4 ppt, 5.7 $\pm$ 1.6 ppt, and 6.9 $\pm$ 1.3 ppt, at the components of P$_{\rm cyc}$, P$_{\rm cyc}/2$, P$_{\rm cyc}/3$, and P$_{\rm cyc}/4$, respectively. These results remain roughly consistent for all subsequent models (the timescales are dominated by the photometric data). We measure a stellar rotation period of 84.35 $\pm$ 0.90 days, with a timescale of evolution of 151 $\pm$ 41 days. The GP shows amplitudes of 20.8 $\pm$ 1.8 ppt, and 10.1 $\pm$ 1.2 ppt, at the components of P$_{\rm rot}$ and P$_{\rm rot}/2$, respectively. We measure the variations in the FWHM to be anti-correlated with the variations in flux, and positively correlated with the gradient of the flux. The RV variations show a small positive correlation with the flux, and a large negative correlation with its gradient, but only for the component at P$_{\rm rot}/2$. The result is consistent with the expectations of spot-models \citep{Aigrain2012}. The activity model (trend against CRX and GP) account for an RMS of 84 cm$\cdot$s$^{-1}$, while the polynomial against the BERV account for an RMS of 53 cm$\cdot$s$^{-1}$. Figure~\ref{berv_corr} shows these RV variations against the CRX and the BERV, along with the best-fit polynomial model. The variations correlated with the CRX are most likely caused by some form of activity of Proxima that is not fully captured by the photometry and FWHM, while those correlated with the BERV are likely the result of residual telluric contamination, detector defects, or other effects anchored in the detector reference frame. 

As cross-check, we repeated the analysis using the \texttt{LBL} RVs derived from the spectra reduced with the ESO NIRPS DRS. We obtained very similar RVs, although more noisy, and consistent results. The RV data from the NIRPS DRS showed an RMS of 2.11 m$\cdot$s$^{-1}$, and we derived RV amplitudes of 1.00 $\pm$ 0.15 m$\cdot$s$^{-1}$ for Proxima b, and 37 $\pm$ 20 cm$\cdot$s$^{-1}$ for Proxima d. The results are in agreement with the ones obtained with spectra reduced with the APERO pipeline.

\subsection{Simultaneous HARPS data} \label{gto_data}

After performing the model on the NIRPS data, we add the HARPS data obtained within the NIRPS GTO and repeat the models previously described. The HARPS RV data consists of 135 nightly binned epochs, over a baseline of 604 days, the same as the NIRPS data. The time series shows an RMS of 3.5 m$\cdot$s$^{-1}$, with a typical measurement uncertainty of 1.4 m$\cdot$s$^{-1}$. 

Activity-induced RV variations of VIS and NIR data are expected to show different amplitudes \citep{Carmona2023}. We account for this effect by including different amplitudes for the GP component in the VIS and NIR data. 

\begin{figure}[!ht]
    \centering
	\includegraphics[width=9cm]{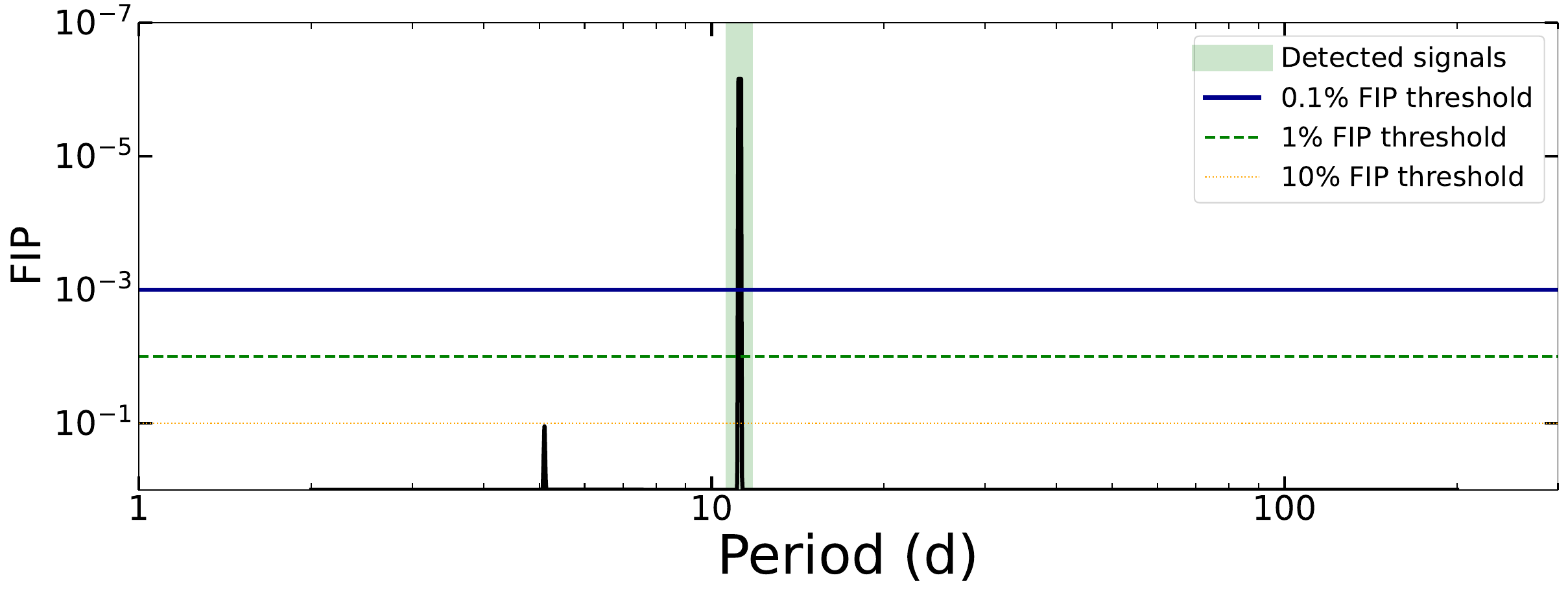}
	\caption{\textbf{FIP Periodogram of the NIRPS+HARPS 2023-2024 model.} The highlighted peak corresponds to the period of 11.19 days of Proxima b. The second weaker peak corresponds to 5.11 days, consistent with the period of Proxima d. }
	\label{fip_periodogram_2}
\end{figure}

As in the NIRPS-only model, we first perform a blind search using the same period limits as before. The inclusion of the HARPS data does not change the significance of the detection of Proxima b, which was already nearly saturated with the NIRPS-only data. It, however, increases the significance of the peak corresponding to Proxima d, which rises up to a FIP of $\sim$10\% (see Fig.~\ref{fip_periodogram_2}). This value of the FIP is in between the optimistic (50\%) and conservative (1\%) thresholds suggested by \citet{hara2022a}. It can be considered, at least, a tentative detection. 

\begin{figure}[!ht]
    \centering
	\includegraphics[width=9cm]{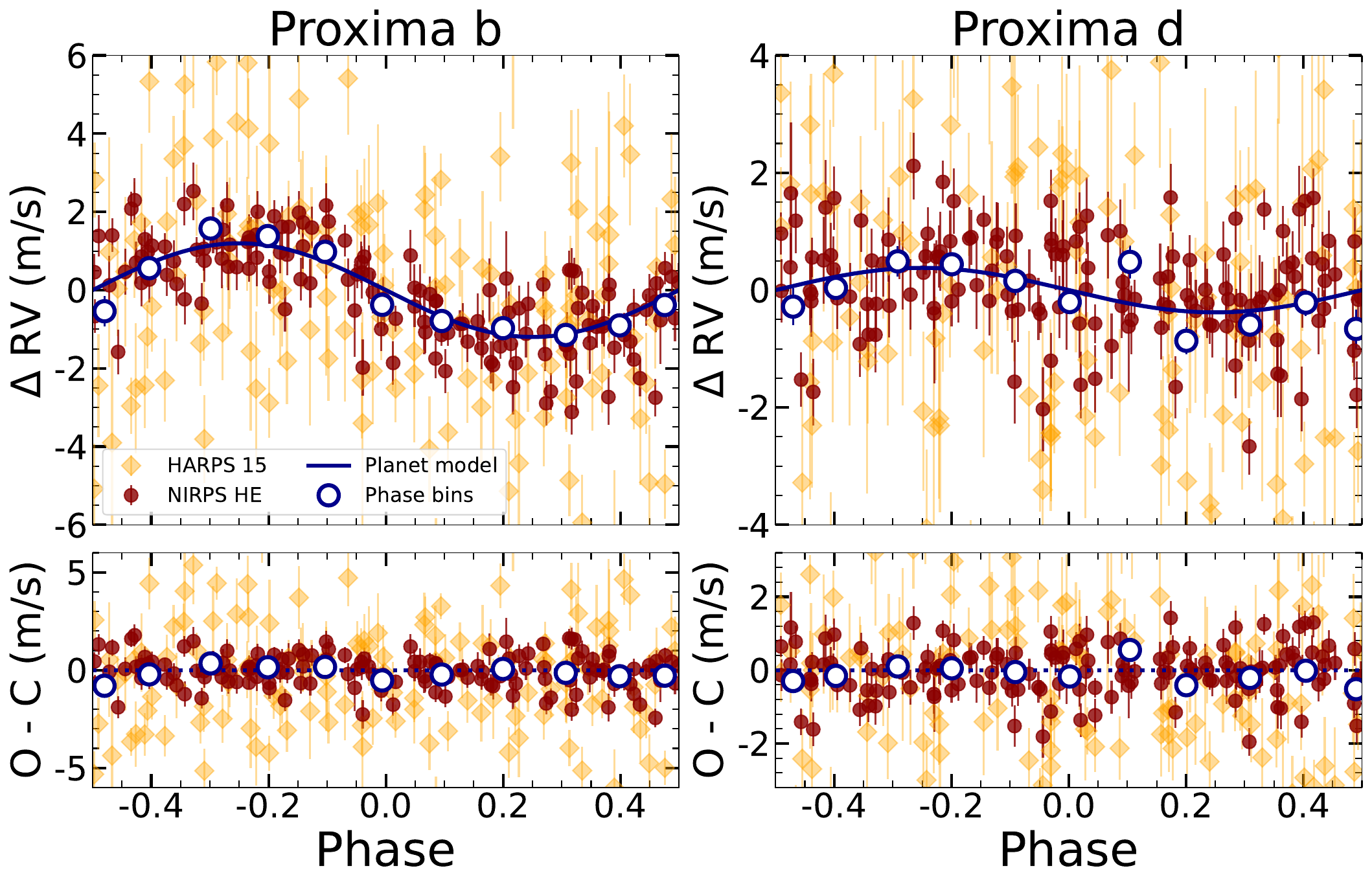}
	\caption{\textbf{Phase-folded plots of the planetary-induced RV signals in NIRPS and HARPS.} Left panels show the phase-folded RV variations induced by Proxima b (top), and the residuals after the fit (bottom). Right panels show the same for Proxima d.}
	\label{model_nirps_harps_phase}
\end{figure}

We repeat our guided search, with the same priors for the planets as before. With the combined GTO dataset we measure amplitudes of 1.198 $\pm$ 0.091 m$\cdot$s$^{-1}$ for the signal of Proxima b, and 37.9 $\pm$ 9.9 cm$\cdot$s$^{-1}$ for the signal of Proxima d. The signal at 5.12 days is detected at a > 3-$\sigma$ level. The ephemerides remain consistent. The residuals after the fit of the NIRPS data have an RMS of 0.84 m$\cdot$s$^{-1}$, while those of HARPS are of 2.77 m$\cdot$s$^{-1}$. Figure~\ref{model_nirps_harps_phase} shows the phase-folded plot of the HARPS+NIRPS RV variations of Proxima b and d.

We repeated both models using only the HARPS data from the NIRPS GTO. These data correspond to approximately the same baseline and exposure time as the NIRPS data, and can act as a good comparison of the performance of NIRPS. In the blind search, we did not have a significant detection of any signal. When performing the guided search, we measured an amplitude of 1.46 $\pm$ 0.31 m$\cdot$s$^{-1}$ for Proxima b, and an amplitude < 1.52 m$\cdot$s$^{-1}$ (95\% confidence) for Proxima d.  

\subsection{Full spectroscopic dataset} \label{full_spec_data}

When including the full spectroscopic dataset, we construct a time series of 722 RV measurements, and 645 FWHM measurements, over a baseline of 24.5 years. The time series includes data of UVES, HARPS, ESPRESSO, and NIRPS. The RMS and median uncertainty of different time series is as follows: 2.00/0.64 m$\cdot$s$^{-1}$ for UVES, 3.20/0.84 m$\cdot$s$^{-1}$ for HARPS, 1.98/0.15 m$\cdot$s$^{-1}$ for ESPRESSO, and 1.69/0.55 m$\cdot$s$^{-1}$ for NIRPS. The combined dataset has an RMS of 2.67 m$\cdot$s$^{-1}$, a median cadence of one point per 1.1 nights, and an average cadence of one point every twelve nights.

\begin{figure}[!ht]
    \centering
	\includegraphics[width=9cm]{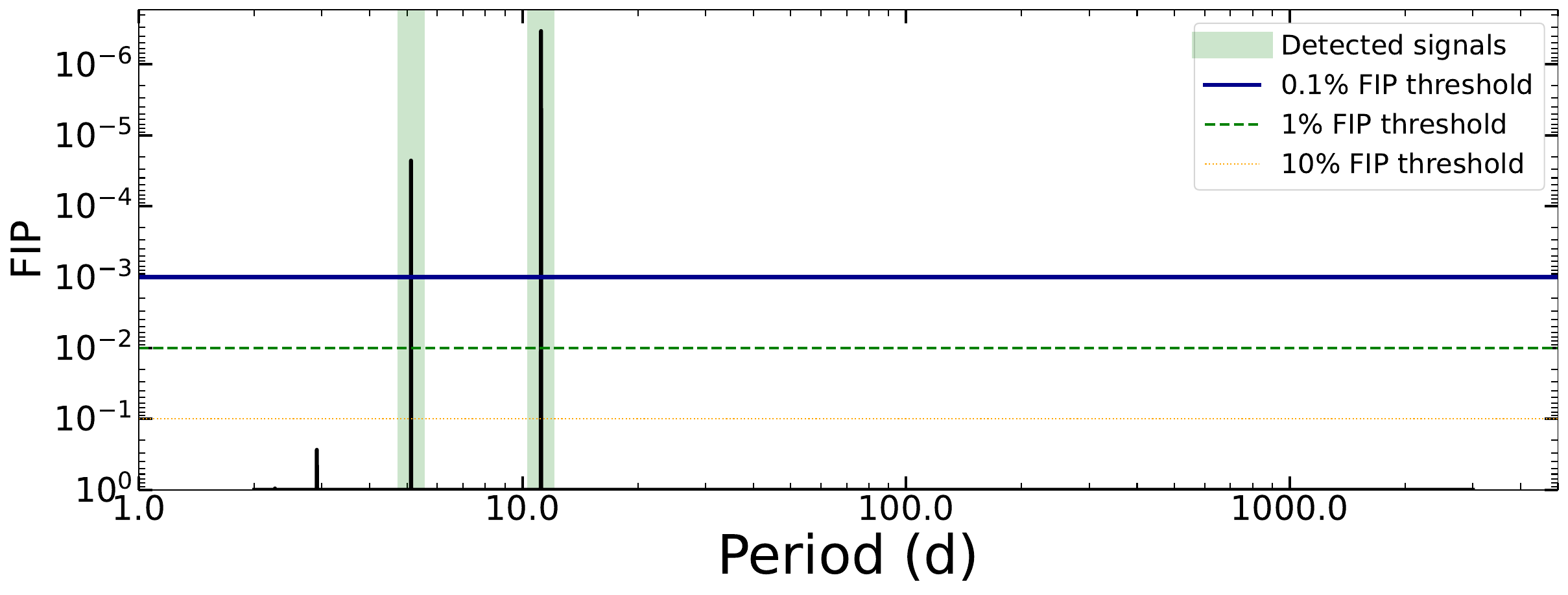}
	\caption{\textbf{FIP periodogram of the full dataset.} The highlighted peaks corresponds to the period of 5.11 (Proxima d), and 11.19 (Proxima b).}
	\label{fip_periodogram_all}
\end{figure}

Following the same steps as with previous dataset, we start by performing a blind search. In this case, as the baseline is long enough, we test a four-planet model to account for the potential signal of the candidate Proxima c and an additional potential planetary signal present in the data. Figure~\ref{fip_periodogram_all} shows the FIP periodogram of the data, under the assumption of a model with up to four planets. We detect two significant peaks, at periods of 11.19 days, and 5.12 days, corresponding to the signals of Proxima b, and Proxima d. The detection of both signals is very significant, according to the FIP framework. Both peaks comfortably cross the 1\% conservative threshold. In fact, the signals are present in virtually all the samples of the posterior distribution of the nested sampling run. We do not find additional significant peaks.There is one additional peak at 2.91 days, which does not reach the 10\% FIP threshold. We investigated the posterior distribution to search for the presence, even marginal, of the signal of the candidate Proxima c. The combined posterior of the frequencies of the four planetary signals included in the model only has a few hundred samples (< 0.1\%) at periods within 3$\sigma$ of the published period of Proxima c. We measure a period of 11.18482 $\pm$ 0.00028 days and an amplitude of 1.233 $\pm$ 0.037 m$\cdot$s$^{-1}$ for Proxima b. For Proxima d, we measure a period of 5.12335 $\pm$ 0.00025 days and an amplitude of 36.8 $\pm$ 3.5 cm$\cdot$s$^{-1}$. We repeated the fit with a more simple three-planet model, with no significant changes.

\begin{figure}[!ht]
    \centering
	\includegraphics[width=9cm]{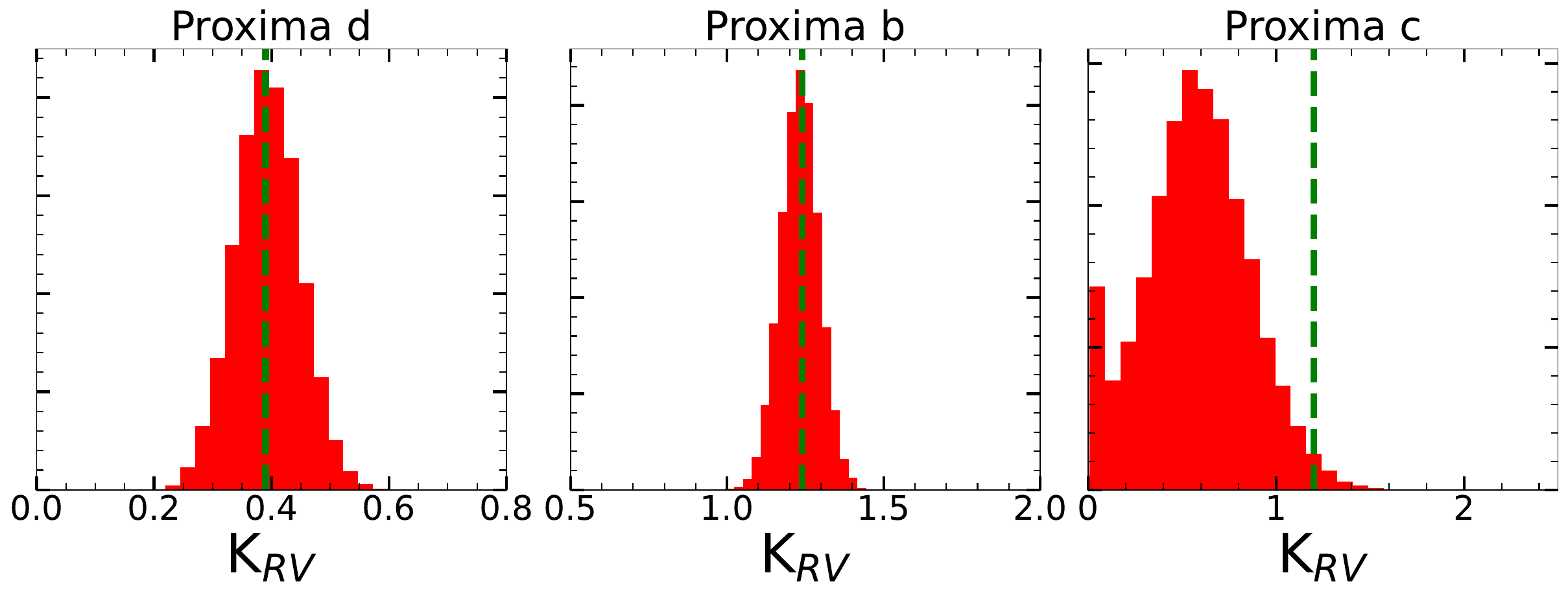}
	\caption{\textbf{RV amplitudes.} Posterior distributions of the amplitudes of the RV signals. The vertical dashed lines show the previously reported amplitudes of Proxima d, b, and c.}
	\label{rv_amplitudes}
\end{figure}

Following the same steps as with previous datasets, and in an attempt to find the signal of the candidate Proxima c, we proceeded with a guided search. We set up a 3-planets model. We maintained the priors of $\mathcal{N}[11.1868,0.1]$ days for Proxima b, and $\mathcal{N}[5.122,0.05]$ days for Proxima d, and used a prior $\mathcal{N}[1900,200]$ days, following \citet{Damasso2020}. We measure an RV semi-amplitude of 39.2 $\pm$ 5.5 cm$\cdot$s$^{-1}$ for Proxima d, an orbital period of 5.12332 $\pm$ 0.00033 days, and a phase of 0.519 $\pm$ 0.031, which corresponds to a time of inferior conjuction $T_{c} = 10557.54 \pm 0.16$ (BJD -- 2450000). For Proxima b we measure an RV semi-amplitude of 1.234 $\pm$ 0.062 m$\cdot$s$^{-1}$, an orbital period of 11.18472 $\pm$ 0.00051 days, and a phase of --0.020 $\pm$ 0.011, corresponding to a $T_{c} = 10548.59 \pm 0.12$ (BJD -- 2450000). We do not significantly detect the signal of Proxima c. We measure a semi-amplitude of 58 $\pm$ 28 cm$\cdot$s$^{-1}$ (< 1.4 m$\cdot$s$^{-1}$, 99.7\% confidence). We recover an orbital period of 1800$^{+100}_{-57}$ days, and a phase of 0.17$^{+0.17}_{-0.10}$, corresponding to a $T_{c} = 9060^{+250}_{-120}$ (BJD -- 2450000). This hints at the presence of a long period signal with similar parameters, although not fully compatible, to those reported of the candidate Proxima c. Figure~\ref{rv_amplitudes} shows the posterior distribution of the RV amplitude of the three signals. 

As the cycle model could have an effect on the determination of the parameters of long period planets, we repeated the 3-planets model in the case of no-cycle model in RV, a simpler cycle model (1-sinusoidal), and a cycle model with a shorter period (following the results of \citet{Wargelin2024}). All models provided upper limits to the presence of the signal consistent with that obtained using the default model.

From the two last models, we can confirm the detection of two signals, at periods of 5.123 days, 11.185 days. We cannot confirm the signal of the candidate Proxima c. However, we find hints of a long period signal that might exists with not-too-different parameters. This signal, however, would have a significantly lower RV amplitude. 

\subsubsection{Stability of the detected signals} \label{section_stability}

To test the stability of the planetary signals over time, we performed a new guided model on the two signals attributed to Proxima b and d, using apodised signals \citep{Gregory2016, Hara2022b}. We define the apodised signals as: 

\begin{equation} \label{eq_circular}
    y(t)=-K \cdot \sin(2 \pi \cdot (t - t_{0})/P_{pl}) \cdot G(\mu,\sigma)
\end{equation}

\noindent where $G$ is Gaussian function with centre $\mu$, width $\sigma$, and height 1. This model tests whether the intensity of the signal is uneven over the baseline of observations, and has proven effective in the past \citet{Dumusque2017}. As planetary signals are stable over time, we would expect to retrieve an undefined $\mu$ and a large $\sigma$. However, if we obtain a well-defined Gaussian, it would mean the signal is only present in part of the dataset. For this test, we used the same priors for the planetary parameters as stated before in the guided inferences. For the priors of the Gaussian, we used $\mathcal{U}[1634,10560]$ (BJD -- 2450000) for $\mu$ to cover the full baseline, and $\mathcal{U}[3, 15]$ for ln~$\sigma$ (20d to 3 $\times$ 10$^{6}$ days), to allow for widths of the Gaussian much larger than the observation baseline, without biasing the results toward large widths. 

The signals of Proxima b and d show strong stability, with a centre consistent with the full dataset, and median values of $\sigma$ of 180 000 days, and 120 000 days, respectively. These values are much larger than the observing baseline. 

\begin{figure}[!ht]
    \centering
	\includegraphics[width=9cm]{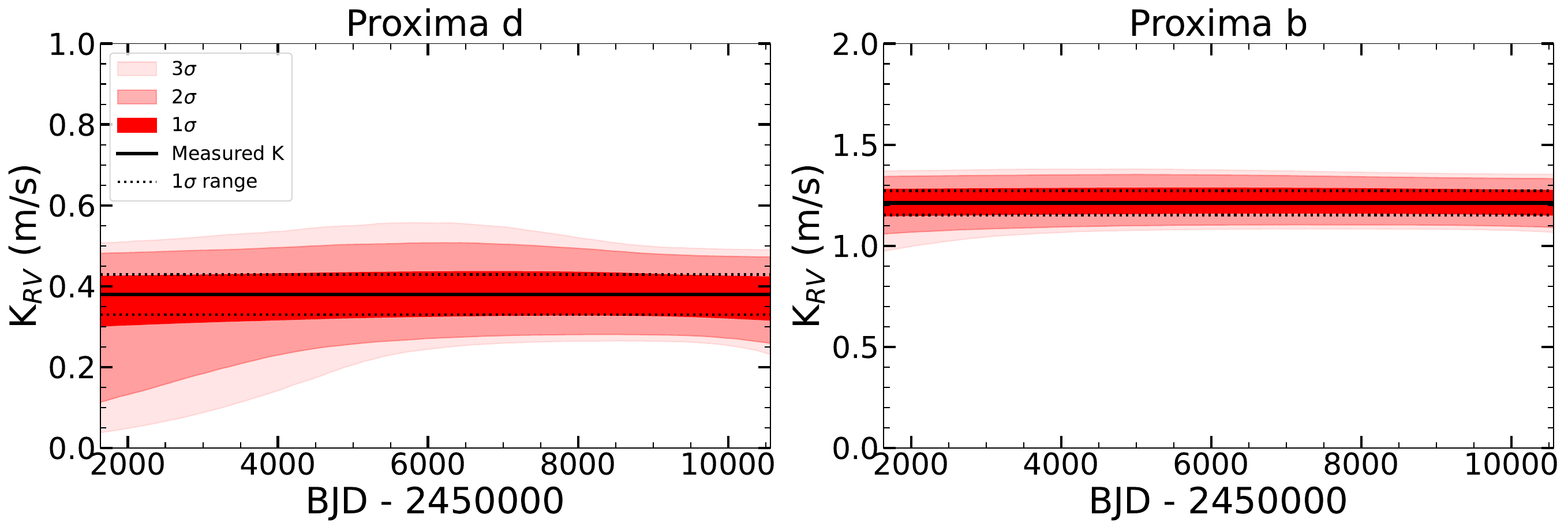}
	\caption{\textbf{Distribution of apodised amplitudes.} The red shaded area shows the confidence intervals of the amplitudes of the apodised signals over time. The horizontal black lines show the median value and 1-$\sigma$ range of the static model.}
	\label{apod_amplitudes}
\end{figure}

Figure~\ref{apod_amplitudes} shows the measured evolution of the amplitude over time. For Proxima b and Proxima d, the distribution is consistent with the final values of the global model throughout the complete baseline of observations. The range is wider at the beginning, coinciding with the data with the lowest observing cadence. Nevertheless, the signals are always present and remain at consistent amplitude with the median amplitude measured in the static model. 

\subsubsection{Consistency between instruments} \label{consistency_test}

We performed a consistency test between the instruments by keeping the same global activity model and using independent sinusoidal signals for the planetary components in all the instruments. Since we do not expect all signals to be detectable in all instruments (e.g., Proxima d at 37 cm$\cdot$s$^{-1}$), we kept the same priors of the guided search, with the aim of testing the differences in semi-amplitude and phase. For planetary signals, we expect all instruments to provide consistent results. For potential non-detections we expect to obtain upper limits that are consistent with the measurements in those instruments that provide a detection. Detections in a specific instrument, incompatible with upper limits from other instruments, would be indicative of instrumental effects.

\begin{figure}[ht]
    \centering

    \includegraphics[width=9cm]{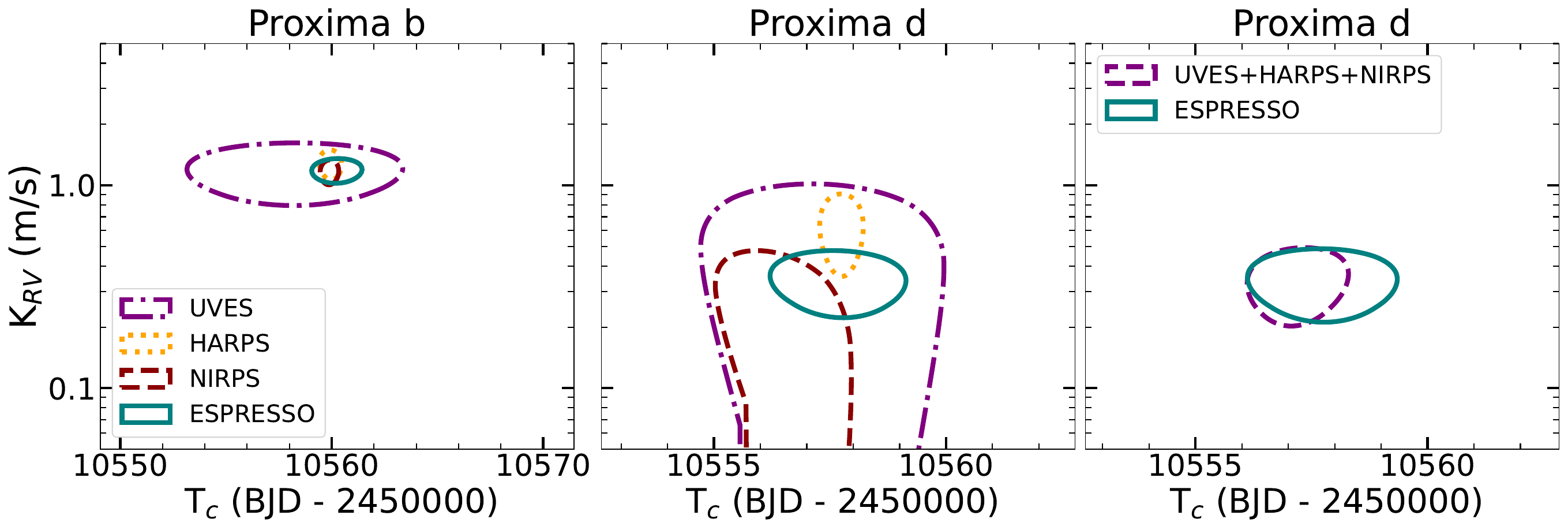}

	\caption{\textbf{Comparison of the parameters between instruments.} RV semi-amplitude vs. time of inferior conjunction for the solutions of individual instruments for Proxima b (left) and d (center). The right panel shows the same comparison for ESPRESSO-only and all other instruments combined.  The lines encapsulate the 95\% confidence interval.}
	\label{inst_comp}
\end{figure}

Figure~\ref{inst_comp} shows the amplitude versus the time of inferior conjunction of the two signals for the different instruments. We found the properties of the signals of Proxima b and d to be consistent across all instruments. In the case of Proxima d, we find consistent parameter estimates in the data of ESPRESSO and HARPS. The amplitude measured in the HARPS data is slightly larger, but consistent within their respective confidence intervals. The constraints imposed by NIRPS and UVES are consistent with the measurement from ESPRESSO. These results are within the expectations of having a stable signal present in datasets of varying precision. In fact, the parameters obtained from all instruments are 1-$\sigma$ compatible. 

We performed one additional test of consistency between instruments. We compared the results obtained for Proxima d with ESPRESSO, against the results obtained using all other instruments combined. While these instruments, independently, struggled to provide a significant measurement, the combined dataset is capable of reaching it. Figure~\ref{inst_comp}, right panel, shows the comparison between the parameters obtained with the ESPRESSO data, and those obtained by combining all other instruments. The parameters are fully consistent, showing that the presence and parameters of Proxima d are instrument-independent.

\subsubsection{Adopted model} \label{adopted_model}

Following the results of sections~\ref{section_stability} and~\ref{consistency_test}, we adopt a model that includes two sinusoidal signals to account for planets Proxima b and d. To establish the final parameters of these signals, we run a model using $\mathcal{U}$ priors for the periods, centered at the known period and with a range of $\pm$ 40\% of the period. The rest of the priors for all parameters remain the same. We compared circular and Keplerian models and the results were consistent with each other. We measure eccentricities < 0.1 for Proxima b and < 0.25 for Proxima d (95\% confidence). The rest of the parameters were consistent within 1$\sigma$. The circular model was statistically favoured, with a $\Delta$ ln Z = + 3 in favour of the simpler model. We adopt the circular model. The full set of priors and parameters of the adopted model is available in Appendix~\ref{append_tables}.

We obtain parameters that are fully consistent with previous measurements. We measure Proxima b to have an orbital period of 11.18465 $\pm$ 0.00052 days and induce an RV semi-amplitude of 1.226 $\pm$ 0.062 m$\cdot$s$^{-1}$. We measure a phase of --0.020 $\pm$ 0.011, which corresponds to a T$_{c}$ = 10548.59 $\pm$ 0.12 (BJD - 2450000). We measure Proxima d to have an orbital period of 5.12338 $\pm$ 0.00035 days and induce a semi-amplitude of 39.2 $\pm$ 5.7 cm$\cdot$s$^{-1}$. We measure a phase of 0.522 $\pm$ 0.031, which corresponds to a T$_{c}$ = 10557.55 $\pm$ 0.16 (BJD - 2450000). 

Over the full dataset, we measure the stellar cycle to have a period of 6560 $\pm$ 85 days. It induces an RMS of 1.55 m$\cdot$s$^{-1}$ in the FWHM$_{VIS}$ data and 0.56 m$\cdot$s$^{-1}$ in the RV$_{VIS}$ data, with a peak-to-peak amplitude of 1.7 m$\cdot$s$^{-1}$. In the RV$_{VIS}$, most of the detectable signal is present at half the cycle. We measure the stellar rotation period to be 83.2 $\pm$ 1.6 days, with a timescale of evolution of 60 $\pm$ 12 days. We measure an RV jitter of 57.5 $\pm$ 7.0 cm$\cdot$s$^{-1}$ in the NIRPS data, 1.64 $\pm$ 0.21 m$\cdot$s$^{-1}$ in the HARPS-03 data, 1.53 $\pm$ 0.12 m$\cdot$s$^{-1}$ in the HARPS-15 data, 92 $\pm$ 21 cm$\cdot$s$^{-1}$ in the UVES data, 42.5 $\pm$ 9.5 cm$\cdot$s$^{-1}$ in the ESPRESSO-18 data, and 30 $\pm$ 7.0 cm$\cdot$s$^{-1}$ in the ESPRESSO-19 data (95\% confidence). The residuals after the fit do not highlight the presence of any significant periodic signal and show an RMS of 0.81 m$\cdot$s$^{-1}$ in the NIRPS data, 1.51 m$\cdot$s$^{-1}$ in the HARPS-03 data, 1.87 m$\cdot$s$^{-1}$ in the HARPS-15 data,  0.59 m$\cdot$s$^{-1}$ in the UVES data, and 0.20 m$\cdot$s$^{-1}$ in the ESPRESSO data.

\begin{figure*}[!ht]
    \centering
	\includegraphics[width=16cm]{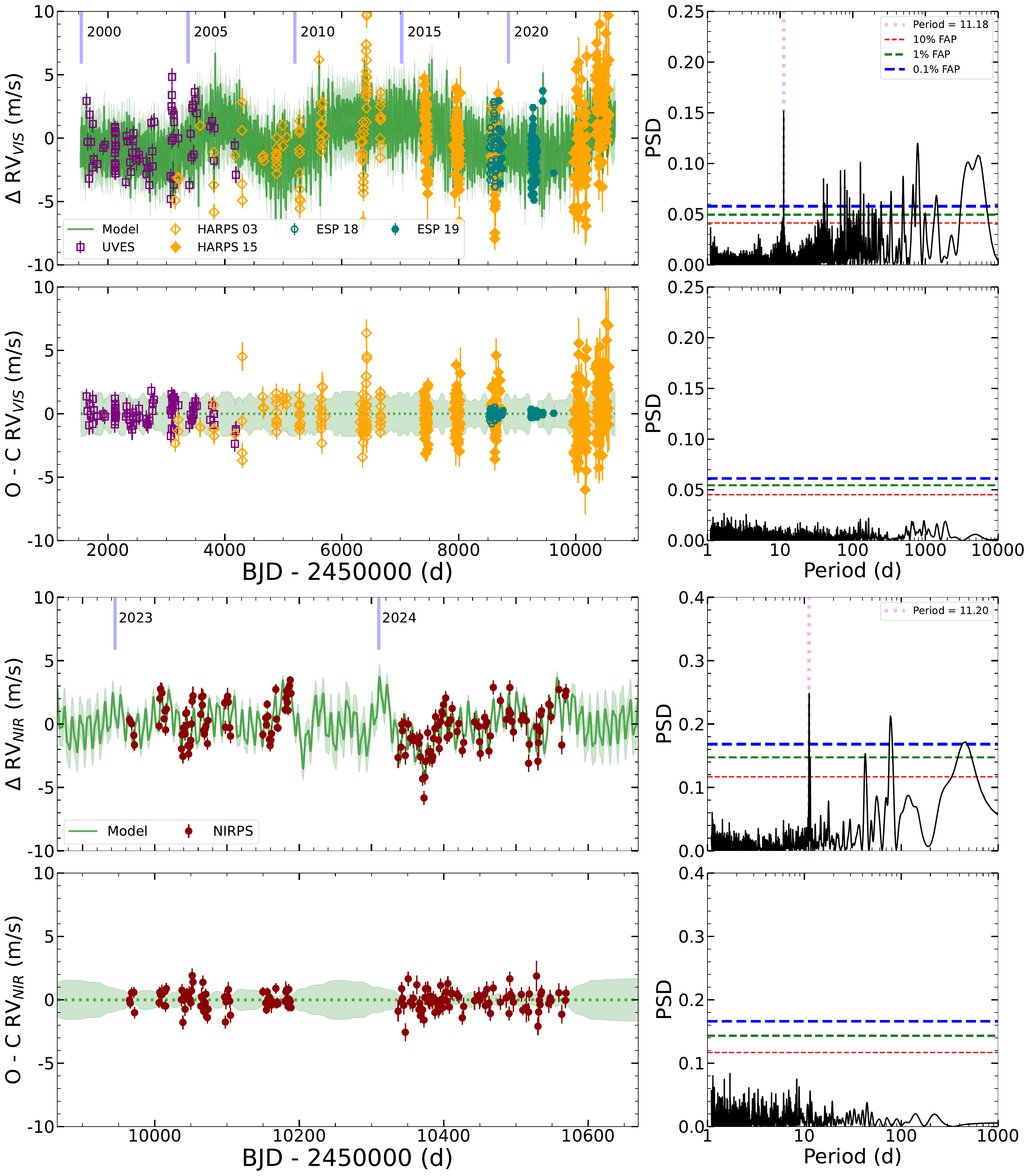}
	\caption{\textbf{RV model using the full dataset.} The two top panels show the VIS RV data (detrended from CRX), with the best model fit (top), and the residuals after the fit (bottom), along with the periodograms of both (right). The two bottom panels show the same for the NIR RV data.}
	\label{model_all}
\end{figure*}

\begin{figure}[!ht]
    \centering
	\includegraphics[width=9cm]{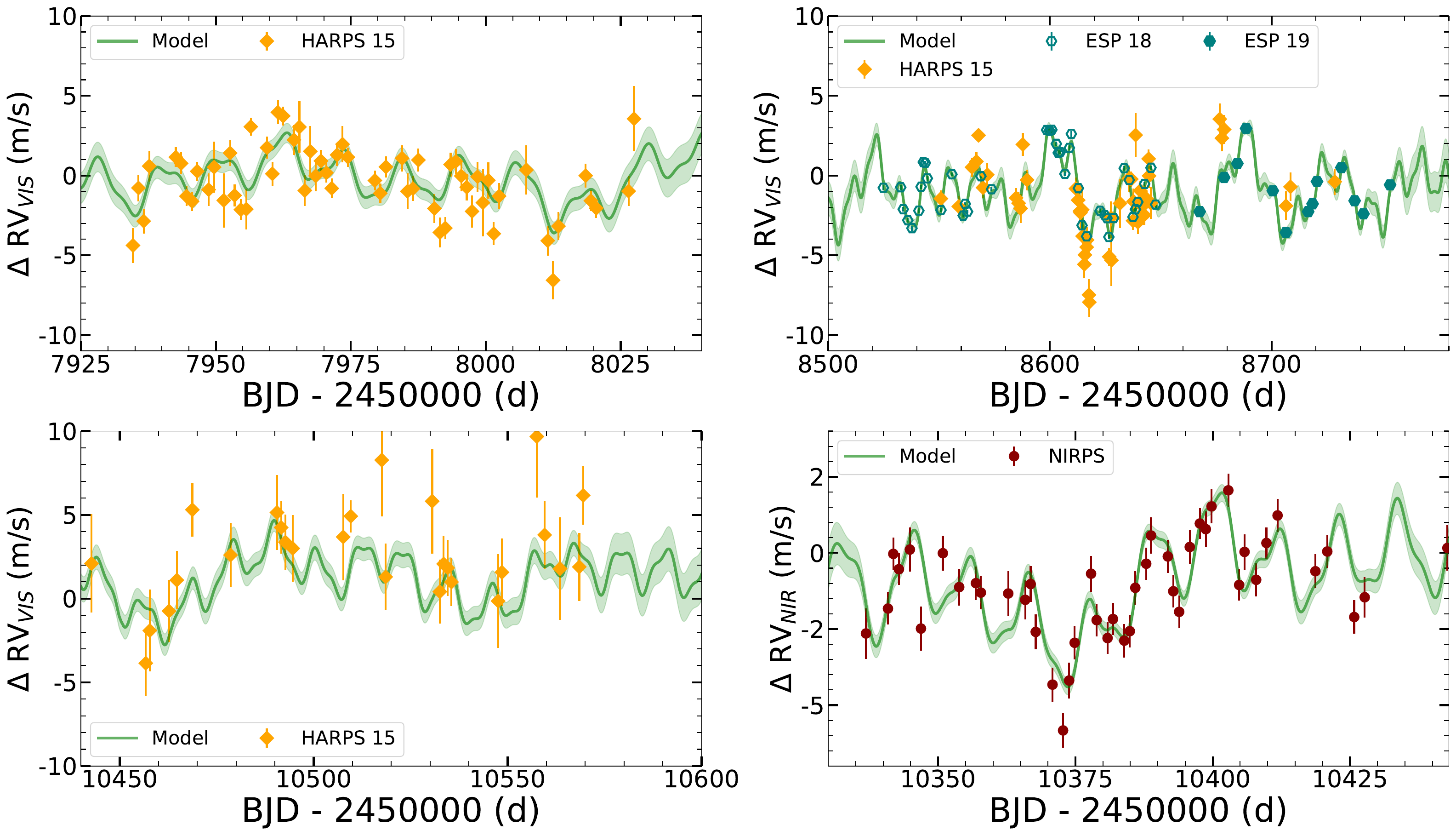}
	\caption{\textbf{Zoom to selected observing campaigns.} RV data of HARPS, ESPRESSO, and NIRPS, detrended from CRX, with the best model fit.}
	\label{model_all_zoom}
\end{figure}

\begin{figure}[!ht]
    \centering
	\includegraphics[width=9cm]{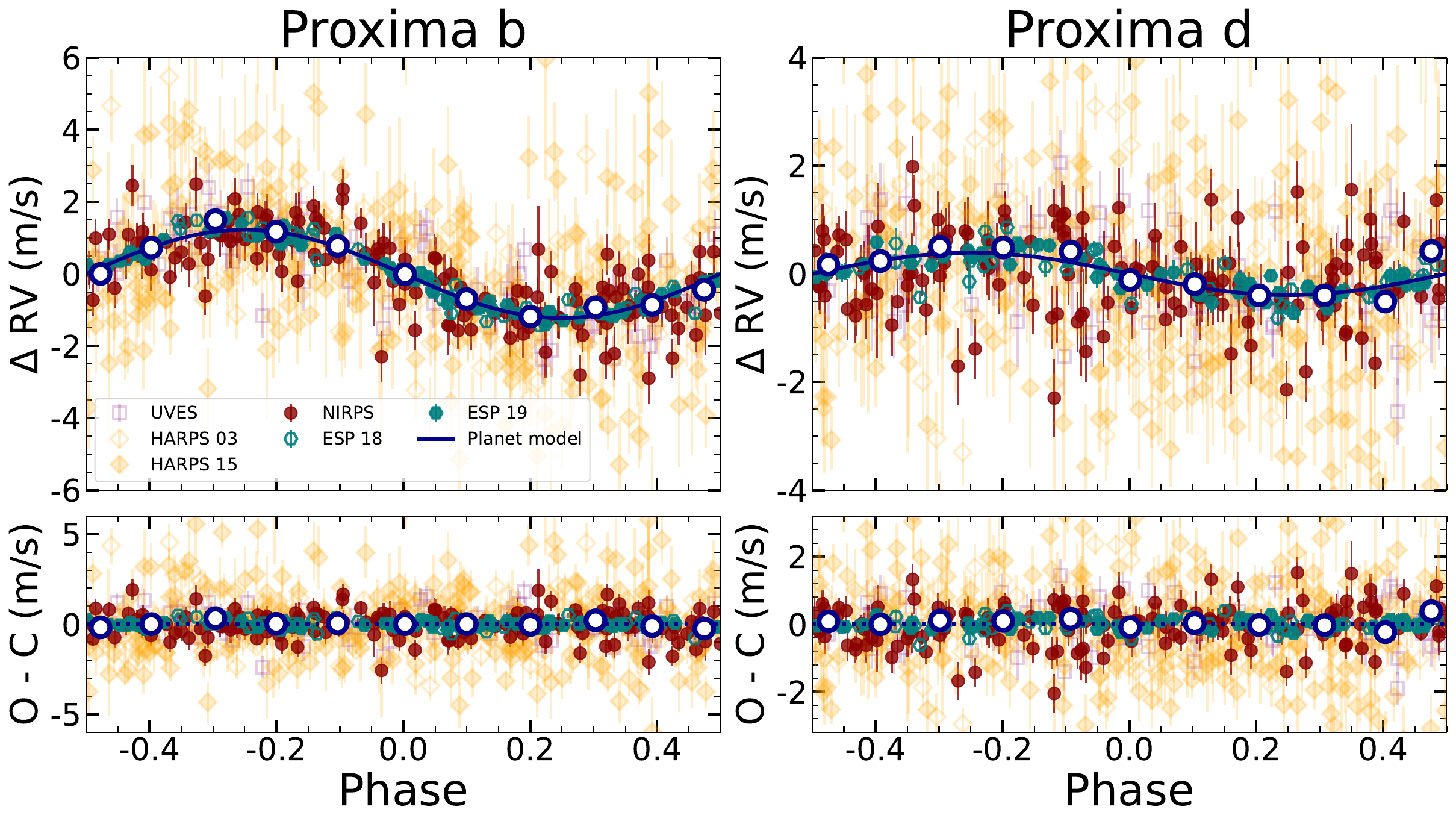}
	\caption{\textbf{Phase-folded plots of the planetary-induced RV signals.} RV variations induced by Proxima b and d with the best model fit (top panels), and the residuals after the fit (bottom panels). The blue circles with white filling  show the average at every phase bin.}
	\label{model_all_phase}
\end{figure}

Figure~\ref{model_all} shows the full RV dataset, along with the best-fit model, the residuals, and their respective periodograms. Table~\ref{table_adopted} shows the full set of priors, and the measured parameters, of the adopted model. Figure~\ref{model_all_zoom} provides a zoomed-in view of specific observing seasons. We separated the NIR and VIS data due to the different amplitudes of the activity model. Figure~\ref{model_all_phase} shows the phase-folded plots of the planetary-induced RV signals. Appendix~\ref{append_fwhm} contains the equivalent figures for the FWHM data and model.

\section{Discussion} \label{sec_disc}

\subsection{Performance of NIRPS} 

We studied the system of Proxima using the NIRPS spectrograph, while obtaining simultaneous HARPS spectra, providing us with a dataset in which NIRPS and HARPS share the same baseline, cadence, and exposure time. We measured the NIRPS data to have typical uncertainties of 55 cm$\cdot$s$^{-1}$, compared to \hbox{1.4 m$\cdot$s$^{-1}$} for the HARPS data. The NIRPS data showed a significantly lower RMS of 1.68 m$\cdot$s$^{-1}$, compared to 3.54 m$\cdot$s$^{-1}$ for HARPS. Most of this can be attributed to the significantly higher S/N data obtained with the NIRPS spectrograph at the same exposure time, while part of it can be ascribed to intrinsic instrumental stability. After modelling the combined GTO data (see sect.~\ref{gto_data}), and subtracting the models, we measured an RMS of the residuals of 0.81 m$\cdot$s$^{-1}$ in NIRPS, and 1.74 m$\cdot$s$^{-1}$ for HARPS. 

\begin{figure}[!ht]
    \centering
	\includegraphics[width=9cm]{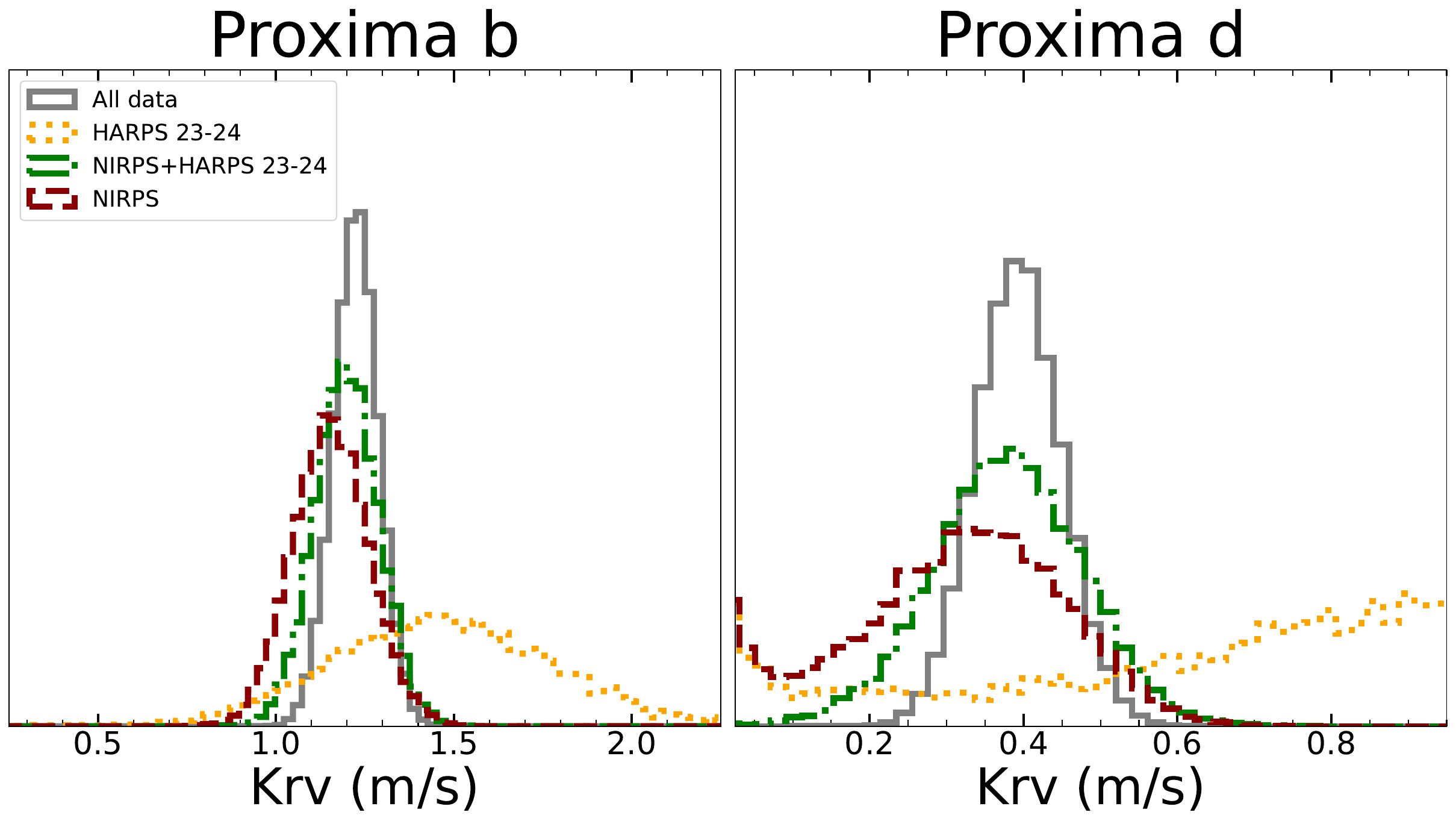}
	\caption{\textbf{Comparison of the amplitudes of NIRPS with other instruments.} Posterior distributions of the RV amplitudes of Proxima b and d with NIRPS, HARPS, NIRPS+HARPS, and the full spectroscopic datasets.}
	\label{rvamp_comp}
\end{figure}

With this dataset, we provided a significant detection of Proxima b in the NIRPS dataset, while only a marginal detection in the HARPS data. Figure~\ref{rvamp_comp} shows the comparison of the amplitudes of the signals of Proxima b and d in the NIRPS, HARPS 2023-2024 data, the combined NIRPS+HARPS dataset, and the full spectroscopic dataset. Using the NIRPS data, we obtain a $\sim$ 11-$\sigma$ determination of the amplitude of Proxima b, which is accurate with the parameters obtained using ESPRESSO data. Using the HARPS 2023--2024 data, we achieved a $\sim$ 4-$\sigma$ determination. The combination of both datasets reached a $\sim$ 12-$\sigma$ determination, and the full dataset a $\sim$ 21-$\sigma$ determination. In the case of Proxima d, the NIRPS dataset alone cannot provide a significant detection. Using the NIRPS data, we obtained a $\sim$ 2-$\sigma$ determination of the amplitude. Using the HARPS 2023--2024 data, we can barely reach $\sim$ 1-$\sigma$ determination. The combination of both reached a $\sim$ 4-$\sigma$ determination, and the full dataset a $\sim$ 7-$\sigma$ determination. While the amplitude measurement of the signal in the NIRPS dataset cannot be considered significant, the peak of the posterior distribution is compatible with the posterior distribution obtained using the full dataset. In contrast, for the HARPS 2023--2024 dataset, the posterior distribution is mostly flat. 

These results show that, in the case of Proxima, and at equal baseline and cumulative exposure time, NIRPS performs significantly better than HARPS. It would be expected that, with a larger dataset, NIRPS alone could be able to provide a fully independent detection of Proxima d, and the detection of other planetary signals inducing sub-m$\cdot$s$^{-1}$ amplitude. It has to be said that, for the most recent dataset, the exposure time was optimised for NIRPS observations. Proxima is significantly brighter in the NIR than in the VIS, which permitted the use of shorter exposure times for the NIRPS observations. With longer exposure times it would be expected to obtain higher quality HARPS data. 

\begin{figure}[!ht]
    \centering
	\includegraphics[width=9cm]{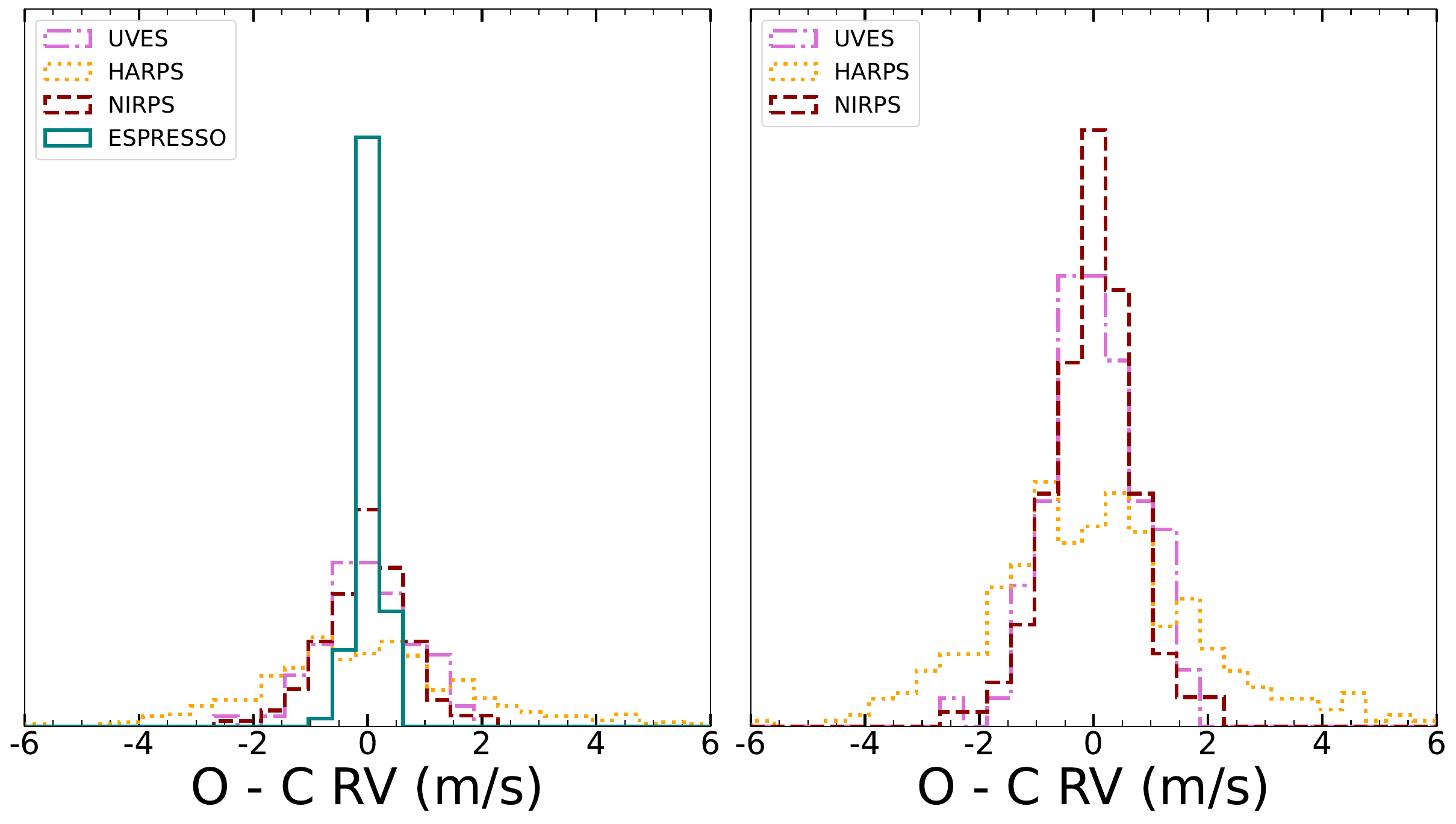}
	\caption{\textbf{Comparison of RV residuals.} Distribution of the residuals of the RV data after subtracting the best fit. The left panel includes all instruments. The right panel excludes ESPRESSO, for an easier visualization of the rest.}
	\label{res_comp}
\end{figure}

Figure~\ref{res_comp} shows the distribution of the RV residuals after the fit for all instruments, using the adopted model from section~\ref{adopted_model}. Here, once again, NIRPS compares favourably with HARPS. NIRPS residuals show an RMS of 80 cm$\cdot$s$^{-1}$, comparable to UVES (80 cm$\cdot$s$^{-1}$), which is mounted on an 8-meter telescope. HARPS shows residuals of 1.5 m$\cdot$s$^{-1}$, while ESPRESSO stands on a league of its own, with just 20 cm$\cdot$s$^{-1}$ residuals. 

\subsection{The planetary system of Proxima}

Our analysis provided significant detections of the signals of Proxima b and Proxima d. In the case of Proxima d, we provided significant evidence of the signal being present in spectrographs other than ESPRESSO and of its parameters being instrument-independent. For the signals of Proxima b and d we found consistent parameters for parameters of the signal across the different models and instruments. We did not find conclusive evidence of the presence of the signal attributed to Proxima c in the data.

Table~\ref{param_planets} shows the measured planetary parameters. The table provides updated ephemerides for the two planets in the system. Figure~\ref{model_all_phase} shows the phase-folded RV data of two planetary signals. 

\begin{table} 
    \begin{center}
    \caption{Parameters of the planets of the Proxima system, using the adopted model (circular).  \label{param_planets}}
    \begin{tabular}[centre]{l c}
    \hline
    Parameter  & Value \\ \hline
    \textbf{Proxima d} \\
    $T_{0}$ -- 2450000 [d] & 10557.55 $\pm$ 0.16 \\
    $P_{\rm orb}$ [d] & 5.12338 $\pm$ 0.00035  \\
    $m_p$ sin $i$ [M$_{\oplus}$] & 0.260 $\pm$ 0.038 \\
    $m_{p~47^{\circ}}$ [M$_{\oplus}$]$^{1}$ & 0.357 $\pm$ 0.072 \\
    $a$ [au] &  0.02881 $\pm$ 0.00017 \\
    $e$ & 0 (fixed)\\
    Incident flux [$S_{\oplus}$] & 1.814 $\pm$ 0.098  \\
    T$_{\rm eq~A = 0.3}$ [K] & 282 $\pm$ 23  \\ 
    K [cm$\cdot$s$^{-1}$]  & 39.2 $\pm$ 5.7  \\ \\
    \textbf{Proxima b} \\
    $T_{0}$ -- 2450000 [d] & 10548.59 $\pm$ 0.12  \\
    $P_{\rm orb}$ [d] & 11.18465 $\pm$ 0.00053 \\
    $m_p$ sin $i$ [M$_{\oplus}$] &  1.055 $\pm$ 0.055 \\
    $m_{p~47^{\circ}}$ [M$_{\oplus}$]$^{1}$ &  1.44 $\pm$ 0.21 \\
    $a$ [au] &   0.04848 $\pm$ 0.00029\\
    $e$ & 0 (fixed) \\
    Incident flux [$S_{\oplus}$] &  0.641 $\pm$ 0.065 \\
    T$_{\rm eq~A = 0.3}$ [K] & 218 $\pm$ 18 \\ 
    K [cm$\cdot$s$^{-1}$]  & 122.6 $\pm$ 6.2 \\
    \hline
    \end{tabular}    
    \end{center}
    $^{1}$The values $m_{p~47^{\circ}}$ correspondsto the planetary masses assuming orbits coplanar to a rotation axis tilted 47 $\pm$ 7 $^{\circ}$ \citep{Klein2021}.
    \end{table}

\subsubsection{Proxima b}

Proxima b was announced using HARPS data \citep{AngladaEscude2016} and later confirmed using ESPRESSO data \citep{Masca2020}. We confirm its presence once again using near-infrared radial velocities with NIRPS. We obtained a very significant detection within the FIP framework, as well as a measurement of its amplitude that is > 10-$\sigma$ different from zero. The planetary parameters derived from the NIRPS data are fully consistent with previous determinations. 

Taking advantage of the complete dataset, we studied the temporal stability of the signal. We found its properties to be very consistent over the complete baseline of observations. Additionally, we demonstrated that the RV amplitude and phase obtained in the data of different instruments are consistent. We measure Proxima b to have an orbital period of 11.185 days, a minimum mass of 1.055 $\pm$ 0.055 M$_{\oplus}$, and an eccentricity lower than 0.1. It orbits at a distance of 0.04848 $\pm$ 0.00029 au and induces an RV amplitude of 1.226 $\pm$ 0.062 m$\cdot$s$^{-1}$. With the complete dataset, we obtained ephemerides with a precision of 2.88 hours. 

\subsubsection{Proxima d}

The signal attributed to Proxima d was first discussed by~\citet{Masca2020}. The planet was later proposed as a candidate by~\citet{Faria2022}. Both works used ESPRESSO data. We provide a significant detection of Proxima d in our combined dataset, with parameters consistent with those proposed by~\citet{Faria2022}. In addition, we find strong evidence of the stability of the signal over time, and of its presence in instruments other than ESPRESSO. Our results confirm the presence of Proxima d, with an orbital period of 5.12338 $\pm$ 0.00035 days, a minimum mass of 0.260 $\pm$ 0.038 M$_{\oplus}$, and an eccentricity lower than 0.25. It orbits at a distance of 0.02881 $\pm$ 0.00017 au and induces an RV amplitude of 39.2 $\pm$ 5.7 cm$\cdot$s$^{-1}$. We provide ephemerides with a precision of 3.8 hours. 

\subsubsection{Proxima c}

Proxima c was proposed as a candidate by \citet{Damasso2020}, but later challenged by \citet{Artigau2022}. We attempted to detect the signal of Proxima c in our dataset but found no evidence of a significant signal within the FIP framework. In addition, we also did not find conclusive evidence when modelling the data using narrower priors around the period proposed by \citet{Damasso2020}. We found hints of the presence of a signal at a similar period, but the amplitude and time of inferior conjunction were not consistent with those previously reported. By performing simple injection-recovery tests, we confirmed that we are sensitive to signals of 1 m$\cdot$s$^{-1}$ amplitude at periods comparable to the proposed period of Proxima c.

\subsubsection{Estimating the true masses of the planets}

The rotation axis of Proxima has been estimated to be tilted 47 $\pm$ 7 $^{\circ}$ with respect to the line of sight \citep{Klein2021}. In addition, \citet{Anglada2017} reported the detection of a dust belt around the star, with a tilt angle of $\sim$ 45$^{\circ}$, which would make it coplanar to the rotation axis. Assuming the planetary orbits to be coplanar with the rotation axis, the masses of Proxima b, and d, would be 1.44 $\pm$ 0.21 M$_{\oplus}$, and 0.357 $\pm$ 0.072 M$_{\oplus}$, respectively.

\subsubsection{Refinement of the planetary ephemeris}

The planetary system of Proxima is one of the most promising candidates for the detection of atmosphere signatures in reflected light, or thermal emission, in rocky planets. It has been identified as a key target for RISTRETTO~\citep{Blind2022, Blind2024}, 
ANDES~\citep{Palle2023} and LIFE~\citep{Quanz2022}. Maintaining precise, and up-to-date, ephemeris of Proxima will be key to maximise the potential of detecting the planet signature and, potentially, information on the composition of its atmosphere. The addition of the NIRPS GTO data provides a significant improvement on the planetary ephemeris. 

Using the previously publicly available data, we can measure the period of Proxima b with an uncertainty of 0.00076 days (65 seconds), and the time of inferior conjunction with an uncertainty of 0.16 days (3.8 hours). The addition of the NIRPS GTO data reduced the uncertainty in the period to 0.00053 days (46 seconds), and the uncertainty in the time of inferior conjuction to 0.12 days (2.9 hours). Moreover, the propagation of the uncertainty meant that a potential observation taken during 2026 would have had an uncertainty on the ephemeris of $\sim$ 6 hours, which with the new data is reduced to $\sim$ 3 hours. 

In the case of Proxima d, with previously available data it was possible to constraint the period to an uncertainty of 0.0011 days (95 seconds), and the time of inferior conjunction to a precision of 0.26 days. With the NIRPS GTO data, we reduced the period uncertainty to 0.00035 days (30 seconds), and the time of inferior conjunction to 0.16 days (3.8 hours). A potential observation performed in 2026 would have had an uncertainty on the ephemeris of $\sim$ 13.5 hours, that after the inclusion of the new data is reduced to $\sim$ 4 hours. 

\subsubsection{Compatibility limits} \label{det_lims}

\begin{figure}[!ht]
    \centering
	\includegraphics[width=9cm]{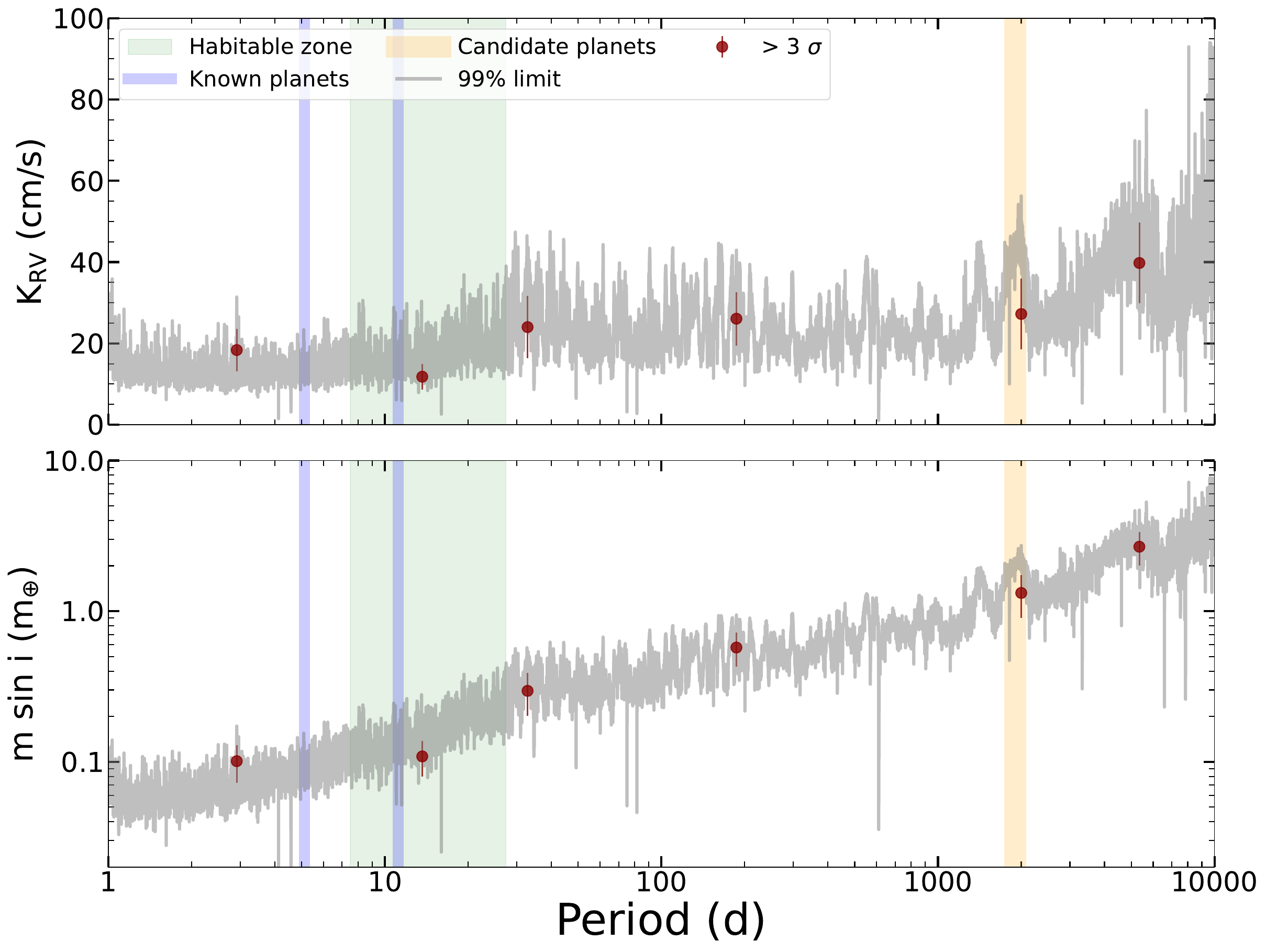}
	\caption{\textbf{Compatibility limits.} The upper panel shows the RV amplitude limit (99\%) as a function of orbital period. The blue shaded bars show the periods of the confirmed planets, Proxima b and Proxima c. The orange shaded line shows the period of the candidate Proxima c. The green shaded area shows the habitable zone. The red points shows those cases in which we obtained a > 3 $\sigma$ determination of the RV amplitude. The lower panels shows the same, but for planetary masses.}
	\label{fig_det_lims}
\end{figure}

Using the results from the adopted model on the full dataset, we measured the compatibility limits at a wide range of orbital periods of Proxima. We froze most of the parameters of the model (trends, cycle, GP, planets) and left free only the white noise and zero point RV components. We included a third sinusoid in the model, with only two parameters, RV amplitude and phase (parametrised as before). We fixed the orbital period to the period we wanted to test. This reduced the model to 12 parameters (5 jitters, 5 zero-points, 2 sine parameters), which could be evaluated in a few seconds. Then randomly swept over the full range of orbital periods (1 - 10~000 days). From the posterior distribution of each try, we computed the 99\% upper limit in RV amplitude, its median value, and standard deviation. Figure~\ref{fig_det_lims} shows these 99\% RV limits as a function of orbital period, both in RV amplitude, and transformed to planetary mass. In addition, it shows the few cases in which the median value of the RV amplitude was more than 3-$\sigma$ different from zero. We found that, for periods shorter than ~10 days, we can exclude the presence of any additional signal with an amplitude larger than 20 cm$\cdot$s$^{-1}$ (minimum masses 0.08 -- 0.15 M$_{\oplus}$). Within the habitable zone, we can exclude, for the most part, the presence of signals with amplitudes between 20 and 30 cm$\cdot$s$^{-1}$ ($m$ sin $i$ 0.1 -- 0.3 M$_{\oplus}$). At periods between the edge of the habitable zone (30d) and 100 days, we can exclude the presence of additional signals beyond 30 -- 40 cm$\cdot$s$^{-1}$ ($m$ sin $i$ 0.3 -- 0.6 M$_{\oplus}$). Between 100 and 1000 days, the limit remains similar (0.5 -- 1.0 M$_{\oplus}$). Beyond 1000-day orbital period, the limits raise up to $\sim$ 60 cm/s ($\sim$ 4 M$_{\oplus}$).

While we did not formally detect the signal of Proxima c in our previous analysis (we only found an upper limit on a guided model), there is an RV amplitude excess at the period of the candidate consistent with the upper limit we previously measured. There seems to be an RV signal around 2000 days orbital period. This signal, however, needs to be of significantly lower amplitude than reported by ~\citet{Damasso2020}. Our limit here is somewhat lower than what we derived when running a full 3-planet model. This is likely due to the activity model changing to accomodate the signal in the case of the full model. In this case, the cycle model is fully frozen and, while the GP shape can still change slightly, its parameters cannot. 

We encountered a few situations in which we measured amplitudes more than 3-$\sigma$ different from zero.  These signals have periods of 2.91d, 13.6d, 33d, 186d, 2000d, and 5350d. Their amplitudes are 18 cm$\cdot$s$^{-1}$, 12 cm$\cdot$s$^{-1}$, 25 cm$\cdot$s$^{-1}$, 26 cm$\cdot$s$^{-1}$, 27 cm$\cdot$s$^{-1}$, and 40 cm$\cdot$s$^{-1}$, respectively. None of these were detected under a rigorous detection exercise. However, they could serve as hints that could guide future studies. The 2.91d in particular coincides with a low-level peak in the FIP periodogram of the full dataset (see Figure~\ref{fip_periodogram_all}).

\subsection{Stellar activity}

In conjunction with the planetary model, we characterised the activity variations of Proxima. We modelled the magnetic cycle in photometry, its induced variations in the width of the spectral lines, and the RV-induced variations. We modelled the stellar rotation, using multi-dimensional Gaussian processes regression. We measured the amplitude of the variations induced in photometry, in the width of spectral lines, and in RV, including the difference in amplitude between the visible and near-infrared data.  

\subsubsection{Magnetic cycle}

For a long time, fully convective M dwarfs (i.e., M3.5 and later) were believed to be unable to support the magnetic dynamos necessary to support magnetic cycles. In recent years, however, several fully convective M dwarfs have shown to have magnetic cycles of lengths comparable to the solar cycle \citep{Masca2016}. In particular, a 7-8 year magnetic cycle was identified in Proxima using long-term photometric monitoring and X-ray data \citet{Jason2007, Masca2016, Wargelin2017,Wargelin2024}.

For this work, we combined ground-based photometric data spanning 23.7 years of observations. The analysis of our data suggests the cycle is significantly longer than previously thought. Assuming we have observed the full cycle, we obtain a measurement of its length of 17.96 $\pm$ 0.23 years (6560 $\pm$ 85 days). The photometric variations due to the cycle are of $\sim$ 100 ppt peak-to-peak, comparable to those caused by the stellar rotation. This suggests that the change in the number of spots is not as significant as in a star like the Sun. Even at its quietest phase, the surface of Proxima maintains enough spot coverage to create $\sim$ 100 ppt peak-to-peak variations. The shape of the cycle is non-sinusoidal, which creates power at several of the harmonics of the main period. We modelled it as a combination of four sinusoidals, with periods $P_{cycle}$,$P_{cycle}/2$ ,$P_{cycle}/3$, and $P_{cycle}/4$. This asymmetry, combined with the shorter baseline of observations used in previous works, likely explains most previous estimates. 

While the parameters obtained in the fit are well constrained, and the model is visually compelling, the short baseline of the data compared to the measured period makes it difficult to be sure that this is the true period of the cycle. The results of Figure~\ref{diff_rot} hint at this period being at least close to the true period, by showing that the evolution of the photometric period shows a solar-like evolution when phase-folded with this cycle period. However, the difference between a cycle of 18 years and a cycle of half the period combined with a longer-term non-periodic variation is not obvious to disentangle with our dataset. The continuation of the phototometric monitoring over the next decades will be crucial to confirm the true period of Proxima's cycle. 

\begin{figure}[!ht]
    \centering
	\includegraphics[width=9cm]{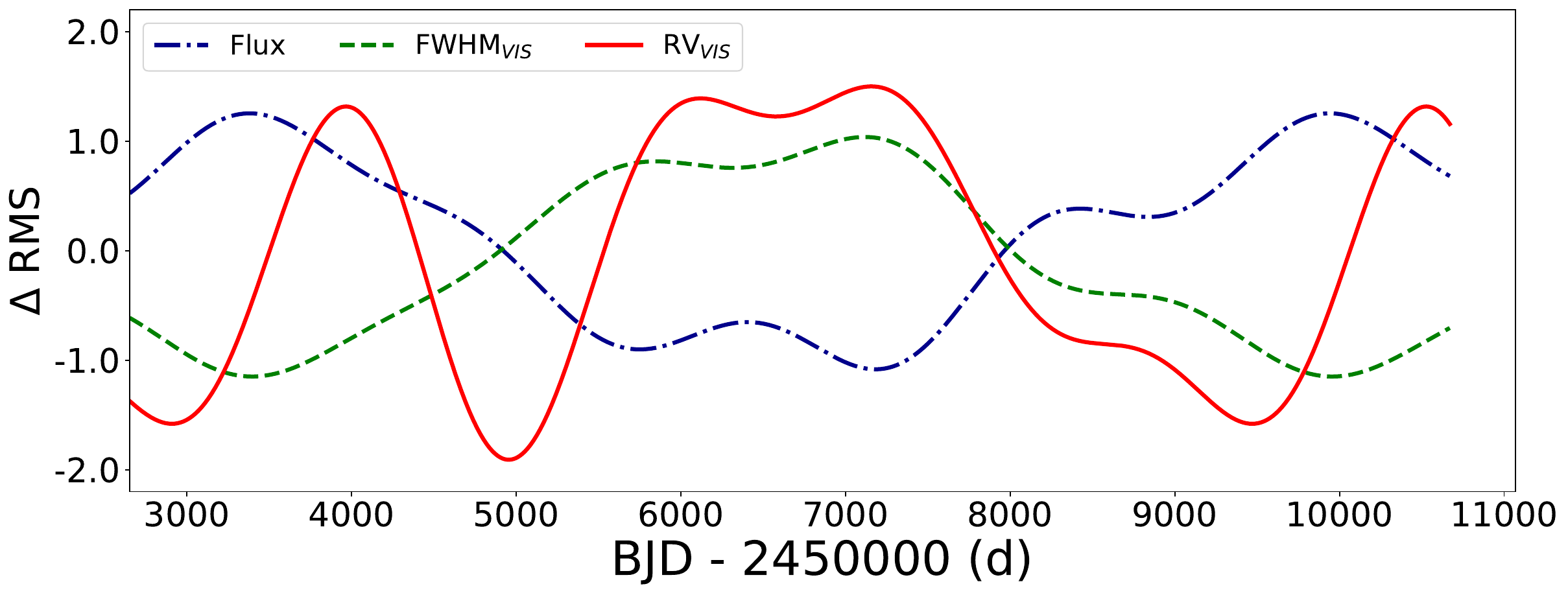}
	\caption{\textbf{Comparison of the cycle models.} Cycle model in Flux, FWHM, and RV, showing the relationship between the components.}
	\label{compare_cycles}
\end{figure}

By modelling the cycle simultaneously in photometry, FWHM, and RV, we can compare its signature in all of them. Figure~\ref{compare_cycles} shows the cycle models in flux, FWHM, and RV, scaled to their own RMS. We find that the cycle in photometry and FWHM behaves in a very similar way, with only small differences in its shape. The opposition in sign shows that when the star becomes brighter, the line width becomes narrower, which is consistent with a spot-dominated stellar surface. In the case of the RV, the variations are somewhat more abrupt than in photometry or FWHM, with even a sign change when the slope of the flux variations changes at around BJD $\sim$ 5000. 

The measured period is approximately twice as long as the one measured by \citet{Wargelin2024}, using a very similar photometric dataset. The difference between both periods arises from differences in data treatment and modelling assumptions. By performing a source-by-source detrending, and assuming a simpler cycle model, we are able to reproduce their results. In our analysis, however, the inclusion of the spectroscopic data reinforces the push towards the longer period. With the current data, it is difficult to firmly establish which of the two sets of assumptions provide a more accurate result. Confidently disentangling both scenarios would require a longer baseline of photometric and spectroscopic observations. 

\subsubsection{Differential rotation} \label{diff_rot}

Stars are not rigid bodies, which means not every part of their surface rotates at the same speed. This effect is primarily driven by the turbulent convection currents within the star. As hot plasma moves from the core to the surface, it carries angular momentum, resulting in different rates of rotation at different latitudes \citep{Spruit2002}. Typically, the angular velocity decreases with increasing latitude, but that is not the case for all stars \citep{Kitchatinov2004}. In the case of the Sun, over the solar cycle, spots appear at high latitudes at the beginning of the cycle, migrating later to the equator as the cycle progresses. In photometry, this latitudinal movement corresponds to a slower rotation period at the beginning of the cycle that gradually increases until the end, before suddenly slowing down again. \citet{Wargelin2017} showed evidence of differential rotation in Proxima. However, the incompleteness of the data made it difficult to fully characterise its evolution.

\begin{figure}[!ht]
    \centering
	\includegraphics[width=9cm]{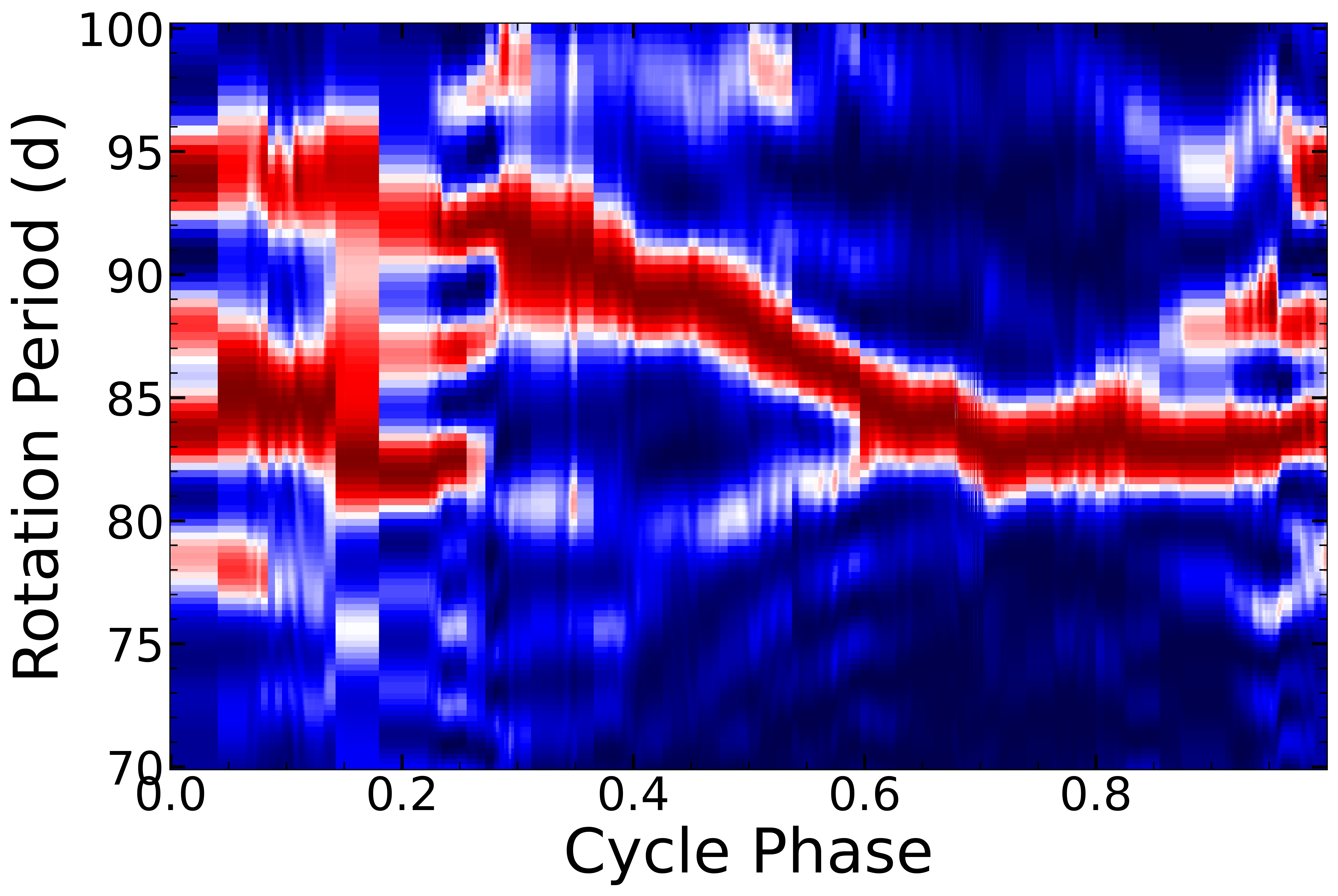}
	\caption{\textbf{Evolution of differential rotation.} Colormap of the evolution of the periodogram power at different periods as a function of the cycle phase. The red shaded regions corresponds to the periods of high periodogram power. The blue shaded regions to the regions of low periodogram power.}
	\label{diff_rot}
\end{figure}

Using the photometric data detrended from the magnetic cycle, with the parameters obtained from the full global model that included all available data, we attempted to measure variations in the photometric period.
We computed a sliding periodogram over the data using a 2000 days window, using a frequency grid between 70 and 100 days. Figure~\ref{diff_rot} shows the map of the periodogram power as a function of the phase of the photometric cycle. We find that the photometric period varies smoothly as the cycle progresses, with a progression that resembles the butterfly diagram of the Sun. We measured multiple periods, between 83 and 95 days, at the beginning of the cycle, when the star is fainter. As the star becomes brighter, only the longer period survives, and it smoothly decreases until stabilising at $\sim$ 83 days for the last third of the cycle.  With these data, we measure a fractional rate of differential rotation of 0.16.

The evolution of the photometric rotation period resembles that of the Sun, with active regions appearing at high latitudes and then migrating towards more equatorial regions as the cycle progresses. However, it is important to remain cautious interpreting the data. The available time series cover less than two periodicities of the cycle, making it difficult to accurately phase-fold the data. The conclusion is based on photometric flux, rather than number of spots present in the disk, which makes it difficult to establish the true beginning of the cycle. In addition, we do not know the inclination of the system. A pole-on stellar alignment might hide the variations related to equatorial spots. 

We attempted the same exercise phase-folding with the period measured by \citet{Wargelin2024}. In this case we did not find a coherent evolution over the cycle phase. We interpret this as evidence of the cycle being closer to 18 years.

\subsubsection{Chromatic variations}

\begin{figure}[!ht]
    \centering
	\includegraphics[width=9cm]{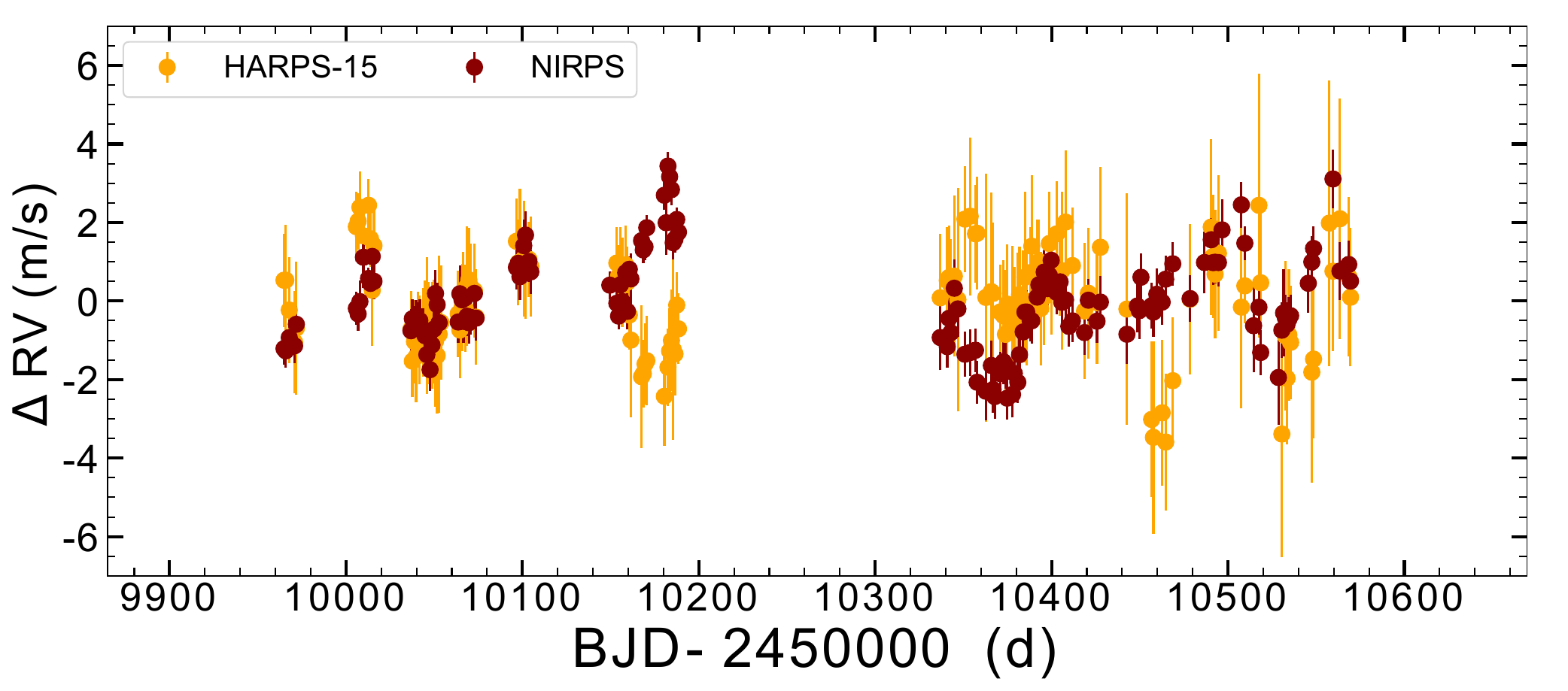}
	\caption{\textbf{Activity-induced RV variations.} Activity-induced RV data from NIRPS and HARPS, after subtracting all planetary and instrumental effects. }
	\label{fig_activity_vis_nir}
\end{figure}

We measured the stellar rotation-induced RV signal of Proxima to have a very small difference in amplitude between the visible and near-infrared data. The rotation variations in HARPS RV data taken simultaneously with the NIRPS data showed an RMS of 1.3 m$\cdot$s$^{-1}$, compared to an RMS of 1.2 m$\cdot$s$^{-1}$ in the near-infrared data. Figure ~\ref{fig_activity_vis_nir} shows the activity component of the RV data in HARPS and NIRPS overplotted. The shape of the variations is slightly different, with some sections in opposition of phase (BJD $\sim$ 10170), and others in phase (BJD $\sim$ 10550). The NIR data shows a clear peak in the periodogram close to the measured rotation period, while the VIS data shows a more complex pattern. The difference in RV amplitude is smaller than what might be expected when extrapolating from past results on active stars ~\citep{Carmona2023}. However, the activity-induced variations of Proxima are much smaller than those of any star when significant chromatic effects have been demonstrated. The results we obtain are closer to those of \citet{CortesZuleta2023}, which found that the stellar activity variations of Gl 205 in the visible and near-infrared are very similar in amplitude, although not neccesarily in shape. 

\subsubsection{Differential temperature variations}

\begin{figure}[!ht]
    \centering
	\includegraphics[width=9cm]{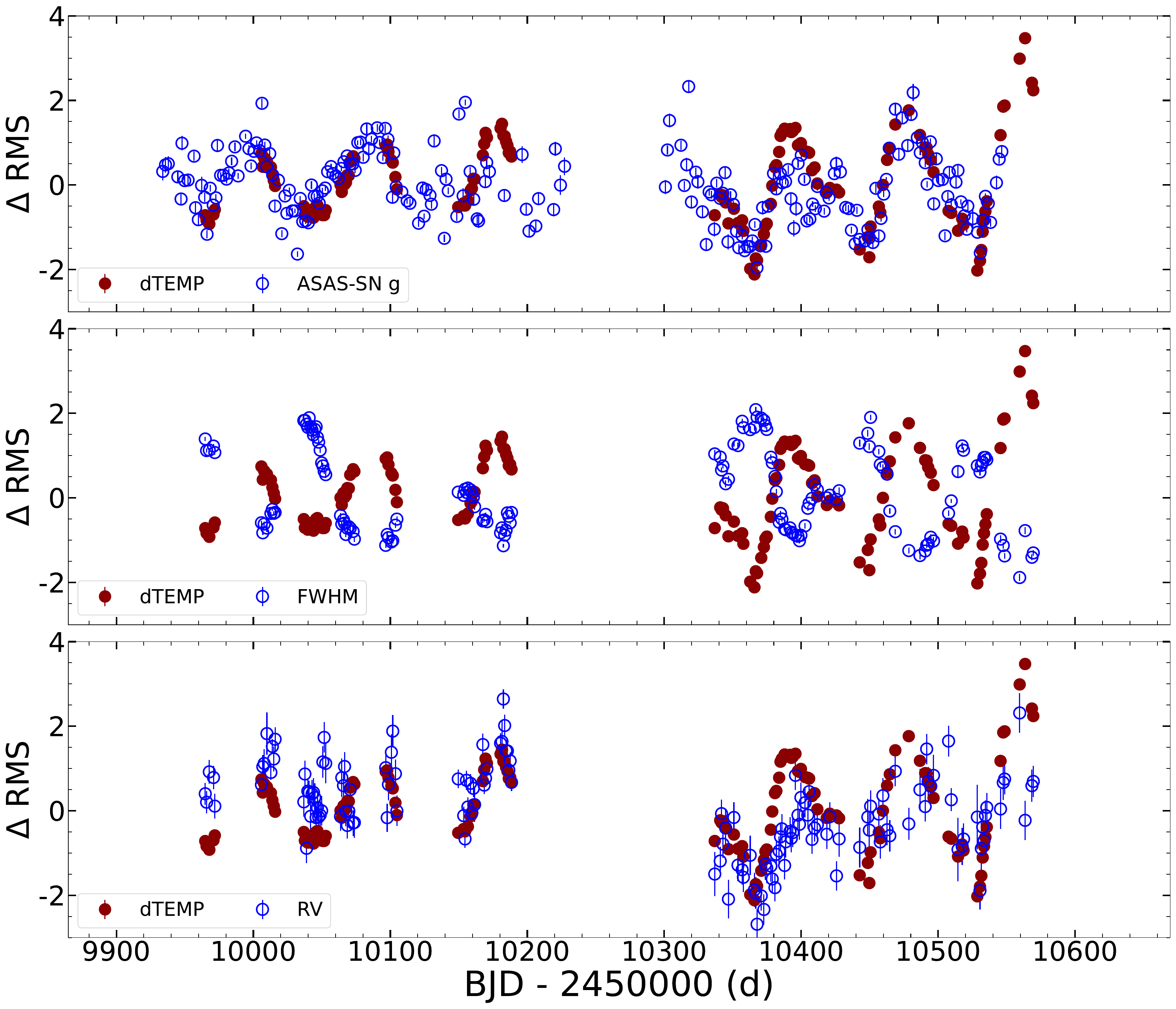}
	\caption{\textbf{Scaled dTEMP variations, compared to other activity proxies.} The top panel shows the variations of the dTEMP compared to the variations in photometric flux. The middle panel shows the variations of the dTEMP compared to the variations in the FWHM. The bottom panel shows the variations of the dTEMP compared to the variations in the RV. All data and errors are scaled to the respective RMS of the displayed time series.}
	\label{fig_dtemp}
\end{figure}

\begin{figure*}[!ht]
    \centering
	\includegraphics[width=18cm]{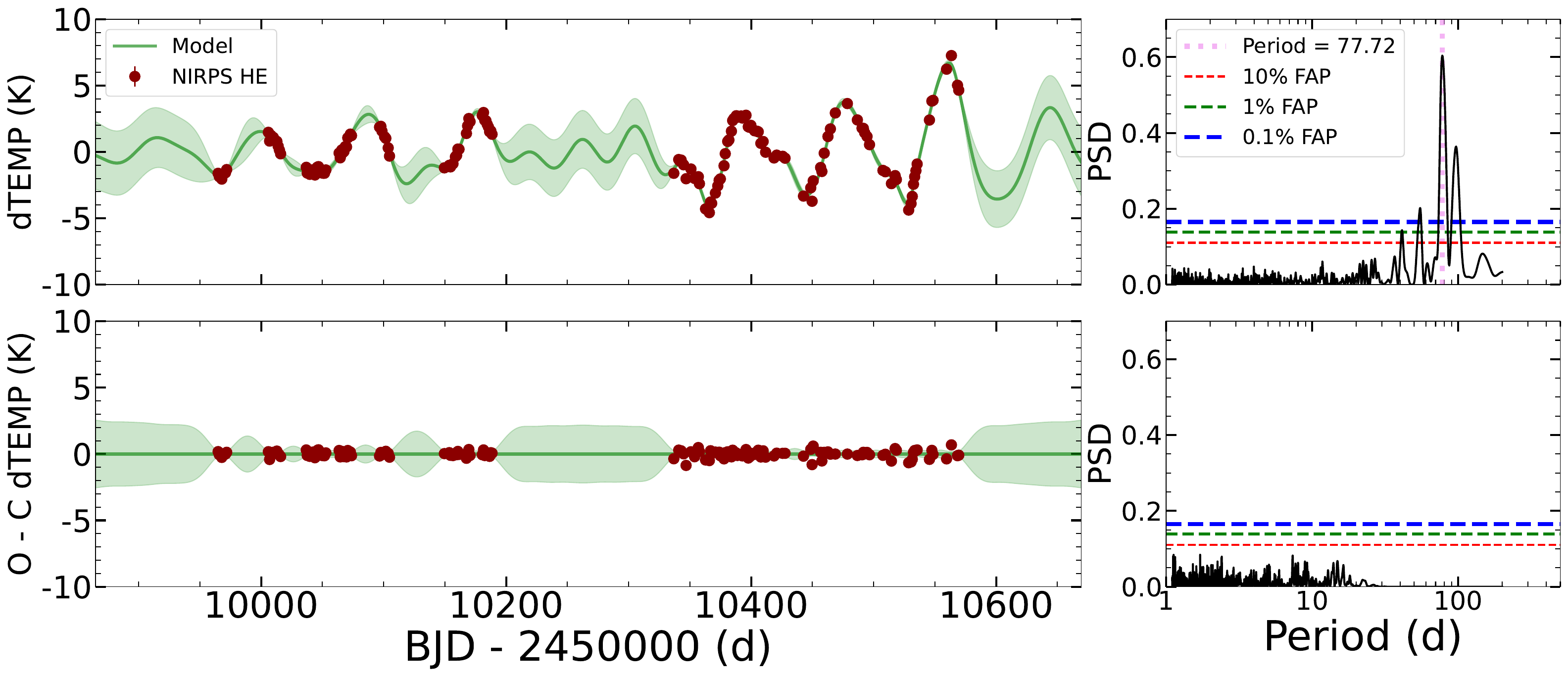}
	\caption{\textbf{NIRPS dTEMP model.} The top row shows the NIRPS dTEMP data, along with the best fit model, and the GLS periodogram of the data. The lower row shows the residuals after the fit, along with their periodogram.}
	\label{fig_model_dtemp}
\end{figure*}

\citet{Artigau2024} introduced a measurement of the differential change in effective temperature, based on small changes in the profile of stellar lines in high resolution spectroscopy, as part of the \texttt{LBL} framework \citet{Artigau2022}. We computed this differential temperature (dTEMP) for the NIRPS data, finding a typical variation of 2.11 K RMS, with a median uncertainty of 0.12 K. The variations show a very stable pattern following the rotation of the star. We compared them with the variations seen in photometry, FWHM, and RVs, corrected from planetary signals. Figure~\ref{fig_dtemp} shows the comparison of these data, all of them scaled to their own RMS. The dTEMP variations show a very similar pattern to the photometric variations, with the same periodicity and almost the exact same shape. At higher temperature we see the star becoming brighter, both implying a smaller filling factor. For Proxima, the dTEMP works as a very direct proxy of stellar flux. Compared to the FWHM, the dTEMP variations are in opposition, but still showing a very similar pattern. This emphasises that the PSF of NIRPS is stable enough that its changes are dominated by stellar variability. As in the case of ESPRESSO, NIRPS FWHM variations work as an inverse proxy of stellar flux \citep{Masca2020}. The activity-induced variations in RV follow these variations in dTEMP with the same periodicity, although the shape is slightly different. The RV variations seem to show a smaller slope over a rotation cycle. However, they are also noisier, making the comparison a bit more difficult. We modelled the dTEMP variation using the same GP-Kernel as used troughout the article. Figure~\ref{fig_model_dtemp} shows the best-fit model, and the residuals after the fit. We recovered similar parameters, although with larger parameter uncertainties, likely due to the lower amount of data. The residuals after the fit show an RMS of 0.25 K.

\section{Conclusions}

We analysed the NIRPS RV data of Proxima, acquired within the context of the NIRPS GTO, the simultaneously obtained HARPS data, and the complete spectroscopic historical dataset spanning 24.5 years of observations. We complemented our analysis with 23.7 years of archival ground-based photometry. We performed a joint model that combined information from photometry, the variations in the width of stellar lines, RVs, variations in the chromatic RV slope, and BERV, into the multi-dimensional Gaussian process framework. 

NIRPS provided an independent detection of Proxima b and evidence for Proxima d. The parameters obtained for Proxima b are consistent with previous studies, and the amplitude of the signal is constrained at a 10-$\sigma$ level. The residuals after subtracting the best-fit model are $\sim$ 80 cm$\cdot$s$^{-1}$, and show that NIRPS outperforms HARPS in RV precision for at least M6-type stars. Combining NIRPS and HARPS data improved detection significance and parameter precision, demonstrating their capability to detect Earth-mass planets around low-mass stars.

The full dataset confirmed Proxima b and d with extremely high confidence (FIP < 0.001\%). Proxima b has a period of 11.18465 $\pm$ 0.00053 days, an RV amplitude of 1.226 $\pm$ 0.062 m$\cdot$s$^{-1}$, and a minimum mass of 1.055 $\pm$ 0.055 M$_{\oplus}$. Proxima d has a period of 5.12338 $\pm$ 0.00035 days, an RV amplitude of 39.2 $\pm$ 5.7 cm$\cdot$s$^{-1}$, and a minimum mass of 0.260 $\pm$ 0.038 M$_{\oplus}$, consistent with previous findings. Combining these results with previous measurements of the system inclination, and assuming coplanarity, we estimate that the true masses of Proxima b, and d, could be 1.44 $\pm$ 0.21 M$_{\oplus}$, and 0.357 $\pm$ 0.072 M$_{\oplus}$, respectively. No conclusive evidence was found for Proxima c, though hints of a lower-amplitude signal with a similar period exist. Taking advantage of the complete dataset, we demonstrate that the parameters of Proxima b and d are consistent between instruments, wavelengths, and over time.

We established compatibility limits for the presence of additional planets in the system. We exclude the presence of planets more massive than $m$ sin $i$ $\sim$ 0.15 M$_{\oplus}$ at periods shorter than 10 days, more massive than 0.3 M$_{\oplus}$ in the habitable zone, 0.6 M$_{\oplus}$ up to 100 days, 1.0 M$_{\oplus}$ up to 1000 days, and to 4 M$_{\oplus}$ up to 10~000 days. Some RV excess at specific periods may warrant further study.

Using 23.7 years of ground-based photometric data, we updated the measurement Proxima's magnetic cycle. We measure a length of 17.73 $\pm$ 0.22 years. The cycle shows significant asymmetries, which create power at several of its harmonics when modelling it by a combination of sine functions. The star's photometric rotation period varies over the cycle, suggesting solar-like differential rotation.

We observe a very small chromatic effect in the activity-induced RV signal of Proxima. The RMS of the activity-induced RVs is $\sim$ 10\% larger in the contemporary HARPS dataset, compared to the NIRPS data.

Last, we find that the newly introduced dTEMP metric correlates well with flux and FWHM measurements, highlighting its potential as an activity proxy.

\begin{acknowledgements} 

We would like to thank Steve Saar, and Erik Mamajek, for useful discussion on Proxima's stellar cycle, and its stellar parameters, respectively.
ASM, JIGH, AKS, RR, FGT, NN, VMP \& JLR  acknowledge financial support from the Spanish Ministry of Science, Innovation and Universities (MICIU) projects PID2020-117493GB-I00 and PID2023-149982NB-I00.\\
\'EA, NJC, RD, TV, CC, PLam, FBa, BB, LMa, RA, LB, AB, AD-B, LD, AL, OL, LMo, JS-A, PV \& JPW  acknowledge the financial support of the FRQ-NT through the Centre de recherche en astrophysique du Qu\'ebec as well as the support from the Trottier Family Foundation and the Trottier Institute for Research on Exoplanets.\\
\'EA, RD, FBa, LMa, TA, J-SM, MO, JS-A \& PV  acknowledges support from Canada Foundation for Innovation (CFI) program, the Universit\'e de Montr\'eal and Universit\'e Laval, the Canada Economic Development (CED) program and the Ministere of Economy, Innovation and Energy (MEIE).\\
XDe, XB, ACar, TF \& VY  acknowledge funding from the French ANR under contract number ANR\-18\-CE31\-0019 (SPlaSH), and the French National Research Agency in the framework of the Investissements d'Avenir program (ANR-15-IDEX-02), through the funding of the ``Origin of Life" project of the Grenoble-Alpes University.\\
TV  acknowledges support from the Fonds de recherche du Qu\'ebec (FRQ) - Secteur Nature et technologies under file 320056.\\
The Board of Observational and Instrumental Astronomy (NAOS) at the Federal University of Rio Grande do Norte's research activities are supported by continuous grants from the Brazilian funding agency CNPq. This study was partially funded by the Coordena\c{c}\~ao de Aperfei\c{c}oamento de Pessoal de N\'ivel Superior—Brasil (CAPES) — Finance Code 001 and the CAPES-Print program.\\
ICL  acknowledges CNPq research fellowships (Grant No. 313103/2022-4).\\
AKS  acknowledges financial support from La Caixa Foundation (ID 100010434) under the grant LCF/BQ/DI23/11990071.\\
SCB, ED-M, NCS, EC, ARCS \& JGd  acknowledge the support from FCT - Funda\c{c}\~ao para a Ci\^encia e a Tecnologia through national funds by these grants: UIDB/04434/2020, UIDP/04434/2020.\\
SCB   acknowledges the support from Funda\c{c}\~ao para a Ci\^encia e Tecnologia (FCT) in the form of a work contract through the Scientific Employment Incentive program with reference 2023.06687.CEECIND.\\
BLCM \& AMM  acknowledge CAPES postdoctoral fellowships.\\
BLCM  acknowledges CNPq research fellowships (Grant No. 305804/2022-7).\\
NBC  acknowledges support from an NSERC Discovery Grant, a Canada Research Chair, and an Arthur B. McDonald Fellowship, and thanks the Trottier Space Institute for its financial support and dynamic intellectual environment.\\
DBF  acknowledges financial support from the Brazilian agency CNPq-PQ (Grant No. 305566/2021-0). Continuous grants from the Brazilian agency CNPq support the STELLAR TEAM of the Federal University of Ceara's research activities.\\
JRM  acknowledges CNPq research fellowships (Grant No. 308928/2019-9).\\
ED-M  further acknowledges the support from FCT through Stimulus FCT contract 2021.01294.CEECIND. ED-M  acknowledges the support by the Ram\'on y Cajal grant RyC2022-035854-I funded by MICIU/AEI/10.13039/501100011033 and by ESF+.\\
XDu  acknowledges the support from the European Research Council (ERC) under the European Union’s Horizon 2020 research and innovation programme (grant agreement SCORE No 851555) and from the Swiss National Science Foundation under the grant SPECTRE (No 200021\_215200).\\
This work has been carried out within the framework of the NCCR PlanetS supported by the Swiss National Science Foundation under grants 51NF40\_182901 and 51NF40\_205606.\\
DE  acknowledge support from the Swiss National Science Foundation for project 200021\_200726. The authors acknowledge the financial support of the SNSF.\\
CMo  acknowledges the funding from the Swiss National Science Foundation under grant 200021\_204847 “PlanetsInTime”.\\
Co-funded by the European Union (ERC, FIERCE, 101052347). Views and opinions expressed are however those of the author(s) only and do not necessarily reflect those of the European Union or the European Research Council. Neither the European Union nor the granting authority can be held responsible for them.\\
RA  acknowledges the Swiss National Science Foundation (SNSF) support under the Post-Doc Mobility grant P500PT\_222212 and the support of the Institut Trottier de Recherche sur les Exoplan\`etes (IREx).\\
LB  acknowledges the support of the Natural Sciences and Engineering Research Council of Canada (NSERC).\\
This project has received funding from the European Research Council (ERC) under the European Union's Horizon 2020 research and innovation programme (project {\sc Spice Dune}, grant agreement No 947634). This material reflects only the authors' views and the Commission is not liable for any use that may be made of the information contained therein.\\
ARCS  acknowledges the support from Funda\c{c}ao para a Ci\^encia e a Tecnologia (FCT) through the fellowship 2021.07856.BD.\\
LD  acknowledges the support of the Natural Sciences and Engineering Research Council of Canada (NSERC) and from the Fonds de recherche du Qu\'ebec (FRQ) - Secteur Nature et technologies.\\
FG  acknowledges support from the Fonds de recherche du Qu\'ebec (FRQ) - Secteur Nature et technologies under file 350366.\\
H.J.H. acknowledges funding from eSSENCE (grant number eSSENCE@LU 9:3), the Swedish National Research Council (project number 2023-05307), The Crafoord foundation and the Royal Physiographic Society of Lund, through The Fund of the Walter Gyllenberg Foundation.\\
AL  acknowledges support from the Fonds de recherche du Qu\'ebec (FRQ) - Secteur Nature et technologies under file 349961.\\
LMo  acknowledges the support of the Natural Sciences and Engineering Research Council of Canada (NSERC), [funding reference number 589653].\\
NN  acknowledges financial support by Light Bridges S.L, Las Palmas de Gran Canaria.\\
NN acknowledges funding from Light Bridges for the Doctoral Thesis "Habitable Earth-like planets with ESPRESSO and NIRPS", in cooperation with the Instituto de Astrof\'isica de Canarias, and the use of Indefeasible Computer Rights (ICR) being commissioned at the ASTRO POC project in the Island of Tenerife, Canary Islands (Spain). The ICR-ASTRONOMY used for his research was provided by Light Bridges in cooperation with Hewlett Packard Enterprise (HPE).\\
CPi  acknowledges support from the NSERC Vanier scholarship, and the Trottier Family Foundation. CPi  also acknowledges support from the E. Margaret Burbidge Prize Postdoctoral Fellowship from the Brinson Foundation.

This work is based on data obtained via the public archive at the European Southern Observatory (ESO). We are grateful to all the observers of the following ESO projects, whose data we are using: 072.C-0488, 082.C-0718, 183.C-0437, 191.C-0505, 096.C-0082, 099.C-0205, 099.C-0880, and 1102.C-0339. We are grateful to the crews at the ESO observatories of Paranal and La Silla. \\
This research has made extensive use of the SIMBAD database, operated at CDS, Strasbourg, France, and NASA's Astrophysics Data System. \\
This research has made use of the NASA Exoplanet Archive, which is operated by the California Institute of Technology, under contract with the National Aeronautics and Space Administration under the Exoplanet Exploration Program. This work makes use of observations from the LCOGT network. \\

The manuscript was written using \texttt{VS Code}. 
Main analysis performed in \texttt{Python3} \citep{Python3} running on \texttt{Ubuntu} \citep{Ubuntu} systems and \texttt{MS. Windows} running the \texttt{Windows subsystem for Linux (WLS)}.
Extensive use of the DACE platform \footnote{\url{https://dace.unige.ch/}}
Extensive usage of \texttt{Numpy} \citep{Numpy}.
Extensive usage of \texttt{Scipy} \citep{Scipy}.
\texttt{AstroimageJ} \citep{Collins:2017}.
\texttt{TAPIR} \citep{Jensen:2013}.
All figures built with \texttt{Matplotlib} \citep{Matplotlib}.
The bulk of the analysis was performed on desktop PC with an AMD Ryzen$^{\rm TM}$ 9 9950X (16 cores, 32 threads, 3.5--4.7 GHz), provided by ASM, and a server hosting 2x AMD Epyc$^{\rm TM}$ 7663 (56 cores, 112 threads, per cpu), provided by the SUBSTELLAR ERC AdG.

\end{acknowledgements}

%
%
\bibliography{biblio_prox}

\begin{appendix}

\onecolumn

\section{Photometric-only Model} \label{append_photonly}

\begin{figure*}[!ht]
    \centering
	\includegraphics[width=18cm]{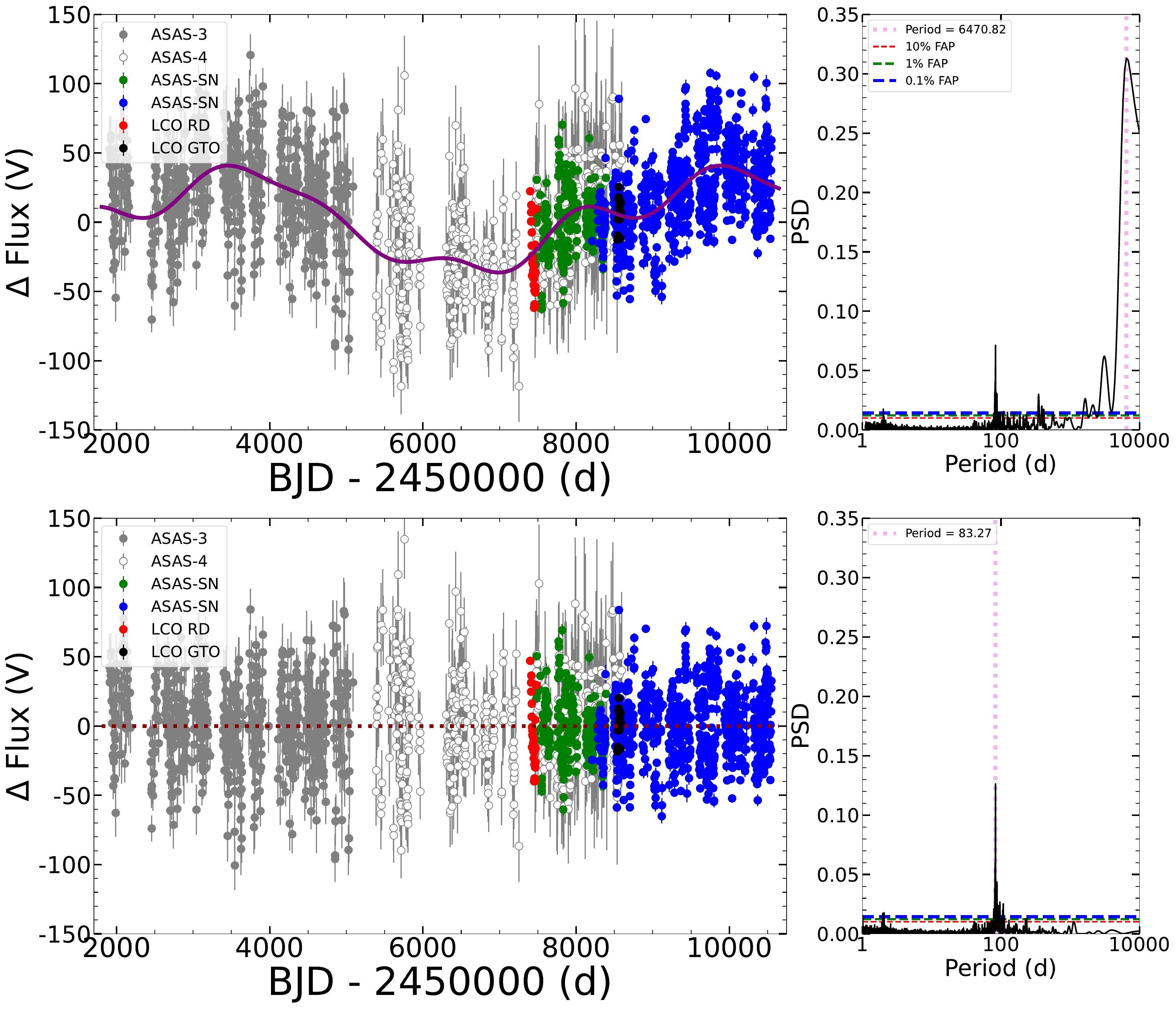}
	\caption{\textbf{Cycle model.} Photometric data with the best-fit model for the cycle, and the residuals after the fit. The right panels show the periodogram of the data and of the residuals. The horizontal lines in the periodograms show the 10\% (red), 1\% (green), and 0.1\% (blue) false alarm probability thresholds.}
	\label{cyc_only}
\end{figure*}

We analysed the photometric data to determine the best model for the cycle and the stellar rotation. We tested several different combinations of polynomials and sinusoidals, until converging to the sum of four sinusoidal signals at periods P$_{\rm cycle}$, P$_{\rm  cycle}$/2, P$_{\rm  cycle}$/3, and P$_{\rm  cycle}$/4. Table~\ref{tab:cyc_comp} shows the comparison of the parameters obtained, and the Bayesian evidence ($\Delta$ Log$Z$), of models using up to five sinusoidals. The model with 4 sinusoidals is preferred ($\Delta$ Log$Z$ > 25 with respect to a single sinusoidal), and all the associated amplitudes are significantly measured. Figure~\ref{cyc_only} shows the best-fit model, the residuals, and the periodograms of both.  Models of lower complexity would leave residual signals with the period of the harmonics of the cycle. The periodogram of the data shows a dominant peak at 6400 days, with the periodogram of the residuals showing a peak at 83 days, corresponding to the stellar rotation. 

\begin{figure*}[!ht]
    \centering
	\includegraphics[width=18cm]{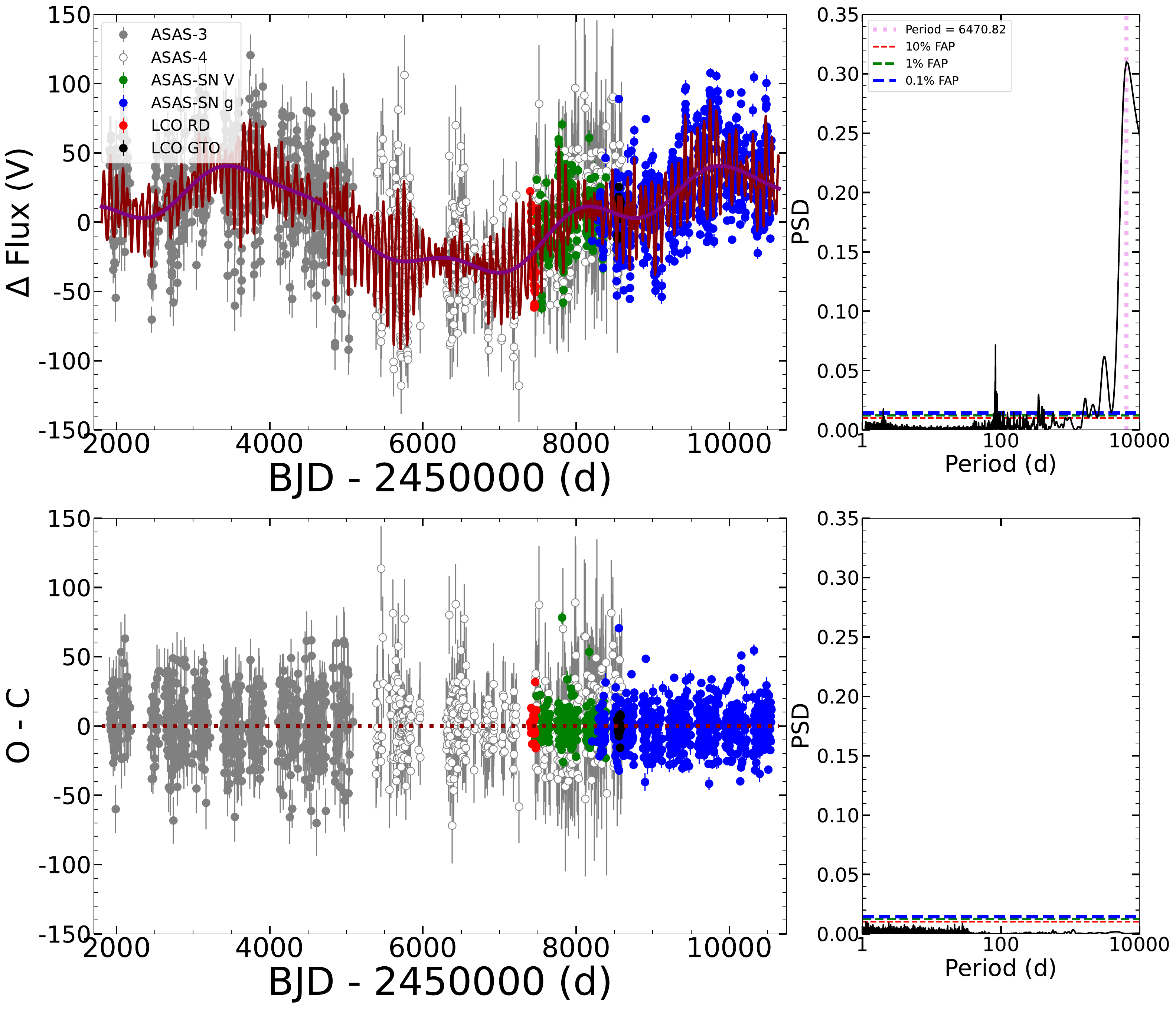}
	\caption{\textbf{Full photometric model.} Photometric data with the best-fit model combining cycle and rotation, and the residuals after the fit. The right panels show the periodogram of the data and of the residuals.}
	\label{gp_phot}
\end{figure*}

We modelled the stellar rotation using Gaussian process regression. We tested different kernels, including several combinations of stochastic harmonic oscillators (SHO), and variations of the Quasi-periodic (QP) kernel. We tried \texttt{S+LEAF}'s MEP, ESP, and ESP with additional harmonic components. We found that all the variations of the QP would supress the cycle signal. The same would happen with combinations of 3 SHO kernels, or more. A single SHO kernel would leave large residuals at the harmonics of the stellar rotation. The combination of 2 SHO kernels, centred at P$_{\rm rot}$, and P$_{\rm rot}$/2, provided a fit that both preserved the amplitudes of the cycle and its harmonic components, and suppressed the rotation signal. We opted for this last case to ensure that the GP model of choice would preserve, as much as possible, the presence of signals at periods longer than the stellar rotation (e.g long period planets in the RV). We performed a blind determination of the cycle period and rotation periods, using Nested Sampling. We used a prior $\mathcal{U}$[2000, 8000] for the cycle, and $\mathcal{U}$[20, 200] for the stellar rotation. 
The best-fit model converted to a cycle period of 6425 $\pm$ 65 days, and a rotation period of 85 $\pm$ 1 days. We used these results, and the five times their uncertainties, as priors in all global models ($\mathcal{N}$[6400,300] days for the cycle, $\mathcal{N}$[85,5] days for the rotation). Figure~\ref{gp_phot} shows the full model, combining the cycle and the GP components.

\begin{table*}
    \begin{center}
        \caption{\textbf{Comparison of cycle complexities} \label{tab:cyc_comp}}
        \begin{tabular}[center]{l l l l l l l l}
            \hline
            N. Sinusoidals & Period [d] & A1 [ppt] & A2 [ppt] & A3 [ppt] & A4 [ppt] & A5 [ppt] & $\Delta$ LogZ*\\
            \hline 
            1 & 6720 $\pm$ 160 & 33.8 $\pm$ 2.3 & &  & & & 0\\
            2 & 6630 $\pm$ 180 & 35.3 $\pm$ 2.2 & 7.0 $\pm$ 1.3 &  & & & 8.4\\
            3 & 6303 $\pm$ 84 & 31.7 $\pm$ 2.0 & 5.4 $\pm$ 1.4 & 7.0 $\pm$ 1.5 & & & 13.9\\
            4 & 6425 $\pm$ 65 & 32.7 $\pm$ 2.0 & 5.6 $\pm$ 1.3 & 6.1 $\pm$ 1.4 & 6.8 $\pm$ 1.1 & & 25.3\\
            5 & 6426 $\pm$ 65 & 32.5 $\pm$ 2.0 & 5.7 $\pm$ 1.3 & 6.1 $\pm$ 1.5 & 6.8 $\pm$ 1.1 & 0.8 $\pm$ 0.8& 23.6\\

            \hline
        \end{tabular}
    \end{center}
    *We take the single-sinusoidal model as reference.
\end{table*}

\clearpage

\section{Presence of the cycle in spectroscopic data} \label{append_cyclespec}

\begin{figure*}[!ht]
    \centering
	\includegraphics[width=16cm]{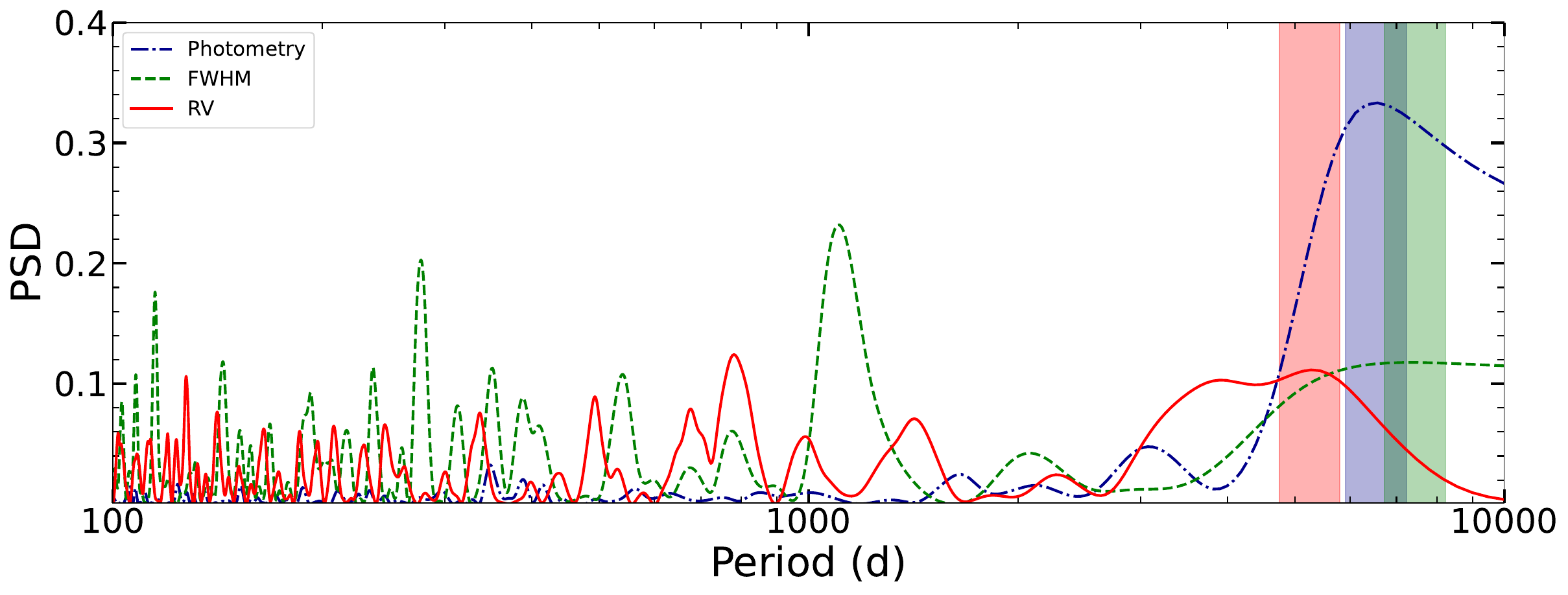}
	\caption{\textbf{GLS periodogram of the photometric and spectroscopic data.} Comparison of the periodograms of the photometric time-series, FWHM, and RV, focused on long periods. The shaded regions show the region of the highest periodogram power at periods longer than 2000 days.}
	\label{cycle_gls}
\end{figure*}

Fig.~\ref{cycle_gls} shows the periodogram of the raw photometric, FWHM, and RV data. The zero-points of the individual datasets have been calculated by matching the median value of contemporary, or nearby, data. The periodograms show peaks, or power excesses, at long periods. These peaks, although poorly defined, are at periods between 5000 and 7000 days. Their origin might be tied to the same cycle seen in photometry. In addition, the periodogram of the FWHM shows several significant peaks (false alarm probability $< $ 0.01) at shorter periods (108d, 270d, 1100d). These peaks correspond to 360d, 120d, and 90d, aliases of the rotation signal. To evaluate the importance of including a model for the stellar cycle in the spectroscopic data, we built a model using the full dataset (photometry, FWHM, and RV) that included all components described in section~\ref{sec_analysis}, except the planetary signals and the cycle model in spectroscopic data. We compared the change of the Bayesian evidence of the models after including the cycle model in FWHM, and then in RV.

We found that including the cycle model in FWHM was favoured ($\Delta$ Log$Z$ $\sim$7), compared to a model in which the cycle only affected the photometric data. We measured a significant amplitude at the main period of the cycle (7.3 $\pm$ 1.8 m$\cdot$s$^{-1}$). For the FWHM, the variations at the harmonic components were not significant. Adding a cycle model in RV is once again favoured against the photometric-only model ($\Delta$ Log$Z$ $\sim$16) and against the model with the cycle in photometry and FWHM ($\Delta$ Log$Z$ $\sim$9). The signal in RV shows significant amplitudes at the main mode (0.86 $\pm$ 0.37 m$\cdot$s$^{-1}$), the third harmonic (1.0 $\pm$ 0.25 m$\cdot$s$^{-1}$) and the fourth harmonic (0.77 $\pm$ 0.21 m$\cdot$s$^{-1}$). In all cases, the measured period remained consistent with the period of the photometry-only model.

With these results, we decided to include the cycle model both in FWHM and RV, following the same approach of the photometric cycle.

\section{Correlation with Chromatic Index} \label{append_crx}

\begin{figure*}[!ht]
    \centering
	\includegraphics[width=18cm]{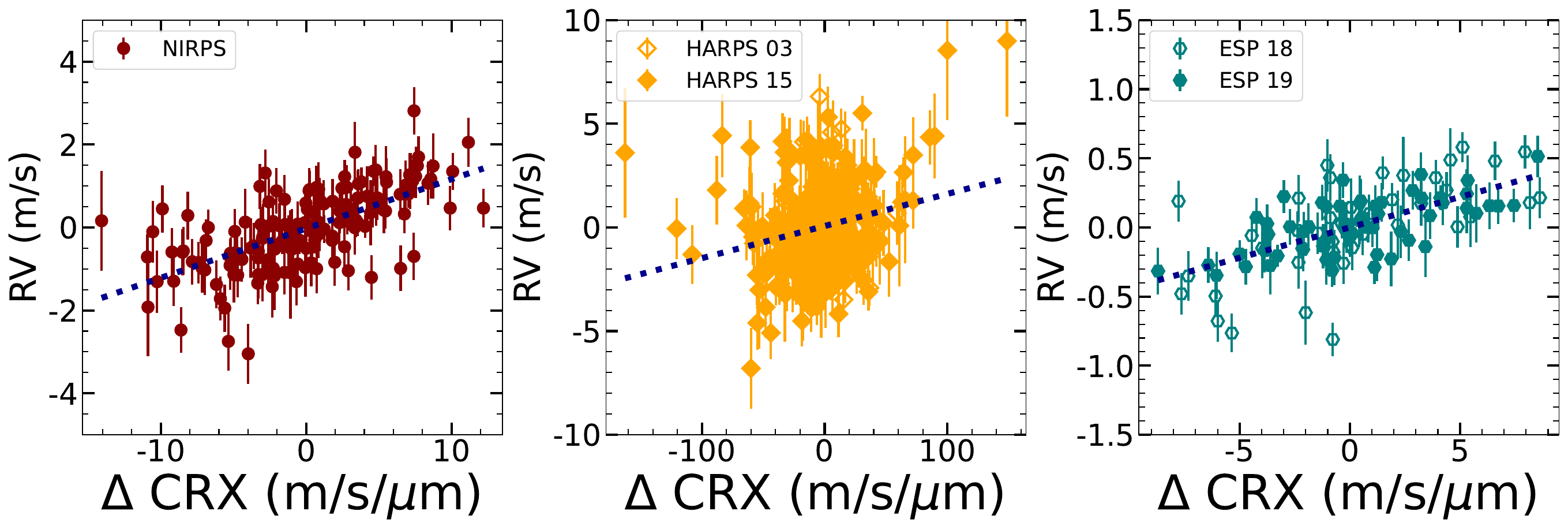}
	\caption{\textbf{RV measurements against CRX.} RV data of NIRPS, HARPS, and ESPRESSO, as a function of the chromatic index.The blue dotted line shows the best linear fit.}
	\label{crx_corr}
\end{figure*}

Most RV data shows a significant trend against the chromatic index. The trend in all cases is positive.Figure~\ref{crx_corr} shows the RV data of the three instruments against their respective CRX. 

\clearpage
\section{Correlation with BERV} \label{append_berv}

\begin{figure}[!ht]
    \centering
	\includegraphics[width=18cm]{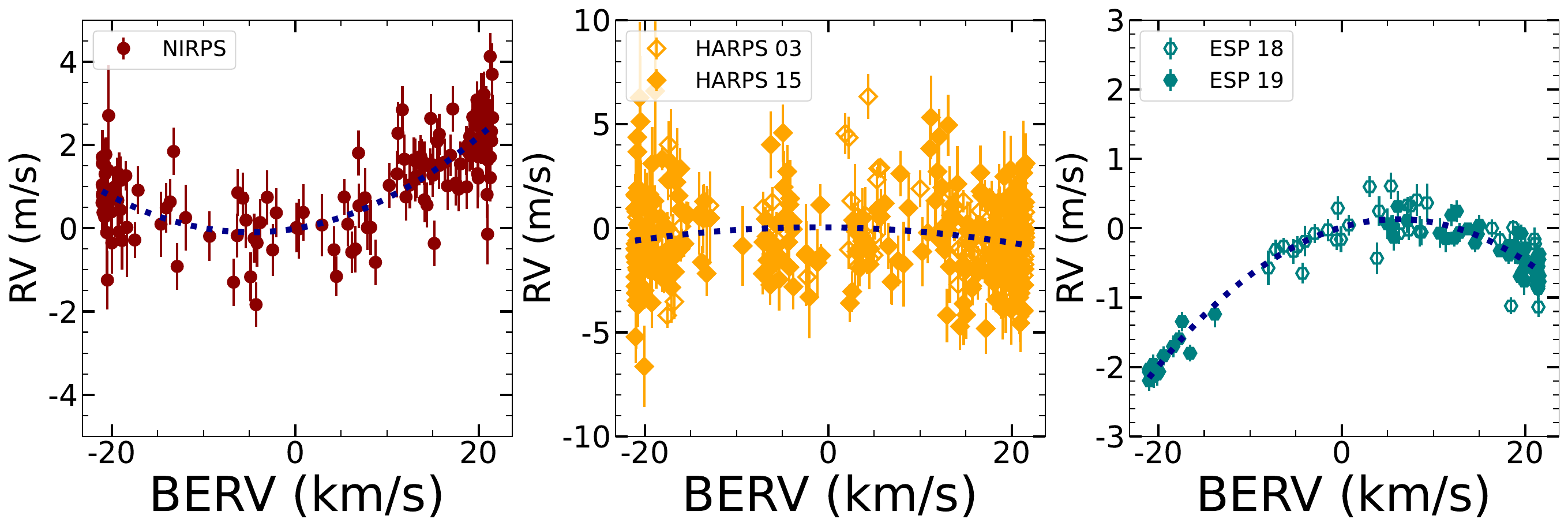}
    \caption{\textbf{RV measurements against BERV.} RV data of NIRPS, HARPS, and ESPRESSO, as a function of the barycentric earth radial velocity.The blue dotted line shows the best linear fit.}
	\label{berv_corr}
\end{figure}

\section{Detection of long period signals}

We could not confirm the detection of the signal attributed to Proxima c. We could only measure an upper limit of the RV amplitude at the proposed period. To evaluate whether this is an indication that the signal does not exists in the data, or of the activity model is suppressing low-frequency signals, we performed simple injection-recovery tests. We conducted two tests in which we injected a long-period signal around the expected period of Proxima c. The injected signals were placed far enough apart that, if the signal of Proxima c exists, it should not interfere significantly with the injected signals. Additionally, the injected periods did not correspond to any harmonics of the stellar cycle. We tested one signal with shorter period (1400 days) and another one with a longer period (2400 days), both with RV amplitudes of 1 m$\cdot$s$^{-1}$ and a $T_{c}$ = 9000 (BJD -- 2450000). We performed a guided search with the exact same setup as before, using priors of $\mathcal{N}[1400,100]$ days, and $\mathcal{N}[2400,200]$ days. In the first case, we recovered a period of 1418 $\pm$ 19 days, a semi-amplitude of 1.17 $\pm$ 0.20 m$\cdot$s$^{-1}$, and a $T_{c}$ = 9030 $\pm$ 50 (BJD -- 2450000).  In the second case, we recovered a period of 2407 $\pm$ 57 days, a semi-amplitude of 1.25 $\pm$ 0.32 m$\cdot$s$^{-1}$, and a $T_{c}$ = 8990 $\pm$ 95 (BJD -- 2450000). With all recovered parameters being 1-$\sigma$ accurate with those injected, it shows we would expect to detect a signal of 1 m$\cdot$s$^{-1}$ at these periods. 

\clearpage

\section{Adopted model - FWHM} \label{append_fwhm}

\begin{figure*}[!ht]
    \centering
	\includegraphics[width=18cm]{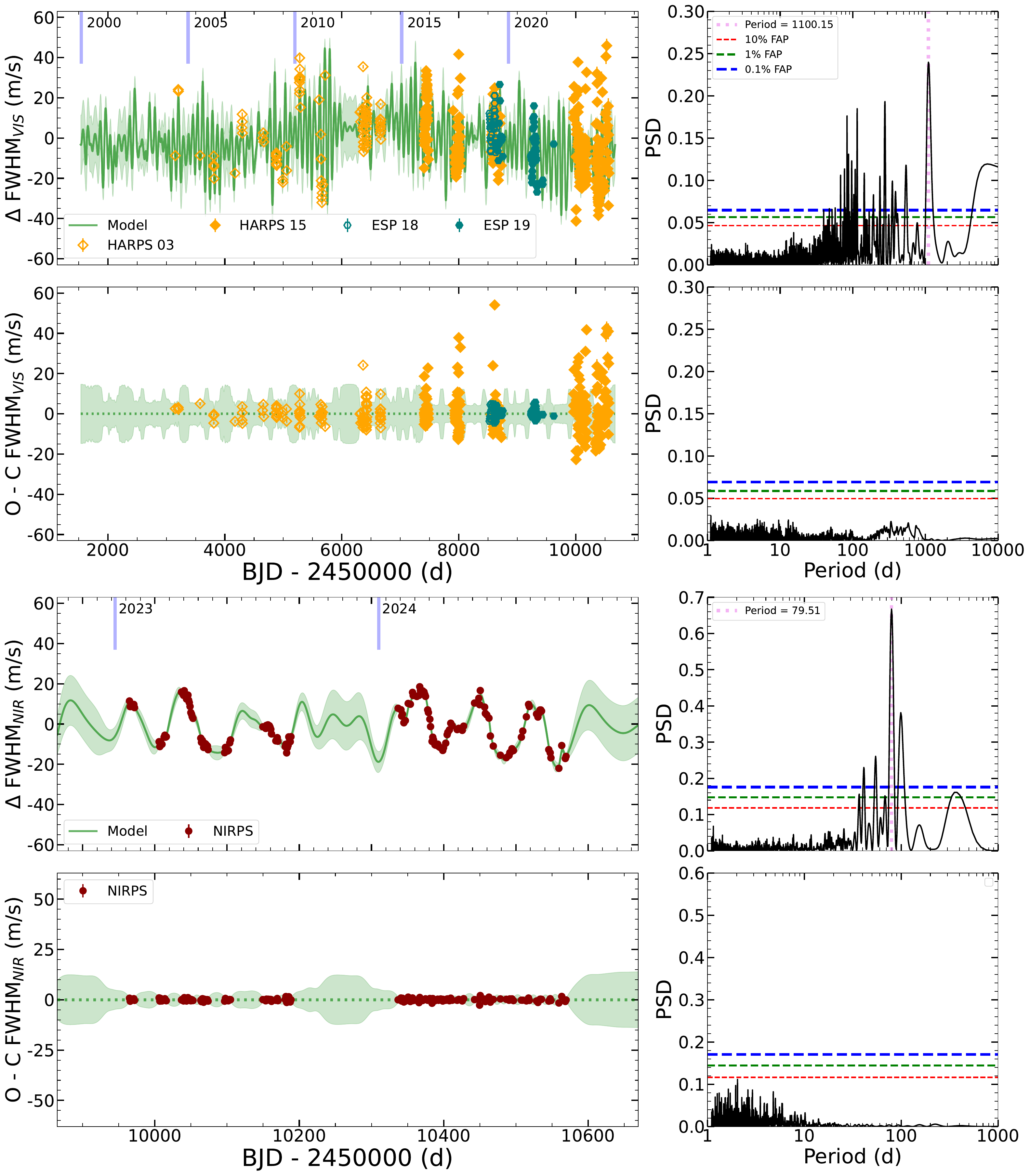}
	\caption{\textbf{FWHM model using the full dataset.} The two top panels show the VIS FWHM data, with the best model fit (top), and the residuals after the fit (bottom), along with the periodograms of both (right). The two bottom panels show the same for the NIR FWHM data.}
	\label{model_all_fwhm}
\end{figure*}

\clearpage

\section{Parameters of the adopted model}\label{append_tables}


\renewcommand{\arraystretch}{1.36}
\begin{longtable}{llc} 
\caption{Priors and measured parameters of the adopted model.} \label{table_adopted} \\
\endfirsthead
\multicolumn{3}{c}{\tablename\ \thetable\ -- \textit{Continued from previous page}} \\
\hline
\endhead
\hline \multicolumn{3}{r}{\textit{Continued on next page}} \\
\endfoot
\hline
\endlastfoot
\hline
Parameter  & Priors & Posterior \\ \hline

\textbf{White noise} \\
ln $\sigma$ Phot$_{ASAS-3}$ [ppt] &  $\mathcal{N}$ (3.5 , 3.5) & 3.082$^{+0.035}_{-0.034}$ \\
ln $\sigma$ Phot$_{ASAS-4}$ [ppt] &  $\mathcal{N}$ (3.5 , 3.5) & 2.840$^{+0.054}_{-0.056}$ \\
ln $\sigma$ Phot$_{LCO~RD}$ [ppt] &  $\mathcal{N}$ (3.5 , 3.5) & 2.38$^{+0.16}_{-0.15}$ \\
ln $\sigma$ Phot$_{LCO~2020}$ [ppt] &  $\mathcal{N}$ (3.5 , 3.5) & 1.90$^{+0.19}_{-0.17}$ \\
ln $\sigma$ Phot$_{ASAS-SN V}$ [ppt] &  $\mathcal{N}$ (3.5 , 3.5) & 2.277$^{+0.063}_{-0.061}$ \\
ln $\sigma$ Phot$_{ASAS-SN g}$ [ppt] &  $\mathcal{N}$ (3.5 , 3.5) & 2.657$^{+0.305}_{-0.308}$ \\
\\ 
ln $\sigma$ FWHM$_{NIRPS}$ [m~s$^{-1}$]  &  $\mathcal{N}$ (2.3 , 2.3) & --0.24$^{+0.25}_{-0.43}$ \\
ln $\sigma$ FWHM$_{HARPS~03}$ [m~s$^{-1}$]  &  $\mathcal{N}$ (2.3 , 2.3) & 1.728$^{+0.088}_{-0.085}$ \\
ln $\sigma$ FWHM$_{HARPS~15}$ [m~s$^{-1}$]  &  $\mathcal{N}$ (2.3 , 2.3) & 2.402$^{+0.049}_{-0.048}$ \\
ln $\sigma$ FWHM$_{ESPRESSO~18}$ [m~s$^{-1}$]  &  $\mathcal{N}$ (2.3 , 2.3) & 1.18$^{+0.14}_{-0.14}$ \\
ln $\sigma$ FWHM$_{ESPRESSO~19}$ [m~s$^{-1}$]  &  $\mathcal{N}$ (2.3 , 2.3) & 0.70$^{+0.13}_{-0.13}$ \\
\\ 
ln $\sigma$ RV$_{NIRPS}$ [m~s$^{-1}$]  &  $\mathcal{N}$ (0.5 , 0.5) & --0.55$^{+0.11}_{-0.13}$ \\
ln $\sigma$ RV$_{HARPS~03}$ [m~s$^{-1}$]  &  $\mathcal{N}$ (0.5 , 0.5) & 0.50$^{+0.12}_{-0.12}$ \\
ln $\sigma$ RV$_{HARPS~15}$ [m~s$^{-1}$]  &  $\mathcal{N}$ (0.5 , 0.5) & 0.428$^{+0.074}_{-0.074}$ \\
ln $\sigma$ RV$_{ESPRESSO~18}$ [m~s$^{-1}$]  &  $\mathcal{N}$ (0.5 , 0.5) & --0.86$^{+0.20}_{-0.22}$ \\
ln $\sigma$ RV$_{ESPRESSO~19}$ [m~s$^{-1}$]  &  $\mathcal{N}$ (0.5 , 0.5) & --1.22$^{+0.21}_{-0.24}$ \\
ln $\sigma$ RV$_{UVES}$ [m~s$^{-1}$]  &  $\mathcal{N}$ (0.5 , 0.5) & -0.08$^{+0.20}_{-0.23}$ \\
\\
\textbf{Zero points} \\
V0 Phot$_{ASAS-3}$ [ppt] &  $\mathcal{N}$ (0 , 100) &   --23.2$^{+2.4}_{-2.4}$ \\
V0 Phot$_{ASAS-4}$ [ppt] &  $\mathcal{N}$ (0 , 100) &   11.0$^{+2.0}_{-2.0}$ \\
V0 Phot$_{LCO~RD}$ [ppt] &  $\mathcal{N}$ (0 , 100) &   42.0$^{+4.9}_{-5.0}$ \\
V0 Phot$_{LCO~2020}$ [ppt] &  $\mathcal{N}$ (0 , 100) &   --10.3$^{+2.9}_{-2.9}$ \\
V0 Phot$_{ASAS-SN V}$ [ppt] &  $\mathcal{N}$ (0 , 100) &   0.41$^{+2.3}_{-2.2}$ \\
V0 Phot$_{ASAS-SN g}$ [ppt] &  $\mathcal{N}$ (0 , 100) &   --23.8$^{+1.9}_{-1.9}$ \\
\\
V0 FWHM$_{NIRPS}$ [m~s$^{-1}$] &  $\mathcal{N}$ (0 , 40) & 2.8$^{+1.9}_{-1.7}$\\
V0 FWHM$_{HARPS~03}$ [m~s$^{-1}$] &  $\mathcal{N}$ (0 , 40) & --3.7$^{+1.8}_{-1.8}$ \\
V0 FWHM$_{HARPS~15}$ [m~s$^{-1}$] &  $\mathcal{N}$ (0 , 40) & 4.4$^{+1.5}_{-1.5}$\\
V0 FWHM$_{ESPRESSO~18}$ [m~s$^{-1}$] &  $\mathcal{N}$ (0 , 40) & --5.8$^{+1.9}_{-1.9}$\\
V0 FWHM$_{ESPRESSO~19}$ [m~s$^{-1}$] &  $\mathcal{N}$ (0 , 40) & 7.2$^{+2.2}_{-2.2}$ \\
\\
V0 RV$_{NIRPS}$ [m~s$^{-1}$] &  $\mathcal{N}$ (0 , 20) & --0.84$^{+0.25}_{-0.26}$ \\
V0 RV$_{HARPS~03}$ [m~s$^{-1}$] &  $\mathcal{N}$ (0 , 20) & 0.74$^{+0.44}_{-0.45}$\\
V0 RV$_{HARPS~15}$ [m~s$^{-1}$] &  $\mathcal{N}$ (0 , 20) & 0.38$^{+0.34}_{-0.34}$\\
V0 RV$_{ESPRESSO~18}$ [m~s$^{-1}$] &  $\mathcal{N}$ (0 , 20) & 1.06$^{+0.34}_{-0.33}$\\
V0 RV$_{ESPRESSO~19}$ [m~s$^{-1}$] &  $\mathcal{N}$ (0 , 20) & 1.83$^{+0.44}_{-0.44}$\\
V0 RV$_{UVES}$ [m~s$^{-1}$] &  $\mathcal{N}$ (0 , 20) & 0.47$^{+0.27}_{-0.27}$\\

\\
\\
\textbf{Polynomial parameters} \\
a CRX$_{NIRPS}$ [\textmu$^{-1}$] &  $\mathcal{N}$ (0 , 0.3) & 0.122$^{+0.019}_{-0.019}$ \\
a CRX$_{HARPS}$ [\textmu$^{-1}$] &  $\mathcal{N}$ (0 , 0.3) & 0.0136$^{+0.0041}_{-0.0042}$ \\
a CRX$_{ESPRESSO}$ [\textmu$^{-1}$] &  $\mathcal{N}$ (0 , 0.3) & 0.041$^{+0.023}_{-0.023}$ \\
\\
a BERV$_{NIRPS}$ [m~s$^{-1}$ / km~s$^{-1}$] &  $\mathcal{N}$ (0 , 0.3) & 0.0360$^{+0.0081}_{-0.0078}$ \\
b BERV$_{NIRPS}$ [m~s$^{-1}$ / km$^{2}$~s$^{-2}$] &  $\mathcal{N}$ (0 , 0.3) & 0.00379$^{+0.00068}_{-0.00065}$ \\
a BERV$_{HARPS}$ [m~s$^{-1}$ / km~s$^{-1}$] &  $\mathcal{N}$ (0 , 0.3) & --0.0023$^{+0.0098}_{-0.0102}$ \\
b BERV$_{HARPS}$ [m~s$^{-1}$ / km$^{2}$~s$^{-2}$] &  $\mathcal{N}$ (0 , 0.3) & --0.00176$^{+0.00096}_{-0.00095}$ \\
a BERV$_{ESPRESSO}$ [m~s$^{-1}$ / km~s$^{-1}$] &  $\mathcal{N}$ (0 , 0.3) & 0.038$^{+0.011}_{-0.011}$ \\
b BERV$_{ESPRESSO}$ [m~s$^{-1}$ / km$^{2}$~s$^{-2}$] &  $\mathcal{N}$ (0 , 0.3) & --0.00306$^{+0.00087}_{-0.00091}$ \\ 

\\
\textbf{Cycle parameters} \\
Cycle period [d] &  $\mathcal{N}$ (6400 , 300) & 6560$^{+85}_{-82}$ \\ 
Phase$_{P}$    & $\mathcal{N}$ (-0.5 , 1) & 0.140$^{+0.012}_{-0.013}$\\
Phase$_{P/2}$    & $\mathcal{N}$ (-0.5 , 1) & 0.154$^{+0.045}_{-0.046}$\\
Phase$_{P/3}$    & $\mathcal{N}$ (-0.5 , 1) & 0.211$^{+0.048}_{-0.043}$ \\
Phase$_{P/4}$    & $\mathcal{N}$ (-0.5 , 1) & --0.217$^{+0.038}_{-0.039}$\\
K Phot$_{P}$ [ppt] &  $\mathcal{U}$ (0 , 50) &  32.7 $^{+2.2}_{-2.3}$\\
K Phot$_{P/2}$ [ppt] &  $\mathcal{U}$ (0 , 50) & 2.8 $^{+2.0}_{-1.7}$ \\
K Phot$_{P/3}$ [ppt] &  $\mathcal{U}$ (0 , 50) & 5.1 $^{+2.1}_{-2.1}$ \\
K Phot$_{P/4}$ [ppt] &  $\mathcal{U}$ (0 , 50) & 7.0 $^{+1.5}_{-1.6}$ \\
\\
K FWHM V$_{P}$ [m~s$^{-1}$] &  $\mathcal{N}$ (0 , 10) &  --6.3$^{+1.7}_{-1.8}$\\
K FWHM V$_{P/2}$ [m~s$^{-1}$] &  $\mathcal{N}$ (0 , 10) & --0.26$^{+1.4}_{-1.4}$  \\
K FWHM V$_{P/3}$ [m~s$^{-1}$] &  $\mathcal{N}$ (0 , 10) & --0.80$^{+1.5}_{-1.5}$ \\
K FWHM V$_{P/4}$ [m~s$^{-1}$] &  $\mathcal{N}$ (0 , 10) & --0.85$^{+1.5}_{-1.4}$ \\
ln Scale FWHM NIR [m~s$^{-1}$] &  $\mathcal{U}$ (-5 , 3) & --2.7$^{+1.6}_{-1.5}$ \\
\\
K RV V$_{P}$ [m~s$^{-1}$] &  $\mathcal{N}$ (0 , 10) & --0.89$^{+0.30}_{-0.30}$ \\
K RV V$_{P/2}$ [m~s$^{-1}$] &  $\mathcal{N}$ (0 , 10) & 0.98$^{+0.20}_{-0.20}$ \\
K RV V$_{P/3}$ [m~s$^{-1}$] &  $\mathcal{N}$ (0 , 10) & 0.63$^{+0.22}_{-0.22}$ \\
K RV V$_{P/4}$ [m~s$^{-1}$] &  $\mathcal{N}$ (0 , 10) & --0.56$^{+0.18}_{-0.18}$ \\
ln Scale RV NIR [m~s$^{-1}$] &  $\mathcal{U}$ (-5 , 3) &  --3.65$^{+1.17}_{-0.92}$ \\

\\
\textbf{GP Parameters} \\
Rotation period [d] &  $\mathcal{N}$ (85 , 5) & 83.2$^{+1.6}_{-1.6}$ \\ 
ln Timescale [d] &  $\mathcal{U}$ (3, 10) & 4.07$^{+0.18}_{-0.17}$ \\ 
A$_{11}$ Phot [ppt] &  $\mathcal{U}$ (0 , 50) & 22.1$^{+1.4}_{-1.3}$ \\
A$_{12}$ Phot [ppt] &  $\mathcal{U}$ (0 , 50) & 8.9$^{+1.2}_{-1.2}$ \\
A$_{21}$ FWHM$_{NIR}$ [m~s$^{-1}$] &  $\mathcal{U}$ (-50 , 50) & -12.2$^{+1.1}_{-1.1}$ \\
A$_{22}$ FWHM$_{NIR}$ [m~s$^{-1}$] &  $\mathcal{U}$ (-50 , 50) & -5.5$^{+1.2}_{-1.3}$  \\
B$_{21}$ FWHM$_{NIR}$ [m] &  $\mathcal{U}$ (-100 , 100) & 39$^{+11}_{-10}$ \\
B$_{22}$ FWHM$_{NIR}$ [m] &  $\mathcal{U}$ (-100 , 100) & --22.7$^{+8.3}_{-7.8}$  \\
A$_{31}$ FWHM$_{VIS}$ [m~s$^{-1}$] &  $\mathcal{U}$ (-50 , 50) & -14.6$^{+1.0}_{-1.0}$ \\
A$_{32}$ FWHM$_{VIS}$ [m~s$^{-1}$] &  $\mathcal{U}$ (-50 , 50) & -0.29$^{+1.1}_{-1.1}$\\
B$_{31}$ FWHM$_{VIS}$ [m] &  $\mathcal{U}$ (-100 , 100) & --28$^{+12}_{-12}$ \\
B$_{32}$ FWHM$_{VIS}$ [m] &  $\mathcal{U}$ (-100 , 100) &  8.9$^{+5.0}_{-5.1}$\\
A$_{41}$ RV$_{NIR}$ [m~s$^{-1}$] &  $\mathcal{U}$ (-50 , 50) & 0.59$^{+0.22}_{-0.22}$ \\
A$_{42}$ RV$_{NIR}$ [m~s$^{-1}$] &  $\mathcal{U}$ (-50 , 50) & 1.33$^{+0.24}_{-0.21}$\\
B$_{41}$ RV$_{NIR}$ [m] &  $\mathcal{U}$ (-100 , 100) &  --11.4$^{+2.2}_{-2.3}$\\
B$_{42}$ RV$_{NIR}$ [m] &  $\mathcal{U}$ (-100 , 100) & --0.9$^{+1.9}_{-1.9}$ \\
A$_{51}$ RV$_{VIS}$ [m~s$^{-1}$] &  $\mathcal{U}$ (-50 , 50) & 0.32$^{+0.15}_{-0.14}$  \\
A$_{52}$ RV$_{VIS}$ [m~s$^{-1}$] &  $\mathcal{U}$ (-50 , 50) & --1.05$^{+0.27}_{-0.26}$ \\
B$_{51}$ RV$_{VIS}$ [m] &  $\mathcal{U}$ (-100 , 100) & --8.6$^{+2.2}_{-2.2}$ \\
B$_{52}$ RV$_{VIS}$ [m] &  $\mathcal{U}$ (-100 , 100) & --8.3$^{+1.4}_{-1.4}$ \\

\\
\textbf{Planets} \\
Period$_{b}$ [d] &  $\mathcal{U}$ (9.0,13.5) & 11.18465$^{+0.00052}_{-0.00053}$ \\
ln K$_{b}$ [m~s$^{-1}$] &  $\mathcal{U}$ (-5 , 2) & 0.204$^{+0.050}_{-0.052}$\\
Phase$_{b}$  &  $\mathcal{U}$ (-0.25 , 0.75) & --0.020 $^{+0.011}_{-0.011}$ \\
Period$_{d}$ [d] &  $\mathcal{U}$ (4.1,6.1) & 5.12338$^{+0.00035}_{-0.00035}$ \\
ln K$_{d}$ [m~s$^{-1}$] &  $\mathcal{U}$ (-5 , 2) & --0.94$^{+0.14}_{-0.15}$ \\
Phase$_{d}$  &  $\mathcal{U}$ (-0.25 , 0.75) & 0.522 $^{+0.031}_{-0.031}$ \\

\hline
\end{longtable}

\clearpage

\section{No transits in the TESS data}\label{append_tables}

\begin{figure}[!ht]
    \centering
    \begin{minipage}{0.4\textwidth}
	    \includegraphics[width=7.4cm]{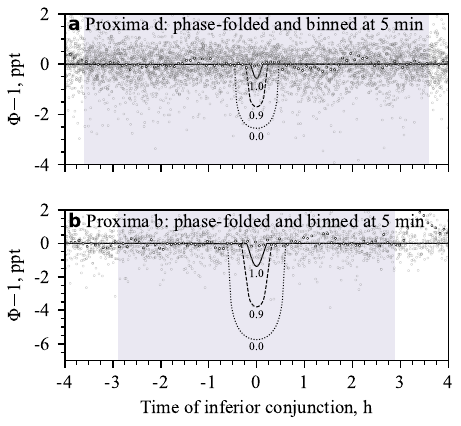}
    \end{minipage}
    \begin{minipage}{0.59\textwidth}
        \includegraphics[width=10.7cm]{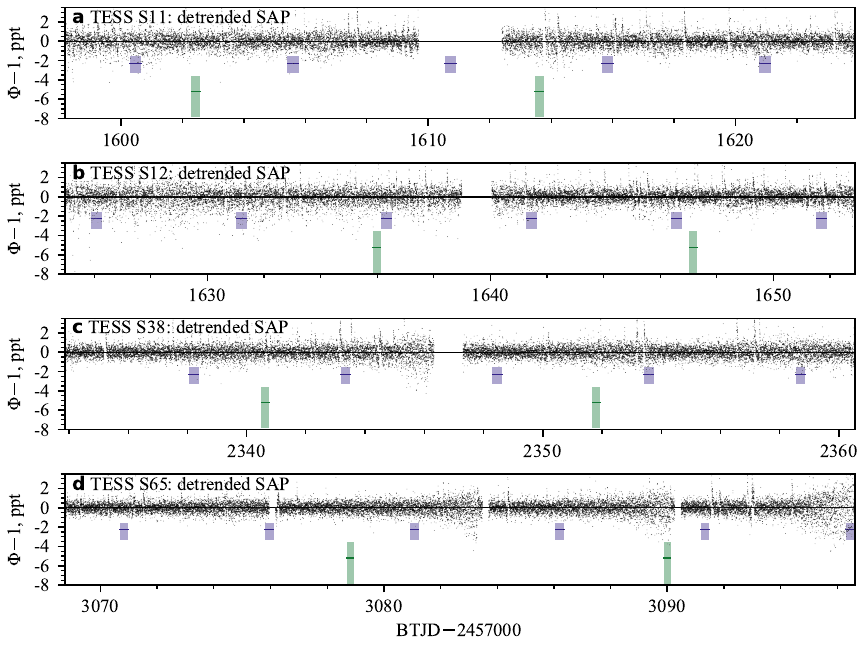}
    \end{minipage}
	\caption{\textbf{Phase-folded TESS data.} Left panel shows the detrended SAP photometry folded with the periods, and ephemerides, of Proxima d (top panel), and Proxima b (bottom panel), with the expected transit signatures for different impact parameters. The shaded region shows the uncertainty in the ephemeris. Right panel shows the TESS SAP flux of the four available sectors, with the dates and predicted depths of the potential transiting events highlighted. The green horizontal lines, and shaded regions, mark the potential transits of Proxima b. The lilac lines, and shaded regions, mark the confidence interval of the potential transits of Proxima d.}
	\label{tess_folded}
\end{figure}

Since the discovery of Proxima b, there have been several attempts at detecting one of its transits. \citet{Jenkins2019} first reported a non-detection, using Spitzer observations. Later, \citet{Gilbert2021} reported the same using data from the first available TESS \citep{Ricker2015} sectors. Using the four TESS sectors currently available (11, 12, 38, and 65) we repeat the exercise to confirm whether it is possible to detect a  transit of Proxima b, or Proxima d, in the data. Following \citep{Stefanov2025}, we downloaded the lightcurves processed by the Science Processing Operations Center (SPOC) pipeline \citep{Jenkins2016}. We took all 2 min SAP measurements from all four sectors, using the hard cadence quality bitmask. Then, we smoothed the time series of each sector independently with the Tukey's biweight time-window slider \citep{Mosteller1977} through the \texttt{W{\={o}}tan} library \citep{Hippke2019wotan}. We used a window length of 6 h, a break tolerance of 12 h and no edge cut-offs. After detrending each sector individually and then combining them together, we could find no evidence of transits. Figure~\ref{tess_folded}, left panel, shows the phase-folded plot of the detrended TESS photometry, along with the expected transit signatures of both planets, for different impact parameters. Figure~\ref{tess_folded}, right panel, shows the time series of the detrended TESS photometry, with the dates of the expected transits highlighted. To this date, TESS has observed during eight potential transit dates of Proxima b, and 21 of Proxima d. In addition, we analysed the data looking for potential transit events with a period of 2.91d, as that signal appears both in the FIP periodogram of section~\ref{full_spec_data} and ~\ref{det_lims}. We did not find any event consistent with this period. 

\section{Effect of changing the GP Kernel}

In addition to the combination of two SHO kernels at P$_{\rm Rot}$, and P$_{\rm Rot}$/2, as presented through all the text, we re-run the adopted model using a single SHO (P$_{\rm Rot}$), three SHOs (P$_{\rm Rot}$, P$_{\rm Rot}$/2, and P$_{\rm Rot}$/4), the MEP kernel, the ESP kernel, and the ESP kernel with 4 harmonic components. The MEP and ESP Kernels are approximations of the traditional Quasi-Periodic kernel, written in the form of semi-separable matrices for increased computation efficiency (see ~\citealt{Delisle2022}). Except for the case of the single SHO, which would significantly underfit the data, the results of the parameters of planets b and d were consistent (within 1$\sigma$) in all cases. The biggest difference when using the combination of three SHOs, or the MEP/ESP kernels, was that they would suppress most low-frequency variations, independently of their origin. The most obvious effect would be in a reduction of the measured amplitude of the cycle, as the GP would absorbe part of that variation and to incorporate it into its own shape. 

\end{appendix}
\label{lastpage}

\end{document}